\documentclass[manuscript,screen]{acmart}
\usepackage{subfig}
\usepackage{caption}
\usepackage{hyperref}
\usepackage{multirow}
\usepackage{makecell}
\usepackage{mathrsfs}
\usepackage{amsmath,amsfonts}
\usepackage{algorithmic}
\usepackage{algorithm}
\usepackage{graphicx}
\usepackage{tabularx}
\usepackage{booktabs}
\usepackage{longtable}
\usepackage[most]{tcolorbox}

\AtBeginDocument{%
  }

\setcopyright{none}
\renewcommand\footnotetextcopyrightpermission[1]{}
\newcommand{\eg}{\textit{e.g.},~}

\begin{document}

\title{Triage in Software Engineering: A Systematic Review of Research and Practice}

\author{Yongxin Zhao}
\email{zyx\_nkcs@mail.nankai.edu.cn}
\affiliation{%
  \institution{Nankai University}
  \country{China}
}

\author{Shenglin Zhang}
\authornote{Shenglin Zhang and Minghua Ma are the corresponding authors.}
\email{zhangsl@nankai.edu.cn}
\affiliation{%
  \institution{Nankai University}
  \country{China}
}

\author{Yujia Wu}
\email{2120240813@mail.nankai.edu.cn}
\affiliation{%
  \institution{Nankai University}
  \country{China}
}

\author{Yuxin Sun}
\email{2111498@mail.nankai.edu.cn}
\affiliation{%
  \institution{Nankai University}
  \country{China}
}

\author{Yongqian Sun}
\email{sunyongqian@nankai.edu.cn}
\affiliation{%
  \institution{Nankai University}
  \country{China}
}

\author{Dan Pei}
\email{peidan@tsinghua.edu.cn}
\affiliation{%
  \institution{Tsinghua University}
  \country{China}
}

\author{Chetan Bansal}
\email{chetanb@microsoft.com}
\affiliation{%
  \institution{Microsoft}
  \country{USA}
}

\author{Minghua Ma${\footnotemark[1]}$}
\email{minghuama@microsoft.com}
\affiliation{%
  \institution{Microsoft}
  \country{USA}
}

\renewcommand{\shortauthors}{Yongxin Zhao et al.}

\begin{abstract}

As modern software systems continue to grow in complexity, triage has become a fundamental process in system operations and maintenance. Triage aims to efficiently prioritize, assign, and assess issues to ensure the reliability of complex environments. The vast amount of heterogeneous data generated by software systems has made effective triage indispensable for maintaining reliability, facilitating maintainability, and enabling rapid issue response. Motivated by these challenges, researchers have devoted extensive effort to advancing triage automation and have achieved significant progress over the past two decades. This survey provides a comprehensive review of 234 papers from 2004 to the present, offering an in-depth examination of the fundamental concepts, system architecture, and problem statement. By comparing the distinct goals of academic and industrial research and by analyzing empirical studies of industrial practices, we identify the major obstacles that limit the practical deployment of triage systems. To assist practitioners in method selection and performance evaluation, we summarize widely adopted open-source datasets and evaluation metrics, providing a unified perspective on the measurement of triage effectiveness. Finally, we outline potential future directions and emerging opportunities to foster a closer integration between academic innovation and industrial application. All reviewed papers and projects are available at \url{https://github.com/AIOps-Lab-NKU/TriageSurvey}.


\end{abstract}

\ccsdesc[500]{General and reference~Surveys and overviews}
\ccsdesc[500]{Computer systems organization~Maintainability and maintenance}
\ccsdesc[500]{Computing methodologies~Artificial intelligence}

\keywords{Triage, software engineering, bug reports, incident tickets, alerts}


\maketitle

\section{Introduction}

\begin{figure}
    \centering
    \includegraphics[width=0.9\linewidth]{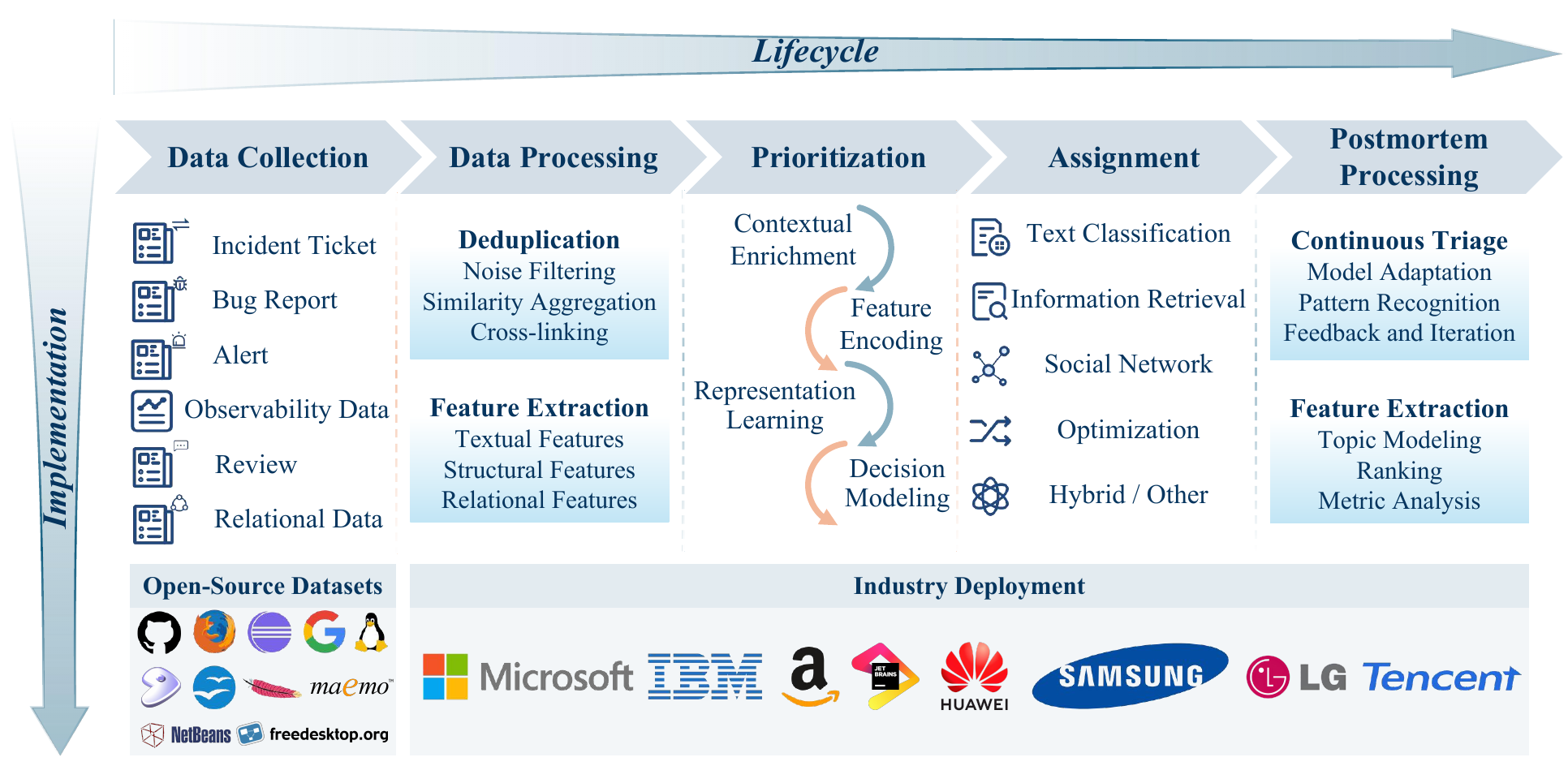}
    \caption{A conceptual map of automated triage research and practice.}
    \vspace{-0.5cm}
    \label{fig:overall}
\end{figure}

As modern software systems and large-scale online services continue to evolve in complexity, the volume of software bugs, incidents, and alerts has grown exponentially. If not addressed promptly, such issues can lead to severe operational disruptions and substantial financial losses~\cite{zhao2020understanding, wang2024comet, wei2025improving}. Effective issue management has thus become indispensable to ensuring the stability and reliability of large-scale systems. When an issue report is submitted, it initiates a critical, multi-step process known as \textit{triage}. Broadly defined, triage encompasses a sequence of analytical activities aimed at efficiently managing the lifecycle of an issue. As shown in Figure~\ref{fig:overall}, the process involves identifying duplicates, prioritizing the issue's urgency, classifying the issue's type (\eg bug, feature request, or security vulnerability), and routing the issue to the most appropriate entity for resolution. This entity may be a specific developer, a component team, or an automated analysis pipeline.

Manual triage, however, is both time-consuming and labor-intensive, particularly given the sheer volume of reports in large-scale projects~\cite{wang2024comet}. For example, the Eclipse project receives approximately 91 bug reports per day~\cite{uddin2024novel}, while Mozilla handles around 300 bug reports daily~\cite{dipongkor2023comparative}. Moreover, triagers in such environments must allocate reports across numerous developers~\cite{lee2022light}, yet it is unrealistic to expect them to possess a comprehensive understanding of each developer's technical expertise, domain familiarity, and current workload~\cite{dao2023automated}. Consequently, manual triage often results in repeated report reassignment, commonly referred to as ``bug tossing'', until a suitable resolver is identified. This process not only delays issue resolution but also diminishes operational efficiency and user satisfaction. Given these challenges, the adoption of automated triage mechanisms has become critical to maintaining the reliability, availability, and scalability of modern software systems. Understanding the role, objectives, and advantages of triage, therefore, provides essential context for developing effective, intelligent solutions that enhance issue resolution efficiency and optimize resource allocation.

\subsection{Why Triage in Software Engineering?}  
In contemporary software engineering (SE), the ever-increasing volume and complexity of incident tickets, bug submissions, and system alerts have rendered triage a cornerstone process for ensuring operational efficiency and system reliability~\cite{chen2019continuous, zhao2020alertrank, su2021reducing, wang2024comet}. Effective triage, spanning incident triage, bug triage, and alert aggregation, delivers several key benefits:

\begin{itemize}  
    \item \textbf{Accelerated issue resolution.} By systematically prioritizing, classifying, and routing reports, triage reduces manual workload and mitigates the ``report tossing'' phenomenon, where reports are repeatedly reassigned until they reach an appropriate resolver~\cite{jeong2009improving, su2021reducing}. This streamlining significantly shortens the time to resolution and minimizes service disruption.
    
    \item \textbf{Enhanced diagnostic accuracy.} Automated triage mechanisms effectively filter out duplicates, irrelevant, or low-quality reports~\cite{liu2023ipack, neysiani2020efficient, catolino2019not}, ensuring that engineers work with concise, actionable information. This improves diagnostic precision and prevents redundant investigations.  

    \item \textbf{Facilitated cross-team collaboration.} By providing a consolidated and structured view of issues, triage promotes effective communication and knowledge sharing among development, operations, and support teams. This is particularly critical in large-scale or community-driven projects, where issues are typically assigned to teams or modules rather than individual engineers~\cite{banitaan2013decoba, jonsson2016automated}. 
    
    \item \textbf{Foundation for intelligent automation.} The structured outputs generated during triage serve as high-quality input for machine learning and knowledge-based reasoning systems. These outputs facilitate the automation of downstream tasks such as root cause analysis, failure prediction, and recommendation of remediation actions~\cite{remil2024aiops, yu2024survey}, thereby enabling proactive and intelligent system management. 
     
\end{itemize} 

Despite these advantages, practical triage remains a complex and challenging endeavor. The heterogeneity of input data, the coexistence of structured and unstructured information, and the intricate dependencies among triage subtasks impose significant constraints on full automation. As a result, the development of systematic, data-driven, and explainable triage approaches has become a pressing need for modern SE practices.

\vspace{-0.3cm}
\subsection{Why a Survey of Triage in Software Engineering?}  
\begin{figure}
    \centering
    \subfloat[Publication distribution over time (2004–Present).\label{fig:sub1}]{\includegraphics[width=0.5\columnwidth]{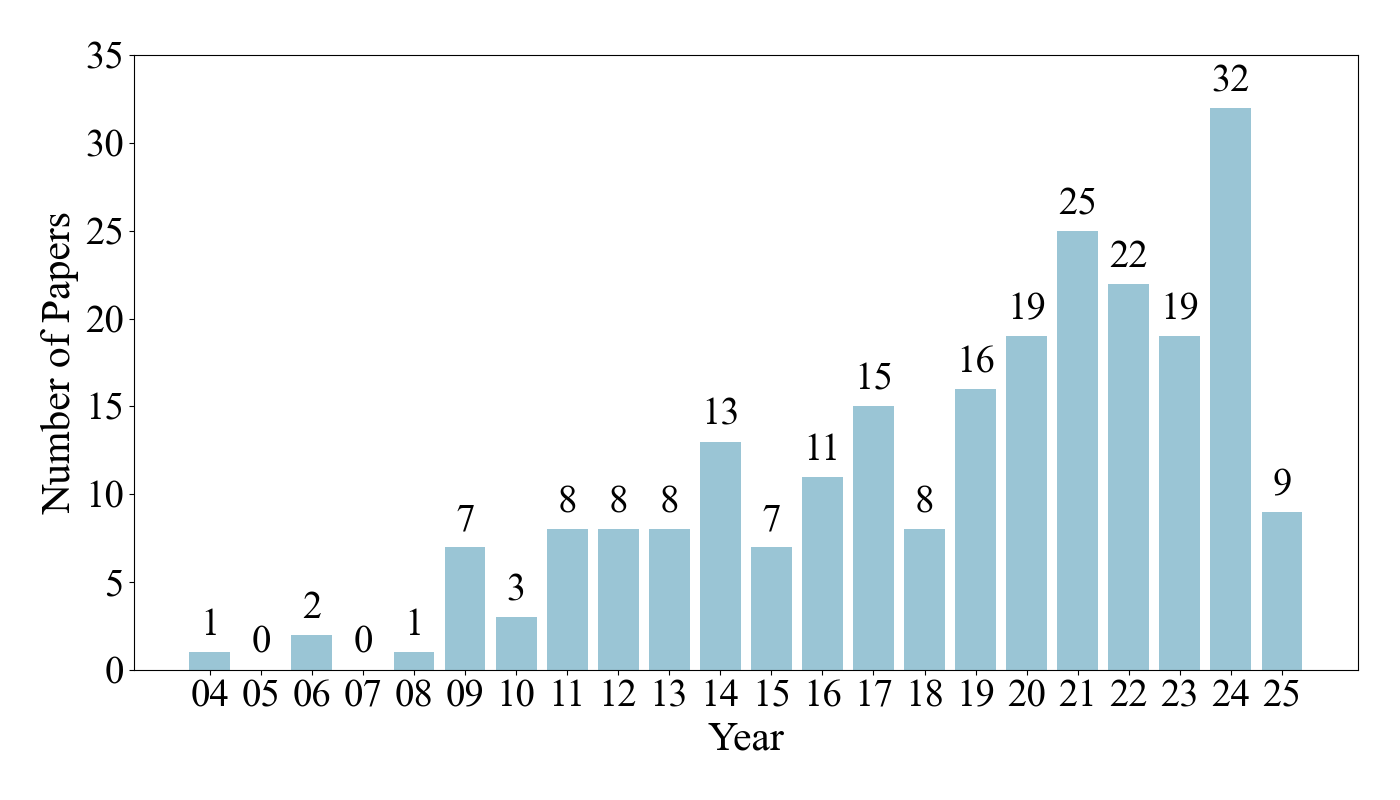}}
    \hfill
    \subfloat[Publication distribution involving industry collaboration (2004–Present).\label{fig:sub2}]{\includegraphics[width=0.5\columnwidth]{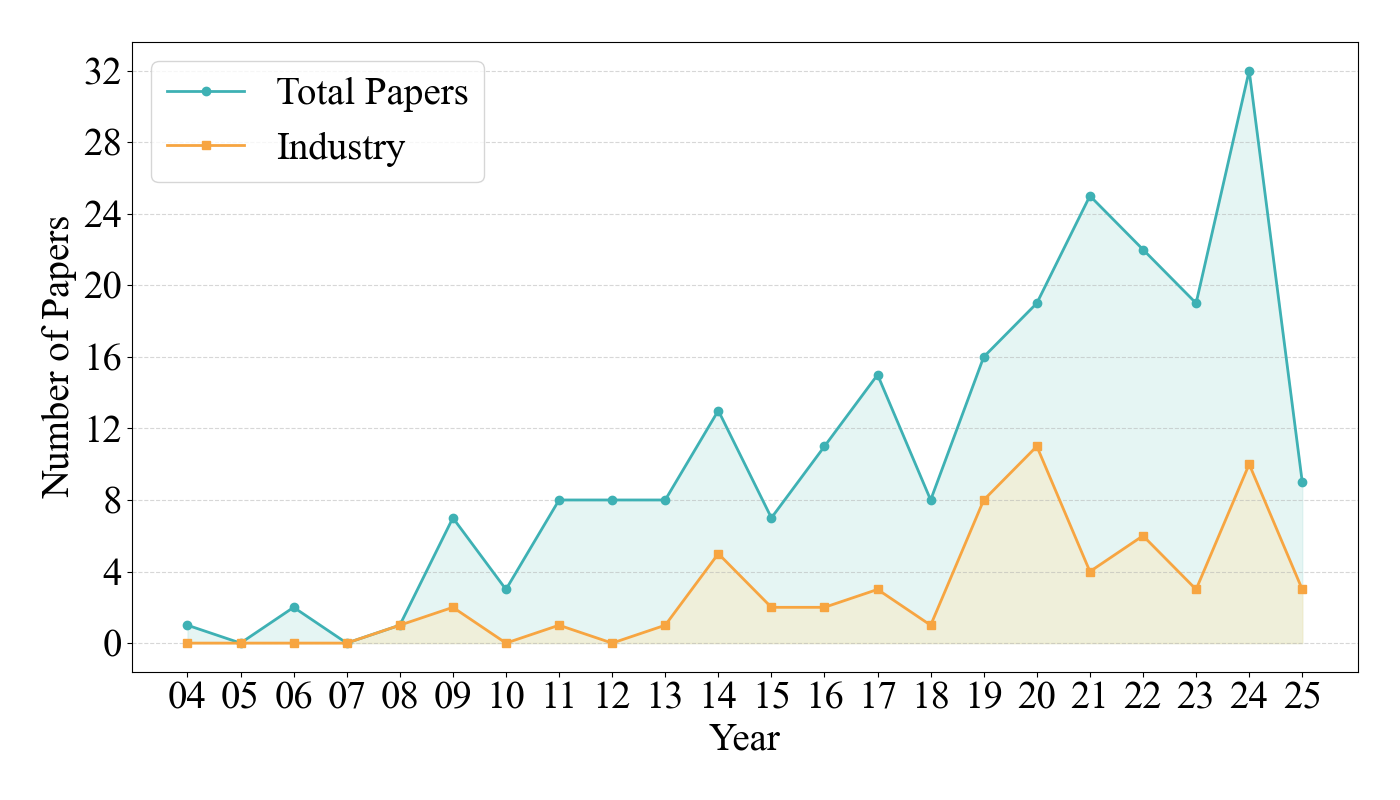}}
    \caption{Analysis of Publication Trends on Triage in Software Engineering.}
    \vspace{-0.3cm}
    \label{fig:pub_trend}
\end{figure}

With the escalating complexity of software ecosystems and the growing prevalence of incidents, research on automated triage has witnessed rapid progress and increasing scholarly attention, as illustrated in Figure~\ref{fig:sub1}. A diverse range of studies has explored various subtasks, such as bug classification, duplicate detection, and incident management, reflecting both the practical importance and the technical challenges inherent in triage. Furthermore, Figure~\ref{fig:sub2} highlights the increasing number of triage-related publications produced in collaboration with industry partners, underscoring the real-world applicability and impact of this line of research.

However, existing surveys often adopt a fragmented perspective, typically focusing on individual subtasks rather than viewing triage as an integrated, end-to-end process. For instance, some reviews focus exclusively on developer assignment~\cite{nagwani2023artificial}, while others examine incident triage within cloud environments~\cite{remil2024aiops, yu2024survey}. This fragmented view overlooks the dependencies and interactions among subtasks, for example, how duplicate detection~\cite{neysiani2020efficient} informs component assignment~\cite{su2021reducing}, or how bug tossing knowledge~\cite{jeong2009improving} enhances developer recommendation~\cite{wu2022spatial}. Such isolation hinders a holistic understanding of triage as a cohesive, interdependent process that spans multiple organizational and technical dimensions.

To bridge this gap, this survey provides a unified and comprehensive examination of triage across the full lifecycle. We systematically review techniques encompassing data preprocessing, prioritization, assignment, and postmortem analysis, offering an integrative perspective on how these components interact. This holistic approach not only reveals methodological trends and open research challenges but also provides actionable insights for practitioners seeking to design or enhance triage pipelines in real-world systems. By consolidating fragmented knowledge into a coherent framework, this survey aims to advance both theoretical understanding and practical implementation of triage in modern software engineering.

\subsection{Research Questions.}
With the growing complexity of modern software systems, triage has become an essential process for effectively managing and prioritizing issues. Over the past two decades, research on triage in software engineering has advanced considerably, spanning diverse contexts such as bug report assignment, incident management, and alert prioritization. Despite this progress, the existing body of work remains fragmented, characterized by inconsistent definitions, heterogeneous methodologies, and varying scopes of application. In light of the rapid proliferation of data-driven and AI-assisted approaches, a comprehensive and systematic review of triage research in software engineering is both timely and necessary.

To this end, we structure our survey around the following research questions, each targeting a fundamental dimension of triage in software engineering:

\begin{itemize}
    \item \textbf{RQ1:} How have existing triage methods evolved across the different stages of the triage lifecycle?
    \item \textbf{RQ2:} What practical challenges and limitations are encountered during the deployment of triage processes in real-world settings?
    \item \textbf{RQ3:} How is the effectiveness of triage approaches evaluated, and what metrics and benchmarks are commonly adopted?
\end{itemize}

These research questions form a coherent taxonomy that captures the conceptual and practical evolution of triage research. RQ1 provides a lifecycle-oriented synthesis of existing methods, mapping the progression of triage techniques across various phases. RQ2 examines the operational and organizational challenges that emerge in real-world deployments, offering insights into gaps between research and practice. Building upon these foundations, RQ3 focuses on evaluation methodologies, highlighting how different metrics and benchmarks are used to assess triage effectiveness. Collectively, this structured framework enables a systematic exploration of triage in software engineering, bridging conceptual understanding, methodological development, and empirical evaluation.

\begin{figure}
    \centering
    \includegraphics[width=0.75\linewidth]{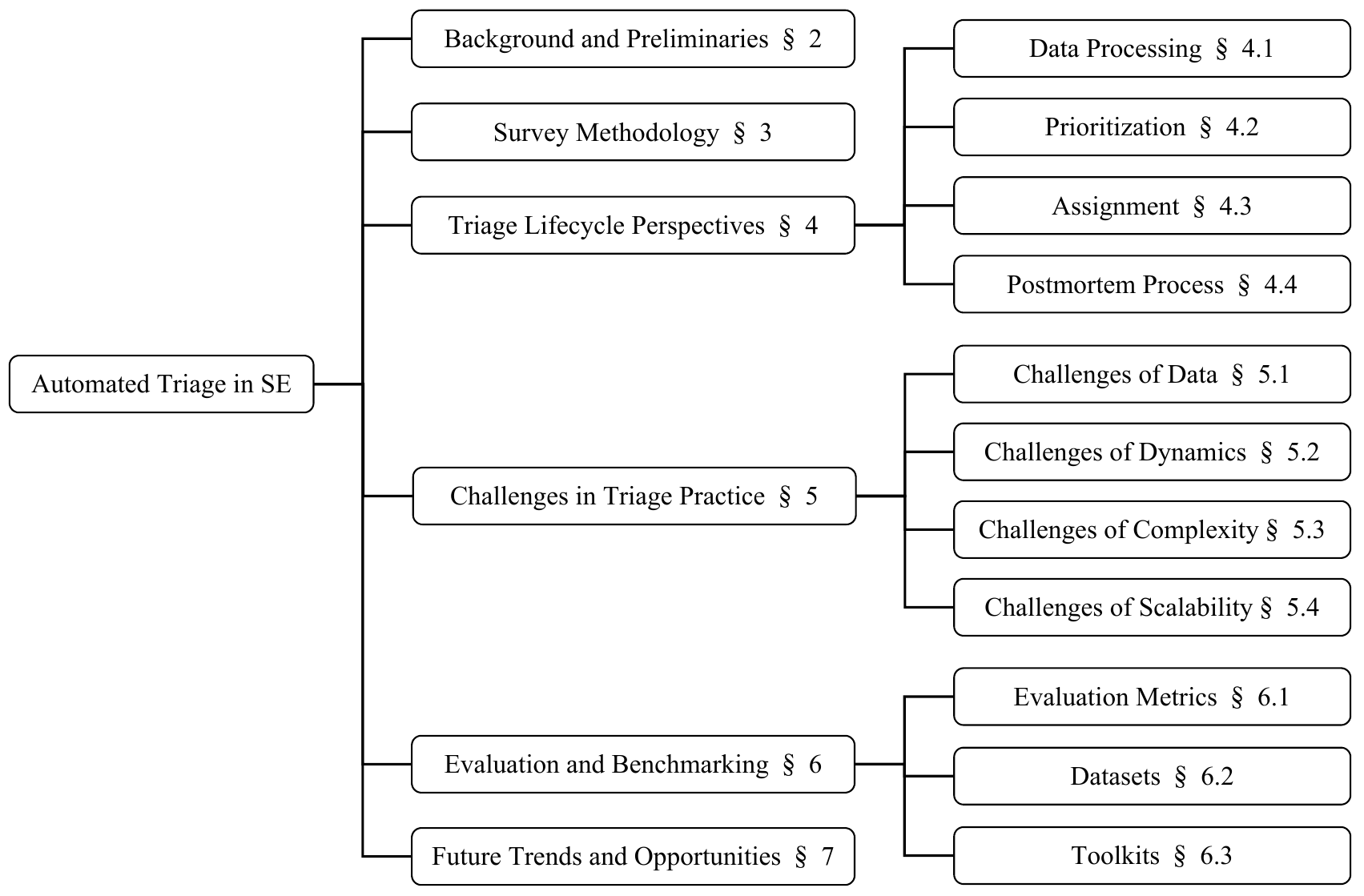}
    \caption{Structure of this survey.}
    \vspace{-0.7cm}
    \label{fig:structure}
\end{figure}

\subsection{Survey Structure.} 
The overall structure of this survey is depicted in Figure~\ref{fig:structure}. Section~\ref{sec_2} introduces essential background concepts, followed by Section~\ref{sec_3}, which details the methodology adopted in this work. Section~\ref{sec_4} presents related studies organized according to the triage lifecycle. Section~\ref{sec_5} discusses practical challenges encountered in real-world triage processes. Section~\ref{sec_6} outlines evaluation metrics and publicly available benchmarks to facilitate empirical studies. Finally, Section~\ref{sec_7} highlights promising directions for future research.

\section{Background and Preliminaries}\label{sec_2}

In this section, we provide the foundational background of triage in software engineering by tracing its historical origins, exploring its adoption, and describing the key data sources that support and inform triage practices. We then establish a conceptual framework for our survey by defining a general triage lifecycle. Finally, we review related surveys to position our work within the broader research landscape.

\vspace{-0.3cm}
\subsection{Background of Triage in Software Engineering}

\textbf{Historical Origins.} The term \textit{triage} originates from the French verb \textit{trier}, meaning ``to sort'' or ``to select''. It was initially introduced within the French medical services as a systematic approach to classifying patients~\cite{robertson2006evolution}. Historically, triage referred to the process of stratifying individuals into categories such as immediate, urgent, and non-urgent during mass casualty incidents, particularly in military contexts. The primary objective was to maximize survival rates under conditions of limited medical resources~\cite{robertson2006evolution, christ2010modern}.

\textbf{Adoption.} With the rapid growth in complexity and scale of software systems, the fundamental philosophy of triage, prioritizing interventions under resource constraints, has found significant relevance beyond traditional medical applications. Within software engineering, triage has become a key practice for managing the overwhelming influx of information, such as bug reports, alerts, and user feedback, that arise during software development and operation. The capacity to efficiently identify, prioritize, and allocate resources to the most critical and high-impact issues is essential for maintaining software quality, ensuring service reliability, and enhancing user satisfaction. To meet these demands, a wide range of automated and intelligent triage approaches have been proposed, drawing upon advances in machine learning, natural language processing, and data mining. These systems aim not only to automate but also to improve upon manual triage processes in terms of both efficiency and accuracy.

\textbf{Data Sources.} A wide spectrum of data sources is utilized in the triage of software engineering, encompassing bug reports, incident tickets, alerts, observability data, reviews, and relational data. Each type of data contributes a unique yet complementary perspective on system behavior, operational conditions, and potential issue contexts.

\begin{itemize}
    \item \textbf{Incident Tickets:}  Incident tickets encompass a range of textual and semi-structured inputs, such as customer-reported incident (CI) tickets, monitor-reported incident (MI) tickets, and raw incident tickets generated within operational environments. These tickets, typically expressed in natural language, contain essential contextual details that support key triage tasks, including incident linking, team assignment, and prioritization. The narrative descriptions often reveal early indicators of system degradation and provide contextual signals that guide subsequent analysis and resolution.  

    \item \textbf{Bug Reports:} Bug reports are typically semi-structured, containing both unstructured narratives and structured metadata, along with historical relational information. The unstructured portion includes summaries and detailed descriptions that often specify reproduction steps, expected and actual outcomes, error messages, stack traces, and embedded code snippets. The structured portion consists of predefined fields such as product, component, operating system, software version, severity, and priority, which provide essential categorical features frequently utilized in triage algorithms. Historical and relational data, including assignment histories and developer activity logs, capture previous decision-making patterns and developer expertise. These data collectively provide valuable context for improving triage accuracy and efficiency.  

    \item \textbf{Alerts:}  Alerts are automatically generated by monitoring systems through predefined rules or anomaly detection mechanisms. They act as direct triggers for incidents and usually contain metadata such as timestamps, source identifiers, severity levels, and brief diagnostic descriptions. Alerts form the initial layer of triage, serving as early warnings that enable proactive system management and timely response to potential issues.  

    \item \textbf{Observability Data:}  Observability data, such as Key Performance Indicators (KPIs), metrics, traces and system logs, offer detailed insights into system behavior and performance dynamics. In triage, these data serve as the analytical foundation for tasks including anomaly detection, root cause analysis, and responsible team assignment. Their temporal nature enables real-time monitoring and supports the correlation of performance deviations with potential system issues.

    \item \textbf{Reviews:} Review data consists primarily of textual information from two sources: users and developers. User data includes ratings and text in app reviews, reflecting their experiences with the product. Developer data consists of commit information and code comments detailing functionality and data transfer. Compared to reports and alerts, review data is larger, less structured, and more conversational, requiring advanced natural language processing (NLP) techniques for analysis.

    \item \textbf{Relational Data:} Relational data encapsulates the historical and contextual interactions among bugs, developers, and software components. In contrast to isolated bug reports, relational data captures the dynamic dependencies and collaboration patterns that emerge throughout the bug resolution process. These data are typically represented as heterogeneous networks or bipartite graphs, where nodes correspond to entities and edges encode their relationships. By modeling such interactions, relational data facilitates the inference of latent expertise, developer collaboration behaviors, and structural dependencies among software modules.
    
\end{itemize}

\subsection{A General Triage Lifecycle}
\begin{figure}
    \centering
    \includegraphics[width=0.9\linewidth]{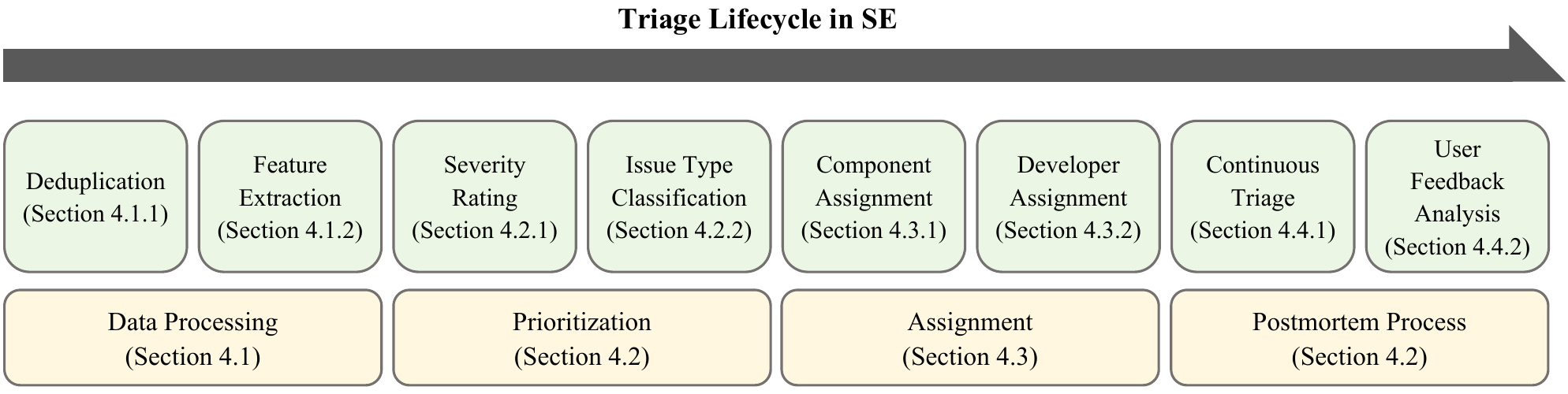}
    \caption{The general lifecycle of triage in SE.}
    \vspace{-0.5cm}
    \label{fig:triage_lifecycle}
\end{figure}

To provide a comprehensive understanding of automated triage, we conceptualize it as a multi-stage lifecycle, as illustrated in Figure~\ref{fig:triage_lifecycle}. This lifecycle models the trajectory of an issue report from its initial submission to its final resolution and the subsequent extraction of actionable knowledge. While not all triage systems implement every stage, this model captures the essential processes that have been extensively studied in the literature. The lifecycle can be broadly divided into four major phases: \textit{Data Processing}, \textit{Prioritization}, \textit{Assignment}, and the \textit{Postmortem Process}.

\subsubsection{Data Processing}
The first phase of the triage lifecycle focuses on transforming raw, unstructured issue reports into structured, feature-rich data that are suitable for automated analysis. The quality of this phase directly influences the performance of subsequent stages. It typically encompasses two main tasks:

\begin{itemize}
    \item \textbf{Deduplication:} Large-scale projects frequently receive multiple reports describing the same underlying defect. Detecting and merging these duplicates is critical for reducing redundant engineering effort. Deduplication is often formulated as a similarity-matching problem, where the objective is to identify duplicate reports based on textual and metadata similarities. The main challenge lies in detecting duplicates that differ in their wording or structure, which requires sophisticated feature extraction methods and similarity metrics~\cite{neysiani2020efficient}.

    \item \textbf{Feature Extraction:} Issue reports usually consist of a mixture of natural language descriptions, code snippets, stack traces, and metadata. The task of feature extraction involves identifying and representing meaningful information from these heterogeneous inputs. A key challenge lies in capturing the semantic relationships between textual and structural elements~\cite{dipongkor2023comparative}. Early studies relied on classical text mining techniques such as Term Frequency–Inverse Document Frequency (TF–IDF) and Latent Semantic Indexing (LSI)~\cite{ahsan2009automatic}. More recent methods incorporate structural analysis by parsing code into Abstract Syntax Trees (ASTs) to represent code semantics separately from textual descriptions~\cite{aung2022multi}. The advent of pre-trained language models (PLMs), including BERT and its variants, has further advanced this area by enabling deep contextual embeddings that capture rich semantic and syntactic information from issue descriptions~\cite{dipongkor2023comparative, lee2022light}.
\end{itemize}

\subsubsection{Prioritization}
After data preprocessing, the next phase involves assessing the urgency and nature of an issue to determine its relative importance. This prioritization phase ensures that limited engineering resources are allocated to the most critical problems. It generally includes the following two tasks:

\begin{itemize}
    \item \textbf{Severity Rating:} The goal of this task is to automatically predict the severity level of a bug or incident (\eg Blocker, Critical, or Minor)~\cite{arshad2024sevpredict}. Accurate severity prediction is essential for effective risk management and for ensuring that system-critical issues receive timely attention. Recent studies have explored the use of large language models, sometimes augmented with sentiment analysis, to infer the implicit urgency embedded in textual descriptions and to improve prediction accuracy~\cite{arshad2024sevpredict}.

    \item \textbf{Issue Type Classification:} In addition to severity, it is often necessary to categorize issues based on their fundamental nature. This classification distinguishes between types such as functional bugs, performance degradations, and security vulnerabilities, thereby supporting more targeted handling. Some studies have proposed taxonomies of issue types and developed automated classifiers to operationalize these distinctions~\cite{catolino2019not}. Other approaches employ multi-task learning frameworks to jointly perform issue classification and developer assignment, enhancing the overall triage efficiency~\cite{aung2022multi}.
\end{itemize}

\subsubsection{Assignment}
The assignment phase is concerned with routing each issue to the most appropriate entity for resolution. The effectiveness of this process directly affects the average Time-To-Resolution. Depending on organizational structure and issue granularity, assignment may occur at different levels:

\begin{itemize}
    \item \textbf{Component Assignment:} In large projects, software is often divided into multiple components, each maintained by a dedicated team. Component assignment aims to identify which component is likely affected by the issue, thereby narrowing the search space for subsequent developer assignment. Research in this area often focuses on learning the relationships between issue descriptions and component-specific information. Recent work has introduced deep learning models that address challenges such as ambiguous or few-shot components, where traditional feature-based methods perform poorly~\cite{su2023still, xu2023method}.

    \item \textbf{Developer Assignment:} This task involves assigning the issue directly to the developer or team best equipped to resolve it. Developer assignment has been a long-standing research focus, evolving from early machine learning models based on text categorization~\cite{anvik2006should} to contemporary deep learning approaches that utilize graph neural networks to capture collaboration structures~\cite{wu2022spatial, dai2023graph}. More recent studies have also leveraged large language models for semantic understanding and context reasoning~\cite{lee2022light}. Advanced systems additionally incorporate dynamic contextual factors such as workload, expertise, and availability~\cite{jahanshahi2022s}. In industrial settings, this task is often adapted to \textit{team-level assignment}, where issues are routed to the most relevant team instead of a single developer to align with collaborative workflows and organizational practices~\cite{jonsson2016automated, sarkar2019improving}.
\end{itemize}

\subsubsection{Postmortem Process}
The triage lifecycle is not a linear or static process. Modern triage systems increasingly incorporate feedback loops and continuous learning mechanisms that allow them to adapt and evolve. This adaptive capability is essential for managing the dynamic nature of software projects and the changing needs of users. Within this ongoing process, two key aspects are particularly significant: continuous triage and the integration of user feedback.

\begin{itemize}
    \item \textbf{Continuous Triage:} Traditional triage models often make a single, static assignment decision when a new report is submitted. In practice, however, the initial information provided by a report may be incomplete, and additional context frequently emerges as engineers discuss and investigate the issue. Continuous triage addresses this limitation by iteratively refining the assignment as new information, such as discussion comments or updated diagnostics, becomes available~\cite{chen2019continuous}. This iterative re-evaluation is particularly important in large-scale online service environments, where incident discussions evolve rapidly and timely adjustments can significantly reduce service downtime.

    \item \textbf{User Feedback Analysis:} In addition to data derived from internal development and monitoring processes, feedback from end-users provides valuable, real-world insights into software performance and emerging issues. Analyzing user feedback, commonly expressed through application reviews or problem reports, serves as a proactive input to the triage pipeline~\cite{gao2018online}. Although the immediate context of user reviews differs from that of internal issue reports, both share common linguistic and semantic characteristics. Consequently, analytical methods developed for processing user reviews can inform and enhance triage practices by offering complementary perspectives on system reliability and user experience.  
\end{itemize}

\subsection{Related Surveys}

\begin{table*}[ht]
    \footnotesize
    \centering
    \caption{
        Comparison of existing triage-related surveys based on their focused domain and analytical perspectives. The abbreviations in the ``Data Sources'' column stand for: \textbf{A} = Alerts, \textbf{OD} = Observability Data (\eg KPIs, Metrics, Logs, and Traces), \textbf{IT} = Incident Tickets, and \textbf{BR} = Bug Reports.
    }
    \label{table:related_surveys}
    \begin{tabular}{ccccc}
        \toprule
        \textbf{Reference} & \textbf{Year}  & \textbf{Focus Domain} & \textbf{Data Sources} & \textbf{Analysis Perspectives}\\
        \midrule
        Remil et al.~\cite{remil2024aiops} & 2024 & Incident management & OD, IT & Fundamental abilities of AIOps \\
        Yu et al.~\cite{yu2024survey} & 2024 & Alert and incident management & A, IT, OD & Workflow of management \\
        Notaro et al.~\cite{notaro2021survey} & 2021 & Failure management & BR, IT, OD & Intervention time window \\
        Zhang et al.~\cite{zhang2016survey} & 2016 & Bug resolution & BR & Lifecycle of bug resolution\\
        Bocu et al.~\cite{bocu2023survey} & 2023 & Bug triage and management & BR & Triage techniques\\
        Uddin et al. ~\cite{uddin2024novel} & 2024 & Bug triage & BR & Triage techniques\\
        Nagwani et al.~\cite{nagwani2023artificial} & 2023 & Bug triage & BR & Triage techniques\\
        Qian et al.~\cite{qian2023survey} & 2023 & Bug triage & BR & Input data source\\
        Akila et al.~\cite{akila2014survey} & 2014 & Bug triage& BR & Workflow of triage \\
        \midrule
        \textbf{Our work} & - & Triage in SE & A, BR, IT, OD & Lifecycle of Triage \\
        \bottomrule
    \end{tabular}
    \vspace{-0.5cm}
\end{table*}
Comprehensive surveys and systematic reviews have extensively examined various subdomains within software maintenance and IT operations. As summarized in Table~\ref{table:related_surveys}, these related studies can be broadly classified into two major categories. The first category, represented by works such as Remil et al.~\cite{remil2024aiops}, Yu et al.~\cite{yu2024survey}, and Notaro et al.~\cite{notaro2021survey}, explores overarching management frameworks. In these studies, triage is considered one component within a broader operational framework, and the primary emphasis lies on the overall system lifecycle or high-level capabilities.

The second category of related surveys focuses specifically on software defects, offering detailed analyses of bug triage techniques. Representative examples include the studies by Zhang et al.~\cite{zhang2016survey}, Nagwani et al.~\cite{nagwani2023artificial}, and Qian et al.~\cite{qian2023survey}. These surveys concentrate almost exclusively on bug reports as the primary data source and investigate various approaches to defect management, classification, and assignment.

This distinction reveals a clear research gap. Although individual subdomains of triage have been studied in depth, no existing survey has systematically synthesized or compared the triage process across different, yet conceptually related, artifacts in software engineering. In particular, there remains a lack of unified examination encompassing alerts, incidents, and bugs. While triage has often been treated as a subordinate activity within larger operational workflows, it in fact constitutes a critical and high-leverage process that warrants focused and systematic investigation.

As modern software systems continue to grow in complexity, the increasing volume, heterogeneity, and urgency of alerts, incidents, and bug reports have made manual triage both inefficient and error-prone~\cite{wang2024comet, dipongkor2023comparative, zhao2020alertrank}. These challenges highlight the necessity for a dedicated and comprehensive examination of automated triage methodologies that transcend specific data types or operational contexts.

In contrast to prior domain-specific reviews, the present survey adopts a holistic perspective by positioning triage itself as the central object of study. The key contribution of this work lies in systematically analyzing software triage as a unified process that spans the entire lifecycle across alerts, incidents, and bugs. This survey offers a comparative analysis of the complete triage pipeline, encompassing data preprocessing, prioritization, assignment, and post-processing, and highlights both the commonalities and distinctive features observed in each domain. To the best of our knowledge, this is the first comprehensive survey that bridges the Artificial Intelligence for IT Operations (AIOps) and IT Service Management (ITSM) community, primarily focused on alert, incident, and operational data, with the software engineering community, which traditionally emphasizes bug data, under a unified triage framework. By integrating insights from these previously separate research areas, this study provides a coherent understanding of software triage and aims to foster future innovation through the convergence of methodologies and perspectives.
\section{Survey Methodology}\label{sec_3}

\subsection{Survey Scope}

The term triage originated in the medical domain, where it refers to the process of prioritizing patients based on the severity of their conditions to ensure that scarce medical resources are allocated efficiently to those in greatest need~\cite{robertson2006evolution}. Over time, this concept has evolved beyond its clinical roots, emerging as a general strategy for managing large volumes of competing tasks under resource constraints. Today, triage serves as a fundamental mechanism in a broad range of domains, including software engineering~\cite{xie2013impact}, cyber security~\cite{lin2018data}, and emergency response systems~\cite{fitzgerald2010emergency}.

Within the context of software development and operations, triage is critical to managing the growing complexity and scale of modern computing systems. It facilitates timely and informed decision-making by prioritizing, classifying, and routing issues, such as bug reports, system alerts, service incidents, and performance anomalies, to the appropriate stakeholders, whether human engineers or automated remediation systems. Effective triage not only reduces the cognitive burden on developers and operators but also enhances system reliability, service responsiveness, and organizational agility.

This survey presents a comprehensive examination of triage practices in the domain of software development and operations. Rather than treating triage as a monolithic or static process, we conceptualize it as a dynamic, multi-phase lifecycle consisting of four interrelated stages: issue intake and preprocessing, prioritization, assignment, and resolution feedback. Each stage introduces distinct technical and organizational challenges, including but not limited to information overload, ambiguity in prioritization criteria, coordination inefficiencies, and the absence of standardized workflows and tools. By systematically analyzing this end-to-end triage lifecycle, we aim to surface common patterns and pain points, critically evaluate existing solutions, and identify opportunities for intelligent automation, human-in-the-loop augmentation, and continuous process improvement.

\subsection{Paper Collection}
Our paper collection process includes two steps: keyword searching and snowballing.

\subsubsection{Keyword Searching}

To ensure a comprehensive and rigorous foundation for this survey, we adopted a systematic literature collection methodology centered on the DBLP Computer Science Bibliography~\cite{dblp}, a widely recognized and authoritative source in the field of computer science. DBLP provides extensive bibliographic metadata for major conferences and journals, and has been frequently used in prior surveys on software engineering and related domains~\cite{chen2020survey, zhang2018empirical, zhang2020machine}. Given its broad coverage, we selected DBLP as our primary indexing platform, noting that papers indexed by other scientific databases (\eg Google Scholar, arXiv) are generally a subset of those available via DBLP~\cite{zhang2020machine}.

We began by identifying a set of high-impact, peer-reviewed publication venues across several relevant areas, including software engineering, artificial intelligence, data mining, and computer systems. Specifically, our selection comprised 13 conferences (\eg ICSE, ASE, ESEC/FSE, AAAI, SIGKDD, SIGCOMM, INFOCOM, NDSS, WWW, ISSRE, CSCW, ICDE, and CIKM) and 4 journals (\eg TSE, TOSEM, TKDE, and JSS). To ensure consistency, three authors independently reviewed the titles, abstracts, and introductions of papers published in these venues, jointly refining a set of search keywords through collaborative discussion. 
The finalized search query was defined as: \textit{(``Triage'') AND (``Incident'' OR ``Bug'' OR ``Alert'')}. 
During the search process for bug triage studies, we observed that most papers~\cite{anvik2006should, ahsan2009automatic, park2011costriage, lee2017applying} treat ``bug triage'' as synonymous with ``bug assignment'',  focusing primarily on developer recommendation tasks. To ensure comprehensive coverage of this subdomain, we therefore extended our query to also include the keyword \textit{(``Bug Assignment'')}.
In parallel, as an emerging field, incident triage has an unclear definition, and there are relatively few papers with it as a keyword. Therefore, we also introduced the following search query to incorporate incident management into consideration, which is \textit{(``Incident Diagnosis'') AND (``Incident Management '')}. In addition, since some triage-related papers~\cite{chen2020icmbrain, gu2020linkcm, chen2019empirical, chen2019continuous} regard review-related methods as customer input, we additionally used the search query \textit{(``user review'') AND (``user comments '')} to search for relevant papers.

We then conducted iterative searches over a 20-year publication window, manually screening all retrieved papers for relevance. This process resulted in an initial corpus of 195 candidate studies deemed pertinent to the scope of triage in software development and operations.

\subsubsection{Snowballing Strategy}

To augment our initial dataset and reduce the risk of publication bias, we employed a snowballing strategy, as recommended by Wohlin et al.~\cite{wohlin2014guidelines}. This involved both backward snowballing (examining references cited by the initial studies) and forward snowballing (identifying studies that cited the initial corpus). In each iteration, we applied a consistent set of criteria to ensure that only papers directly relevant to our research scope were retained.

This iterative snowballing process continued until theoretical saturation was reached, that is, no additional relevant papers were identified in subsequent rounds. Through this method, we identified and incorporated 51 additional studies, resulting in a final corpus of 246 peer-reviewed papers encompassing a broad range of triage-related research.

\subsubsection{Quality Assessment}
To ensure the validity and reliability of our findings, we conducted a structured quality assessment of all candidate studies, in line with best practices for systematic literature reviews~\cite{kitchenham2004procedures}. Each study was evaluated based on a predefined checklist across three critical dimensions~\cite{hall2011systematic, hosseini2017systematic}:

\begin{enumerate}
    \item \textbf{Data Transparency:} Whether the paper clearly describes its dataset(s), including collection methods, characteristics, and availability.
    
    \item \textbf{Methodological Rigor:} Whether the proposed model, framework, or approach is described in sufficient technical detail to allow reproducibility.
    
    \item \textbf{Evaluation Clarity:} Whether the evaluation procedure is robust, and whether results are reported in a transparent and interpretable manner.
\end{enumerate}

Two authors independently assessed the 246 studies identified via keyword search and snowballing. In cases of disagreement, a third author facilitated resolution through discussion and consensus. Based on this process, 12 studies were excluded due to insufficient methodological clarity or reporting rigor, and the remaining 234 studies were included in the final corpus for analysis.

\subsection{Publication Trend and Distributions}

In total, we collected 234 studies related to triage in the context of software engineering and operations. Figure~\ref{fig:sub1} presents the histogram of annual papers. As illustrated, there has been a consistent upward trajectory in research activity, with a particularly marked increase in recent years. This trend underscores the rising importance of triage in managing the complexity, scale, and dynamism of modern software systems, thereby reaffirming the timeliness and relevance of this survey.

\begin{figure}
    \centering
    \includegraphics[width=0.55\linewidth]{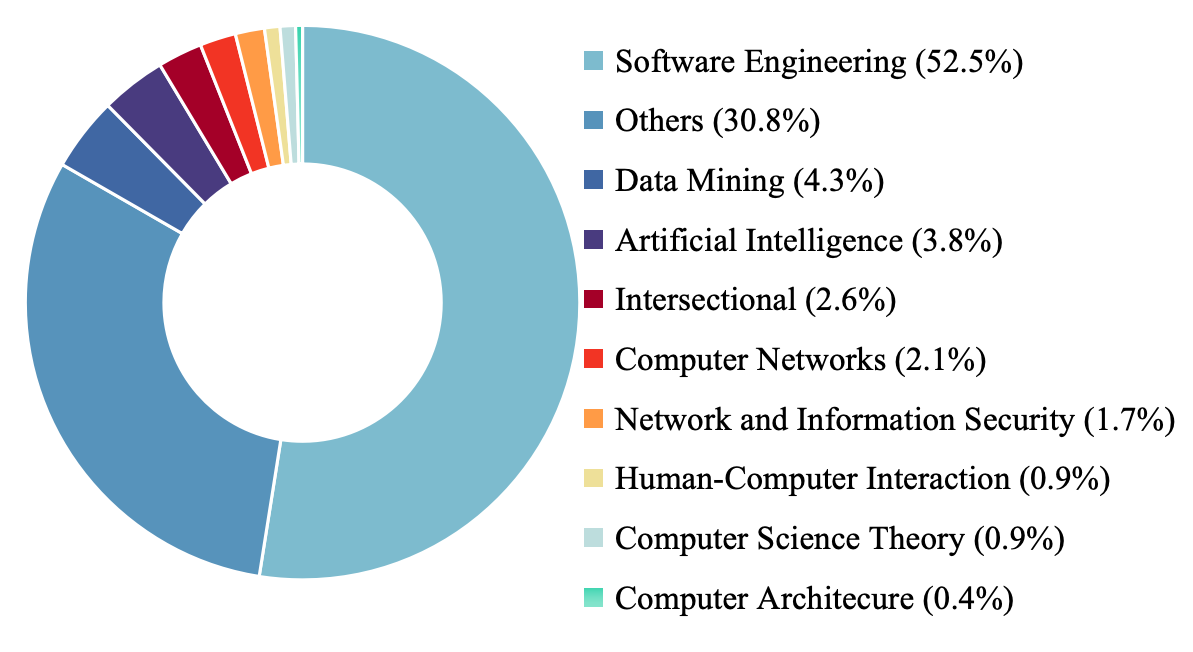}
    \caption{Publication distribution of distinct venues.}
    \label{fig:venue_stat}
\end{figure}

To gain further insight into how triage-related research is disseminated across the academic landscape, we analyzed the distribution of publications by venue. As shown in Figure~\ref{fig:venue_stat}, the majority of the selected studies have been published in leading software engineering venues, such as ICSE, ESEC/FSE, ASE, ISSRE, TSE, and the JSS. These venues have served as primary forums for advancing triage techniques, particularly in the areas of bug report classification, fault prioritization, and incident handling workflows. Beyond the domain of software engineering, triage research has also garnered increasing attention in adjacent venues, including data mining (\eg SIGKDD), artificial intelligence (\eg AAAI), and computer networks (\eg SIGCOMM).

These patterns indicate that triage is inherently multidisciplinary, integrating insights from software engineering, artificial intelligence, data mining, and distributed systems. The broad range of publication venues highlights its wide applicability and cross-domain relevance, spanning both theoretical advancements and practical implementations.

We further investigated the evolution of academic–industrial collaborations in triage research. As depicted in Figure~\ref{fig:sub2}, such collaborations were nearly nonexistent before 2008 and began to emerge gradually in the following years. A noticeable upward trajectory became evident after 2014, with the number of joint studies increasing alongside the overall volume of triage-related publications. The most pronounced surge occurred between 2019 and 2020, marking the peak of industry collaborations within the field. Despite subsequent fluctuations in publication counts, the proportion of industry-affiliated research has remained consistently significant, reflecting the increasing practical orientation and applied relevance of triage research in contemporary software engineering.

This growing trend underscores the strong alignment between academic inquiry and industrial needs, demonstrating that triage is not only a subject of theoretical interest but also a practical necessity for real-world systems. The sustained presence of industry collaborations further highlights the demand for scalable, actionable, and deployable triage solutions capable of supporting complex production environments.

\section{Review from Lifecycle Perspectives}\label{sec_4}

In this section, we organize the collected papers from the perspective of the triage lifecycle. Figure~\ref{fig:overall} presents the common tasks throughout the triage lifecycle.

\subsection{Data Processing}

Data processing constitutes a critical stage within the triage pipeline, transforming raw system data into a structured representation that facilitates subsequent diagnostic analysis. Core tasks in this stage include data deduplication and feature extraction. This section surveys prior studies on processing methodologies, with a particular focus on these key tasks. Given the heterogeneity of data sources, spanning differences in structure, granularity, and reliability, existing studies adopt diverse data processing pipelines. Accordingly, we organize the review by primary data source and summarize the corresponding processing strategies and methodological choices within each category.

\subsubsection{Deduplication}

In large-scale enterprises, report processing systems generate thousands of reports daily, creating substantial challenges for triage engineers~\cite{sun2025tixfusion}. A significant proportion of these reports are repetitive and often refer to the same defect. By enabling collective processing, the aggregation of such duplicate reports can markedly improve the efficiency of classification engineers. The report duplication approach seeks to identify reports that refer to identical bugs, cloud incidents, or system defects, allowing engineers to focus on resolving underlying issues rather than spending effort on redundant inspections.

\textbf{Incident Tickets.} In cloud computing environments, service incidents pose significant risks to both customer satisfaction and business revenue. A major challenge arises from the distributed and uncoordinated nature of incident reporting, often leading to redundant or duplicate tickets~\cite{liu2023ipack}. Consequently, the effective aggregation of related reports has become a prerequisite for efficient incident triage. Existing studies address this challenge through three main perspectives: offline correlation and clustering, information extraction for data filtering, and online graph-based aggregation.

(1) \textit{Offline Correlation and Clustering: }
Early studies explored the use of statistical and semantic associations to cluster related incidents.
Ding et al.~\cite{ding2014mining} established a framework for mining correlated events through concept-based clustering and similarity retrieval.
Subsequent work, such as LinkCM~\cite{gu2020linkcm}, introduced transfer learning and semantic encoding to improve incident linkage across large-scale systems.
LiDAR~\cite{chen2020lidar} further incorporated structural information from component dependency networks, extending beyond purely textual similarity.
More recently, iPACK~\cite{liu2023ipack} integrated multiple data sources, customer tickets, and system incident logs, to achieve more accurate aggregation of duplicate reports.
These studies collectively demonstrate a gradual evolution from static similarity-based clustering to multi-source semantic correlation for improved incident triage.

(2) \textit{Information Extraction for Valid Data: }
Another filtering research line focuses on extracting discriminative information from service tickets to filter relevant data before aggregation.
Lou et al.~\cite{lou2017experience} introduced a rule-based process for identifying effective attribute combinations from customer work orders.
With advances in large language models (LLMs), TixFusion~\cite{sun2025tixfusion} automates this process through iterative extraction of key operational and anomaly-related features from textual tickets, enhancing aggregation precision.
Together, these studies highlight a shift from manual feature engineering toward intelligent and adaptive data understanding.

(3) \textit{Online Graph-Based Aggregation: }
To overcome the latency of offline approaches, GRLIA~\cite{chen2021grlia} proposes an unsupervised graph learning framework for real-time aggregation of correlated incidents. By encoding both topological and temporal dependencies among cascading failures, GRLIA supports scalable and adaptive triage in dynamic environments.

\textbf{Bug Reports.} 
Within the software development lifecycle, especially during the development and maintenance phases, bug reports serve as a major diagnostic resource. Bug Tracking Systems (BTS) such as Bugzilla and JIRA function as centralized repositories that accumulate vast amounts of historical data~\cite{shokripour2013so}. Large-scale open-source projects like Mozilla or Eclipse may contain hundreds of thousands of reports, resulting in considerable redundancy and overlapping information. The unstructured nature of textual descriptions and the variability of reporter expertise further complicate automated analysis.

To address these challenges, researchers have explored various preprocessing strategies to improve the efficiency of downstream triage and duplicate detection tasks~\cite{sabor2017durfex}. Early studies primarily focused on dataset reduction to eliminate redundant or noisy information. Zou et al.~\cite{zou2011towards} applied a joint feature and instance selection approach to construct smaller yet more effective training sets, demonstrating that the order of these two reduction phases significantly affects classification performance. Building on this idea, Xuan et al.~\cite{xuan2014towards} employed a binary classifier to automatically determine the optimal application order of instance and feature selection, thereby enhancing the quality of the reduced datasets used in triage.

More recent work has shifted attention toward feature extraction and representation optimization for duplicate detection. Neysiani et al.~\cite{neysiani2020efficient} proposed an approach that aggregates unigram and bi-gram textual features with term frequency and inverse document frequency, introducing a hybrid metric to evaluate feature effectiveness and construct a compact yet informative feature set.

Collectively, these studies highlight an evolution from dataset reduction to refined feature engineering, emphasizing the critical role of preprocessing in improving the accuracy and scalability of automated bug report analysis.

\textbf{Alerts.} Alerts play a crucial role in maintaining the reliability of modern online service systems by providing early warnings of potential failures. However, due to the intricate interdependencies among system components, a single fault can trigger a large number of cascading alerts, which often overwhelm traditional operational workflows and render manual triage infeasible~\cite{chen2022oas}. Consequently, the automated aggregation and summarization of alerts have become critical to reducing redundancy and enhancing operational efficiency.

(1) \textit{Graph-Based and Statistical Summarization: }
Early studies focused on graph-based and statistical summarization to mitigate alert storms. NoDoze~\cite{nodoze2019} constructs causal dependency graphs of alerts, assigns anomaly scores according to historical frequency, and propagates them through network diffusion to identify representative subgraphs for triage. Similarly, AlertRank~\cite{zhao2020alertrank} conducts an empirical study of large-scale alert storms and combines accurate storm detection with summarization to recommend representative alerts for analysts.

(2) \textit{Deep Representation Learning: }
Subsequent research introduced deep representation learning to improve correlation detection among alerts. Warden~\cite{li2021warden} enhances incident management by automatically grouping correlated alerts and enabling proactive failure detection in large-scale cloud environments. Building on this idea, OAS~\cite{chen2022oas} extracts both semantic and behavioral features from raw alerts and fuses them through a deep learning model to determine inter-alert correlations for effective summarization.

(3) \textit{Large Language Models: }More recent approaches leverage LLMs to integrate contextual reasoning and external operational knowledge. COLA~\cite{kuang2024cola} exemplifies this trend by combining correlation extraction with LLM-based reasoning, guided by standard operation procedures (SOPs) as auxiliary knowledge for online alert aggregation.

Despite their progress, these methods remain constrained by a heavy reliance on supervised training. The requirement for extensive, manually labeled data imposes substantial operational costs and limits its generalization to dynamic or data-scarce environments.

\textbf{Reviews.} After existing triage methods generate initial results, users or engineers often provide feedback to refine system performance~\cite{chen2020icmbrain}. Such feedback can enhance the accuracy and reliability of incident resolution, yet its practical use is challenged by the noisy and unstructured nature of user comments. In particular, app reviews are typically short and contain a large proportion of irrelevant information; only about 30\% of them offer actionable insights for app improvement~\cite{chen2014ar}. Consequently, effective techniques for filtering and prioritizing informative feedback have become an essential research focus.

Early work, such as AR-Miner~\cite{chen2014ar} established a foundational framework for extracting valuable user opinions from large-scale reviews. The approach integrates four key stages: filtering noisy or irrelevant reviews, applying topic modeling to cluster similar feedback, ranking reviews based on informativeness, and visualizing the top-ranked items to support developers' decision-making. Building upon this line, PAID~\cite{gao2015paid} advances the granularity of analysis from reviews to phrases. It extracts and prioritizes key phrases through rule-based selection and maintains a Phrase Bank for developers. To capture the temporal dynamics of user concerns, PAID further applies Dynamic Latent Dirichlet Allocation (dLDA) to model topic evolution across app versions and recommend the most relevant phrases for each identified topic. More recently, IFeedback~\cite{zheng2019ifeedback} extends feedback analysis toward automated fault detection. It constructs Word Combination-based Indicators (WCIs) by pairing words in feedback text and filters them with historical data to retain those most indicative of system faults. These indicators serve as metrics for identifying potential issues in real time.

Overall, the evolution from AR-Miner to PAID and IFeedback reflects a progressive refinement of feedback analysis, from coarse-grained filtering of informative reviews to fine-grained phrase extraction and automated fault detection. These methods collectively aim to enhance the interpretability and utility of user feedback in guiding software maintenance and evolution.

\begin{table*}[ht]
    \footnotesize
    \centering
    \caption{
        Summary of data deduplication methods in triage oprations based on their category, the year, and the data. The abbreviations in the ``Data'' column stand for: \textbf{A} = Alerts, \textbf{R} = Reviews, \textbf{IT} = Incident Tickets, \textbf{L} = Logs, \textbf{KPI} = Key Performance Indicators, \textbf{DN} = Dependency Networks, and \textbf{BR} = Bug Reports.
    }
    \label{table:deduplication}
    \begin{tabular}{ccccl}
        \toprule
        \textbf{Category} & \textbf{Technique}  & \textbf{Year} & \textbf{Data} & \textbf{Core Method}\\
        \midrule
        \multirow{7}{*}{Incident Tickets} 
        & Ding et al.~\cite{ding2014mining} & 2014 & IT & FCA + GVSM\\
        & LinkCM~\cite{gu2020linkcm} & 2020 & IT & Symmetric Model + BERT + Decomposable Attention Mechanism \\
        & LiDAR~\cite{chen2020lidar} & 2020 & IT, DN & TextCNN + Node2Vec \\
        & iPACK~\cite{liu2023ipack} & 2023 & IT & Incident-Aware Framework \\
        & Lou et al.~\cite{lou2017experience} & 2017 & IT & Extracting Attribute Combinations\\
        & TixFusion~\cite{sun2025tixfusion} & 2025 & IT & LLM\\
        & GRLIA~\cite{chen2021grlia} & 2021 & IT, DN, KPI & Graph Representation Learning \\
        \midrule
        \multirow{3}{*}{Bug Reports}&Zou et al.~\cite{zou2011towards}& 2011 & BR &Training Set Reduction Approach\\
        &Xuan et al.~\cite{xuan2014towards} & 2014 & BR & Instance Selection + Feature Selection + Binary Classifier\\
        &Neysiani et al.~\cite{neysiani2020efficient} & 2020 & BR & Term Frequency + Inverse Document Frequency\\
        \midrule
        \multirow{5}{*}{Alerts}
        & NoDoze~\cite{nodoze2019} & 2019 & A & Causal Dependency Graphs + Network Diffusion\\
        & AlertRank ~\cite{zhao2020alertrank} & 2020 & A & Accurate Storm Detection\\
        & Warden~\cite{li2021warden} & 2021 & A & Automated Alert Grouping + Proactive Failure Detection\\
        & OAS~\cite{chen2022oas} & 2022 & A & Deep Learning Model\\
        & COLA~\cite{kuang2024cola} & 2024 & A, SOP & Integrates Correlation Extraction + LLM-Based Reasoning \\        
        \midrule
        \multirow{3}{*}{Reviews}& AR-Miner~\cite{chen2014ar}& 2014 & R &\makecell[l]{Topic Modeling + Ranking Scheme + Intuitive Visualization Approach}\\
        &PAID~\cite{gao2015paid} & 2015 & R & \makecell[l]{Rule-Based Filtering Strategy + Grouping-Based Ranking Strategy + dLDA}\\
        &IFeedback~`\cite{zheng2019ifeedback} & 2019 & R & \makecell[l]{WCIs + Rule-Based Filtering Strategy}\\
        
        \bottomrule
    \end{tabular}
\end{table*}

\subsubsection{Feature Extraction} 
Reports serve as dense, multimodal data sources, amalgamating structured fields like timestamps with unstructured, free-form text such as failure descriptions and operator annotations. The inherent verbosity and heterogeneity of this content pose a significant challenge to automated analysis. Consequently, the ability to accurately and efficiently distill actionable diagnostic information from these reports is paramount for effective triage and subsequent resolution.

\textbf{Incident Tickets.} 
Incident processing methods have progressively evolved from text-based analysis to multimodal integration, requiring different feature extraction strategies to address the increasing diversity of operational data.

(1) \textit{Textual and Sequential Incident Analysis: }Early approaches focused on mining textual and sequential patterns from historical incidents. Shao et al.~\cite{shao2008efficient} employed probabilistic sequence modeling to discover resolution patterns that supported expert recommendation, while the IcM BRAIN framework~\cite{chen2020icmbrain} introduced AI-based processing of incident tickets and customer inputs to improve triage and correlation efficiency in large-scale production environments.

(2) \textit{Semantic and Structural Modeling: }
As incident data grew in scale and complexity, textual analysis alone became insufficient to capture the contextual and relational characteristics of service failures. Consequently, later research emphasized semantic and structural modeling to represent incidents in more meaningful and interconnected ways.
Triangle~\cite{yu2021triangle} addressed this need by enhancing key information extraction through semantic alignment between incident descriptions and domain documents, thereby improving triage accuracy and contextual consistency. Building on this direction, COT~\cite{wang2021fast} further extended semantic modeling to a structural level by constructing correlation graphs that captured dependencies among incidents and services. 

(3) \textit{Multimodal and LLM-Driven Approaches: }
With the growing availability of heterogeneous operational data such as metrics, logs, traces, and textual reports, research has increasingly shifted toward multimodal fusion and LLM-based reasoning to achieve comprehensive situational awareness.
FaultProfIT~\cite{huang2024faultprofit} initiated this direction by structuring textual tickets into key fields, facilitating more accurate context interpretation by LLMs. Building upon the integration of multiple information sources, DiLink~\cite{ghosh2024dilink} combined textual and structural embeddings within a unified framework to enhance multimodal representation learning. Extending this idea further, Goel et al.~\cite{goel2024x} applied LLMs in combination with vector-based retrieval to improve incident summarization and historical similarity matching. 

(4) \textit{Visual-Text Fusion in Ticket Understanding: }In addition, Mandal et al.~\cite{mandal2019improving} explored visual-text fusion by processing screenshots attached to service tickets, combining image and text extraction to enhance entity recognition and context understanding.

Overall, these methods illustrate a clear progression from early sequence-based and textual processing toward semantically enriched, multimodal, and LLM-empowered approaches, reflecting the ongoing shift toward comprehensive and context-aware incident feature extraction.

\textbf{Bug Reports.} 
In bug triage, feature extraction from both unstructured textual content and structured metadata plays a pivotal role in determining accurate developer assignment. Bug reports typically contain summaries, detailed descriptions, and error traces as unstructured text, together with structured fields such as product, component, severity, and version. 

(1) \textit{Text-Based Approaches: }
Early studies primarily adopted statistical classifiers, among which the Naive Bayes (NB) model was the most widely used. Murphy et al.~\cite{murphy2004automatic} applied a Naive Bayes classifier to bug report text for developer prediction, but its performance was constrained by the assumption of word independence and heuristic features. To overcome these limitations, Xuan et al.~\cite{xuan2010automatic} enhanced NB by integrating Expectation–Maximization (EM) and a Weighted Recommendation List, enabling the exploitation of unlabeled data for improved accuracy. Alenezi et al.~\cite{alenezi2013efficient} further combined NB with Chi-Square–based term selection and load-balancing strategies, although the proposed system lacked empirical validation.
With the rise of semantic representation learning, subsequent research shifted toward richer text modeling. Panda et al.~\cite{panda2022topic} transformed unstructured reports into developer–topic relations via LDA topic modeling and applied Intuitionistic Fuzzy Sets to capture uncertainty in developer expertise. Wang et al.~\cite{wang2024empirical} treated bug assignment as a text classification task, systematically comparing combinations of word-embedding models (\eg Word2Vec, GloVe, ELMo, BERT) and deep classifiers (\eg LSTM, Bi-LSTM with attention, TextCNN) for predicting the most suitable developers.

(2) \textit{Metadata-Driven and Code-Driven Approaches:}
Recognizing the limitations of purely textual information, later work explored structured and code-related features. Park et al.~\cite{park2016cost} enhanced triage for non-textual bug reports by computing import-path similarities, using Jaccard and tree-edit distances, to infer bug types and maintain developer profiles via time-decayed repair history. Sun et al.~\cite{sun2017enhancing} proposed EDR\_SI, which incorporated developer habits and experiences through Collaborative Topic Modeling on historical commits, extracting personalized developer–file associations. Xu et al.~\cite{xu2023method} designed a Crash Bug Component Prediction (CBCP) model that mapped crash call-stack functions to components, computing IDF-based statistics to train a Random Forest for component-level fault localization.

(3) \textit{Multi-Field and Contextual Feature Fusion: }
Beyond text and code, several studies investigated additional report fields to enhance feature diversity. Sarkar et al.~\cite{sarkar2019improving} empirically demonstrated that alarm logs and crash dumps did not improve triage accuracy on Ericsson's dataset, and introduced a confidence-threshold mechanism for selective high-confidence predictions. Shokripour et al.~\cite{shokripour2015time} proposed a time-aware TF-IDF method that emphasized temporally relevant terms to better match developers' historical expertise. Sajedi et al.~\cite{sajedi2020vocabulary} refined textual input using a Stack Overflow–based technical vocabulary, retaining only domain-specific terms and weighting developer expertise by recent activity. Nath et al.~\cite{nath2021principal} combined discrete and textual features through PCA and entropy-based keyword filtering, using probabilistic multi-labeling to reflect team-level expertise. Li et al.~\cite{li2021revisiting} conducted a large-scale empirical analysis revealing that traditional textual features can degrade classifier performance due to noisy content such as code snippets and stack traces. Building on this, Li et al.~\cite{li2024automatic} found that nominal features, such as reporter, component, and priority, capture developer preferences more effectively than advanced NLP-derived text embeddings.

\textbf{Observability Data.} Logs provide chronological records of system events and serve as essential evidence for fault detection and diagnosis. Early systems such as SAS~\cite{lou2013sas} automated incident analysis by correlating heterogeneous logs and key performance indicators to generate reports and recommendations, supporting incident triage through human-in-the-loop decision-making.
Subsequent studies have advanced this line of research by leveraging large language models for semantic understanding and retrieval. COMET~\cite{wang2024comet} exemplifies this trend through an LLM-enhanced pipeline that transforms raw logs into compact representations and retrieves semantically similar historical incidents to assist fault management and team prediction. Moreover, ART~\cite{sun2024art} fuses metrics, logs, and traces into unified temporal representations, capturing both cross-modal and temporal dependencies.

Overall, these approaches illustrate a progression from early feature-driven systems to semantically enriched, LLM-augmented frameworks that improve automation and interpretability in incident analysis.

\textbf{Reviews.}
User reviews provide essential insights into user needs and product evolution, making their systematic extraction and interpretation crucial for informed software maintenance. Research in this field has evolved from explicit linkage construction to semantic understanding and automated analysis.

CRISTAL~\cite{palomba2015user} initiated this direction by linking user reviews with issues and commits, enabling traceability between feedback and development actions.
Wang et al.~\cite{wang2017app} extended this idea by quantifying these relationships through topic-based modeling, capturing how user requests and feature updates co-evolve over time.
Building on these foundations, ALLHANDS~\cite{zhang2025allhands} advances toward intelligent feedback understanding by leveraging large language models to transform unstructured reviews into structured semantic representations.

Collectively, these studies reflect a methodological progression from linking feedback to interpreting its semantic implications, highlighting the growing sophistication of user review analysis in supporting adaptive software evolution.

\textbf{Relational Data.} 
In addition to individual bug reports, relational data encodes the historical and contextual interactions among bugs, developers, and components, providing a structural foundation for enhancing bug triage accuracy. A range of approaches have leveraged these relations to extract informative features and represent developer expertise more effectively.

(1) \textit{Constructing Relational Networks:}
One major line of research focuses on constructing relational networks that explicitly capture the connections among developers, components, and bugs.
BugFixer~\cite{hu2014effective} models such relations through a Developer–Component–Bug network, integrating network-derived associations with textual similarity to jointly represent structural and semantic information for developer recommendation.
Building upon this idea, KSAP~\cite{zhang2016ksap} formulates a heterogeneous network and employs meta-path–based relational extraction to model collaboration patterns among developers. By coupling these structured relational features with nearest-neighbor search on historical reports, it strengthens the alignment between textual relevance and collaborative context.

(2) \textit{Representation Learning on Interaction Graphs: }While these methods rely on handcrafted relational structures, subsequent studies emphasize representation learning on interaction graphs, enabling more flexible and expressive modeling of complex dependencies.
PCG~\cite{dai2024pcg} embeds bugs and developers into a unified vector space, where prototype clustering and semantic contrastive learning jointly capture explicit and implicit relational cues.
NCGBT~\cite{dong2024neighborhood} extends this representation paradigm by introducing a bipartite graph formulation and neighborhood contrastive learning, thereby enhancing local structural consistency while preserving semantic proximity.
Complementarily, BPTRM~\cite{wei2025improving} exploits tossing relationships between developers to construct a personalized collaboration graph. By encoding these interaction patterns together with bug content features, it refines developer representations in a way that aligns historical behavior with current bug requirements.

(3) \textit{Temporal Dynamics: } A significant research direction involves incorporating temporal dynamics into the analysis of developer collaboration to model its evolution. 
For instance, Wu et al.~\cite{wu2022spatial} jointly model structural and temporal dependencies through a graph recurrent convolutional network that encodes developers' dynamic interaction preferences. 
Similarly, GCBT~\cite{dai2023graph} unifies spatial and temporal graph convolutions to extract both static and evolving expertise representations, with bug embeddings initialized via pretrained language models to ensure semantic consistency across tasks.

By combining these perspectives, relational data–driven approaches provide a comprehensive understanding of developer–bug interactions, leading to more accurate and context-aware bug triage.

\begin{center}
\footnotesize
\begin{longtable}{p{1.6cm}cccl}
    \caption{
        Summary of extracting feature from data in triage operations based on their category, the year, and the data. The abbreviations in the ``Data'' column stand for: \textbf{R} = Reviews, \textbf{IT} = Incident Tickets, \textbf{L} = Logs, \textbf{KPI} = Key Performance Indicators, \textbf{DN} = Dependency Networks, \textbf{CC} = Code Commits, \textbf{BR} = Bug Reports, \textbf{SC} = Source Code, \textbf{ST} = Stack Traces, and \textbf{MD} = Metadata (\eg historical developer assignments, comments).
    }
    \label{table:feature extraction} \\
    \toprule
    \textbf{Category} & \textbf{Technique}  & \textbf{Year} & \textbf{Data} & \textbf{Core Method} \\
    \midrule
    \endfirsthead
    
    \multicolumn{5}{c}%
    {\tablename\ \thetable\ -- \textit{Continued from previous page}} \\
    \toprule
    \textbf{Category} & \textbf{Technique}  & \textbf{Year} & \textbf{Data} & \textbf{Core Method} \\
    \midrule
    \endhead

    \midrule \multicolumn{5}{r}{\textit{Continued on next page}} \\
    \endfoot

    \bottomrule
    \endlastfoot
    \multirow{10}{*}{Incident Tickets}
    & Shao et al.~\cite{shao2008efficient} & 2008 & IT & \makecell[l]{Variable-Order Markov + VMS  Search} \\
    & IcM BRAIN framework~\cite{chen2020icmbrain} & 2020 & IT, R & AI-Based Techniques \\
    & Triangle~\cite{yu2021triangle} & 2021 & IT, KPI & \makecell[l]{Semantic Alignment + TF-IDF + Team Information Enrichment Mechanism}\\
    & COT~\cite{wang2021fast} & 2021 & IT & \makecell[l]{Historical Meta-Incident Construction + Correlation Graph Construction}\\
    & FaultProfIT~\cite{huang2024faultprofit} & 2024 & IT & Regex/Parsers + MacBERT\\
    & DiLink~\cite{ghosh2024dilink} & 2024 & IT, DN & \makecell[l]{TF-IDF + LSTM + Node2Vec + Graph Attention Networks + Orthogonal \\Procrustes Alignment}   \\
    & Goel~\cite{goel2024x} & 2024 & IT, KPI & GPT-3.5-turbo + FAISS\\
    & Mandal et al.~\cite{mandal2019improving}  & 2019 & IT & \makecell[l]{Contour Detection + Canny Edge Detection + ResNet50 + CNN + Tesseract}\\
    \midrule
    \multirow{17}{*}{Bug Reports}& Murphy et al.~\cite{murphy2004automatic} & 2004 & BR & Naive Bayes Classifier\\
    & Xuan et al.~\cite{xuan2010automatic} & 2010 & BR & \makecell[l]{Naive Bayes Model + Expectation-Maximization + Weighted \\Recommendation List}\\
    & Alenezi et al.~\cite{alenezi2013efficient} & 2013 & BR & Chi-Square Test + Naive Bayes Classifier\\
    & Panda et al.~\cite{panda2022topic} & 2022 & BR & LDA Topic Modeling + Intuitionistic Fuzzy Sets\\
    & Wang et al.~\cite{wang2024empirical} & 2024 & BR & \makecell[l]{Word2Vec + GloVe + NextBug + ELMo + BERT + LSTM + Bi-LSTM + \\TextCNN}\\
    & Park et al.~\cite{park2016cost} & 2016 & BR, SC & Import-Path Similarities\\
    & Sun et al.~\cite{sun2017enhancing} & 2017 & BR, CC & Collaborative Topic Modeling\\
    & Xu et al.~\cite{xu2023method} & 2023 & BR, ST & \makecell[l]{Crash Bug Component Prediction Model + IDF + Random Forest Model}\\
    & Sarkar et al.~\cite{sarkar2019improving} & 2019 & BR & \makecell[l]{Logistic Regression + Confidence-Threshold Mechanism} \\
    & Shokripour et al.~\cite{shokripour2015time} & 2015 & BR & \makecell[l]{Time-Aware TF-IDF + Ranking} \\
    & Sajedi et al.~\cite{sajedi2020vocabulary} & 2020 & BR & \makecell[l]{Stack Overflow-Based Vocabulary Filtering} \\
    & Nath et al.~\cite{nath2021principal} & 2021 & BR & \makecell[l]{PCA + Entropy-Based Keyword Selection + Contribution-Weighted \\Probability Labeling} \\
    & Li et al.~\cite{li2021revisiting} & 2021 & BR & \makecell[l]{VSM + TF-IDF} \\
    & Li et al.~\cite{li2024automatic} & 2024 & BR & \makecell[l]{TextCNN + SVM Classifiers} \\
    
    
    \multirow{3}{*}{Observability Data} 
    & SAS~\cite{lou2013sas} & 2013 & L, KPI & Data-Driven Techniques\\
    & COMET~\cite{wang2024comet} & 2024 & L & Multi-Stage Processing Pipeline + GPT + FastText\\
    & ART~\cite{sun2024art} & 2024 & L, M & \makecell[l]{Transformer Encoder Self-Attention + GRU + GraphSAGE}\\
    \midrule
    \multirow{4}{*}{Reviews}
    &CRISTAL~\cite{palomba2015user} & 2015 & CC, R & \makecell[l]{Issue and Commit Extractor + Information Retrieval Techniques}\\
    & Wang et al.~\cite{wang2017app} & 2017 & R, L & Biterm Topic Model\\
    & ALLHANDS~\cite{zhang2025allhands} & 2025 & IT, R & \makecell[l]{Regular Expressions + HTML Parsers +  LLM-Based Classification +  \\Abstractive Topic Model}\\ 
    \midrule
    \multirow{7}{*}{Relational Data}
    & BugFixer~\cite{hu2014effective} & 2014 & BR, MD & \makecell[l]{Developer-Component-Bug Network + Textual Similarity Search} \\
    & KSAP~\cite{zhang2016ksap} & 2016 & BR, MD & \makecell[l]{Developer-Component-Bug Network + K-Nearest-Neighbor Search} \\
    & PCG~\cite{dai2024pcg} & 2024 & BR, MD & \makecell[l]{Bug–Developer Interactions Graph + Semantic Contrastive Learning} \\
    & NCGBT~\cite{dong2024neighborhood} & 2025 & BR, MD & \makecell[l]{Bug–Developer Interactions Graph + Neighborhood Contrastive Learning} \\
    & BPTRM~\cite{wei2025improving} & 2025 & BR, MD & \makecell[l]{Bug–Developer Interactions Graph} \\
    & Wu et al.~\cite{wu2022spatial} & 2022 & BR, MD & \makecell[l]{Developer Collaboration Networks + JRWalk + GRCNN} \\
    &GCBT~\cite{dai2023graph} & 2023 & BR, MD & \makecell[l]{Bug–Developer Interactions Graph + IR-Based Classifier} \\
\end{longtable}
\vspace{-0.5cm}
\end{center}
\subsection{Prioritization}

Prioritization in triage refers to the systematic process of ranking or ordering issues such as software bugs, system alerts, or performance anomalies according to predefined criteria, including severity, urgency, cost, or relevance. It plays a vital role in enabling efficient resource allocation and ensuring timely issue resolution within software engineering and IT operations. In software development and system maintenance, triage entails analyzing incoming data to identify, classify, and delegate issues to appropriate personnel or automated procedures. Prioritization determines the order in which issues are addressed, allowing critical problems to be resolved promptly, which helps minimize downtime, control operational costs, and enhance user satisfaction.

The effectiveness of prioritization lies in its ability to optimize resource utilization, such as developer time or computing capacity, while improving system reliability and operational efficiency. Its importance becomes particularly evident in large-scale systems, where the volume and heterogeneity of incoming issues can easily exceed manual processing capabilities. Sound prioritization strategies address key challenges such as sparse data, information overload, and diverse issue types. This section synthesizes prior research on prioritization in triage systems, with a focus on categorizing studies based on the output data types.

\subsubsection{Severity Rating} 

In both software development and cloud incident management, severity serves as a critical criterion for classifying and prioritizing bugs and incidents according to their potential impact and the urgency of remediation. Clearly defined severity levels enable development teams, testers, operators, and other stakeholders to systematically assess the significance of addressing specific faults and to allocate resources accordingly.

\textbf{Incident Tickets.} Effective incident prioritization aims to distinguish critical events from incidental or low-impact ones, thereby enabling timely and resource-efficient responses. 

Existing studies approach this challenge from complementary perspectives that progressively enhance interpretability, contextual awareness, and automation.
DeepIP~\cite{chen2021deepip} addresses the problem from a data-driven learning perspective. It employs a neural model to identify incidents that are likely incidental and ranks all reported events according to their predicted relevance, thereby supporting more focused remediation.
Building on this concept of data-guided prioritization, Saurabh et al.~\cite{malgaonkar2022prioritizing} incorporate multi-criteria features and expert knowledge into the prioritization process. Their framework integrates statistical and linguistic indicators through a heuristic weighting scheme and refines the prioritization through regression modeling validated by domain experts, thus enhancing the interpretability and reliability of the results.
Extending these approaches toward richer contextual reasoning, Sadlek et al.~\cite{sadlek2025severity} introduce an attack-graph-based framework that evaluates incident severity in relation to potential kill-chain progressions and asset criticality. By embedding prioritization within a structured representation of cyberattack sequences, this method captures interdependencies among alerts and enables severity assessment grounded in operational context.

\textbf{Bug Reports.} 
A substantial body of research has focused on automating bug severity and priority prediction by leveraging both textual content and metadata from bug reports~\cite{cook2020using,alazzam2020automatic}.
These studies can be broadly divided into two complementary lines: (1) severity estimation and prioritization, which infer the relative importance of bugs, and (2) severity-aware triage optimization, which integrates severity signals into task assignment and scheduling.

(1) \textit{Severity Estimation and Prioritization: }
Within the first line of work, several studies adopt classification and similarity-based learning to infer severity levels.
Kanwal et al.~\cite{kanwal2012bug} utilized supervised classifiers such as SVM and Naïve Bayes to quantify the contribution of different bug report features and establish evaluation metrics for automated prioritization.
Zhang et al.~\cite{zhang2016towards} further incorporated historical case similarity, combining feature-based learning with K-nearest retrieval to improve the reliability of severity prediction and fixer recommendation.

(2) \textit{Severity-Aware Triage Optimization: }
Building on these predictive foundations, subsequent research integrates severity inference into broader triage and scheduling frameworks.
Jahanshahi et al.~\cite{jahanshahi2022wayback} reconstructed historical bug dependency graphs to analyze how severity interacts with bug interdependencies and developer workloads, providing dynamic evaluation metrics for severity-driven triage performance.
Extending this idea, their later work~\cite{jahanshahi2022s} introduced a schedule- and dependency-aware framework that models bug-developer assignment as an optimization problem, incorporating severity, dependency, and workload constraints to improve triage efficiency.

Recent studies have enhanced these severity prediction pipelines through deeper semantic modeling and uncertainty handling.
Arshad et al.~\cite{arshad2024sevpredict} employed a transformer-based architecture to capture contextual and sentiment information within bug reports, enabling semantic reasoning for real-time severity prediction.
Panda et al.~\cite{panda2024software} further addressed uncertainty and class imbalance by integrating topic modeling with intuitionistic fuzzy representations, allowing the model to express multi-level priority associations with soft, probabilistic interpretations.

Collectively, these studies reveal a coherent methodological landscape in which severity estimation serves as a foundational signal for downstream triage optimization.
Traditional classification and similarity learning establish interpretable feature mappings, dependency- and schedule-aware frameworks operationalize severity within realistic coordination constraints, and recent transformer- and fuzzy-based models extend this pipeline toward context-sensitive and uncertainty-resilient prioritization.

\textbf{Alerts.} In the context of alert prioritization, research has progressively advanced from probabilistic modeling to feature-driven learning frameworks.
Lin et al.~\cite{lin2018car} proposed CAR, which captures temporal and content correlations among heterogeneous alerts through a hierarchical probabilistic framework, enabling the ranking of both individual alerts and their underlying patterns.
Subsequently, Zhao et al.~\cite{zhao2020alertrank} developed AlertRank, which integrates textual, temporal, and anomaly-related information into a unified feature representation and applies a learning-to-rank model to assign severity scores.
Together, these methods reflect a shift from statistical dependency modeling toward interpretable, data-driven approaches for effective alert prioritization.

\begin{center}
\footnotesize
\begin{longtable}{p{1.5cm}cccl}
    \caption{
       Summary of severity rating based on their category, the year, and the data. The abbreviations in the ``Data'' column stand for: \textbf{A} = Alerts, \textbf{R} = Reviews, \textbf{IT} = Incident Tickets, \textbf{L} = Logs, \textbf{KPI} = Key Performance Indicators, and \textbf{BR} = Bug Reports.
    }
    \label{table:Severity Rating} \\
    \toprule
    \textbf{Category} & \textbf{Technique}  & \textbf{Year} & \textbf{Data} & \textbf{Core Method} \\
    \midrule
    \endfirsthead
    \toprule
    \textbf{Category} & \textbf{Technique}  & \textbf{Year} & \textbf{Data} & \textbf{Core Method} \\
    \midrule
    \endhead
    \bottomrule
    \endlastfoot
        \multirow{3}{*}{Incident Tickets} & DeepIP~\cite{chen2021deepip} & 2021 & IT & Attention-Based Convolutional Neural Network\\
        &Sadlek et al. \cite{sadlek2025severity} & 2025 & IT, L, KPI & \makecell[l]{MulVAL Attack Graph Generator + MITRE ATT \&CK Techniques}\\
        &Saurabh et al. \cite{malgaonkar2022prioritizing}& 2022 & IT, R & \makecell[l]{Regression-Based Prioritization Technique + Weighted Heuristic}\\
        \midrule
        \multirow{6}{*}{Bug Reports}&Kanwal et al.~\cite{kanwal2012bug} & 2012 & BR & SVM + Naive Bayes Classifier\\
        &Zhang et al.~\cite{zhang2016towards} & 2016 & BR & REP Algorithm + K-Nearest Neighbor Classification\\\
        &Arshad et al.~\cite{arshad2024sevpredict} & 2024 & BR & \makecell[l]{Fine-Tuned Transformer Model}\\
        &Panda et al.~\cite{panda2024software} & 2024 & BR & \makecell[l]{Intuitionistic Fuzzy Sets + Topic Modeling}\\
        &Jahanshahi et al.~\cite{jahanshahi2022wayback} & 2022 & BR & \makecell[l]{Bug Dependency Graph + Developer Load Tracking} \\
        &Jahanshahi et al.~\cite{jahanshahi2022s}& 2022 & BR & \makecell[l]{Schedule and Dependency-Aware Bug Triage + SVM/LDA + Integer Programming}\\
        \midrule
        \multirow{2}{*}{Alerts} & CAR~\cite{lin2018car} & 2018 & A & Hierarchical Bayesian + Entity Embedding-Based Approach \\
        &AlertRank~\cite{zhao2020alertrank} & 2020 & A, KPI & Feature Set + XGBoost Ranking\\
\end{longtable}
\vspace{-0.5cm}
\end{center}

\subsubsection{Issue Type Classification}

In triage, accurately distinguishing between recurring issues and previously unseen problems is crucial, as it enables the reuse of established repair procedures for known issues or the initiation of in-depth investigation for unknown ones. Moreover, the ability to identify anomalous reports that are directly associated with actual faults from a large volume of alerts, particularly during alert floods, allows engineers to prioritize the most critical incidents. This targeted focus not only reduces cognitive load but also enhances the efficiency and timeliness of fault response. Furthermore, such prioritization plays a key role in mitigating operational risks, preventing service degradation, and optimizing resource allocation in large-scale systems.

\textbf{Structure Information.} 
For algorithms with low dependency on historical data, unsupervised and lightweight supervised techniques are commonly employed to classify issue types.
These methods can be broadly categorized into clustering-based anomaly grouping and classification-based issue typing, reflecting the evolution from pattern discovery to semantically guided categorization.

(1) \textit{Clustering-Based Methods: }
Early efforts emphasized unsupervised grouping of anomalies and updates based on statistical and structural similarity.
Lim et al.~\cite{lim2014identifying} introduced a Hidden Markov Random Field combined with EM-based clustering (HMRF-kMedoid-EM) to detect performance anomaly categories. The model discretizes performance metrics, identifies salient attributes, and supports systematic triage through cluster-driven anomaly interpretation.
Lin et al.~\cite{lin2014unveiling} improved clustering precision by detecting connected components and recursively refining large clusters via graph cuts, followed by Non-negative Matrix Factorization (NMF) for dimensionality reduction and hierarchical grouping through KD-tree and linkage clustering.
Wang et al.~\cite{wang2017app} applied k-means to cluster update activities (AUs), using Euclidean similarity and average-distance variation to determine cluster numbers. Each centroid represented an Update Pattern reflecting the evolution of feature demand and responsiveness.
Later, Zhao et al.~\cite{zhao2020understanding} extended clustering-based detection to streaming alerts, combining Extreme Value Theory (EVT) with Isolation Forest denoising and clustering to identify alert storms and select representative alerts, thus reducing diagnostic redundancy while preserving key signals.

(2) \textit{Classification-Based Methods: }
Subsequent studies shifted toward supervised and multi-label formulations to directly predict issue categories.
Catolino et al.~\cite{catolino2019not} established an automated bug classification framework grounded in a nine-class taxonomy (\eg Configuration, Network, GUI, Security), derived through iterative content analysis of 1,280 bug reports.
Building on such taxonomy-driven approaches, Meher et al.~\cite{meher2024deep} developed a deep learning framework that integrates expert-validated keyword sets for eight bug types and heuristically annotated 1.36 M reports using Word Mover's Distance, enabling attention-based models (Transformer, BERT, CodeBERT, DistilBERT) to perform fine-grained classification.
Aung et al.~\cite{aung2022multi} proposed Multi-Triage, a multi-label model jointly predicting issue types and responsible developers. By separating text and code tokens and augmenting contextual data, it improves the robustness of imbalanced classification.
Expanding the label scope, Sepahvand et al.~\cite{sepahvand2023using} distinguished design-related from implementation-related defects via a CNN that fuses textual and code-smell features, effectively linking defect symptoms with design-level anti-patterns such as large classes or high complexity.

Overall, these studies demonstrate a clear methodological evolution, from unsupervised clustering for anomaly grouping toward deep and multi-label classification frameworks that incorporate semantic, contextual, and design-level cues for precise issue typing.

\textbf{Historical Information} 
The utilization of historical data is crucial for issue type classification, as it enables models to capture recurring patterns, contextual dependencies, and developer behavior over time, thereby improving the accuracy of issue categorization and supporting more consistent, data-driven decision-making in software maintenance processes.

In the early stage of incident and bug report analysis, research primarily focused on structural modeling and knowledge-driven reasoning.
Xuan et al.~\cite{xuan2012developer} first leveraged developer social networks to rank contribution priorities and enhance bug triage and severity prediction.
Building on structural similarity, Park et al.~\cite{park2016cost} addressed non-textual bug reports such as crash logs by computing code import path similarity using Jaccard or tree-edit distance and classifying reports with k-nearest neighbors.
Further extending this direction, Zeng et al.~\cite{zeng2017knowledge} introduced Kilo, which encoded expert domain knowledge as hierarchical vectors within a probabilistic graphical model, enabling multi-label reasoning through a sum–product inference mechanism.

Subsequent work shifted toward representation learning and semantic similarity modeling to overcome the limitations of purely structural features.
LinkCM~\cite{gu2020linkcm} proposed a symmetric multi-instance model to learn effective representations from correlated system incidents, updating only fully connected layers with labeled CI–MI linking data for efficient triage.
Haering et al.~\cite{haering2021automatically} further enhanced semantic linkage by encoding problem and bug reports into context-sensitive vector embeddings using DistilBERT and ranking their cosine similarity, thereby improving automatic bug report recommendation and linkage accuracy.

Recent studies have advanced toward hierarchical and large-model–based reasoning frameworks for complex failure understanding.
ART~\cite{sun2024art} adopted Extreme Value Theory and cut-tree clustering to detect anomalies and assign failures by modeling system-level deviations, while FaultProfIT~\cite{huang2024faultprofit} utilized hierarchy-guided contrastive learning over MacBERT and Graphormer encoders to capture multi-level fault taxonomy.
Building on the progress of large language models, ALLHANDS~\cite{zhang2025allhands} integrated in-context learning with retrieval-based prompting and human-in-the-loop refinement for feedback categorization and topic modeling.
Its LLM-based QA agent further decomposes analytical queries, generates executable Python code, and produces multimodal insights, signifying the transition toward intelligent, knowledge-grounded incident analysis.

\begin{center}
\footnotesize
\begin{singlespace}
\begin{longtable}{cccl}
\caption{
Summary of issue type classification based on their category, the year, and the core method.
} 
\label{table:Issue Type Classification} \\

\toprule
\textbf{Category} & \textbf{Technique} & \textbf{Year} & \textbf{Core Method} \\
\midrule
\endfirsthead

\multicolumn{4}{c}%
{\tablename\ \thetable\ -- \textit{Continued from previous page}} \\
\toprule
\textbf{Category} & \textbf{Technique} & \textbf{Year} & \textbf{Core Method} \\
\midrule
\endhead

\midrule \multicolumn{4}{r}{\textit{Continued on next page}} \\
\endfoot

\bottomrule
\endlastfoot
        \multirow{4}{*}{Structure Information} 
        & Lim et al.~\cite{lim2014identifying} & 2014 & \makecell[l]{Hidden Markov Random Field + EM-Based Clustering (HMRF-kMedoid-EM) }\\
        &Lin et al. \cite{lin2014unveiling}& 2014 & Non-Negative Matrix Factorization + KD-Tree\\
        &Wang et al.\cite{wang2017app}& 2017 & K-Means + Euclidean Distance\\
        &Zhao et al.~\cite{zhao2020understanding} & 2020 & \makecell[l]{Extreme Value Theory + Isolation Forest-Based Denoising}\\
        \multirow{4}{*}{\makecell{Structure Information\\(continued)}}&Catolino et al.~\cite{catolino2019not} & 2019 & \makecell[l]{Taxonomy Construction + Classification Model} \\
        &Meher et al.~\cite{meher2024deep} & 2024 & \makecell[l]{Taxonomy Construction + Attention-Based Classification Model} \\
        &Aung et al.~\cite{aung2022multi}& 2022 & \makecell[l]{Contextual Data Augmentation + Multi-Label Classifying} \\
        & Sepahvand et al.~\cite{sepahvand2023using} & 2023 & CNN-Based Model + PMD-Based Analysis \\
        \midrule
        \multirow{10}{*}{Historical Information} 
        & Xuan et al.~\cite{xuan2012developer}& 2012 & Social Network \\
        &Park et al.~\cite{park2016cost} & 2016 & \makecell[l]{SVM  +  LDA Topic Modeling  +  Content-Based Recommendation  +  Content-Boosted \\Collaborative}\\
        &Kilo~\cite{zeng2017knowledge} & 2017 & Domain Expert Knowledge + Sum-Product Algorithm\\
        &LinkCM \cite{gu2020linkcm} & 2020 & Symmetric Model + Fully-Connected Network\\
        & Haering et al. \cite{haering2021automatically} & 2021 & DistilBERT\\
        &ART \cite{sun2024art} & 2024 & Extreme Value Theory  + Cut-Tree Clustering\\
        &FaultProfIT~\cite{huang2024faultprofit} & 2024 & \makecell[l]{MacBERT-Based Incident Encoder + Graphormer-Based Hierarchy Encoder}\\
        &ALLHANDS~\cite{zhang2025allhands} & 2025 & \makecell[l]{Multi-Task In-depth Analysis + Hierarchical Agglomerative Clustering + LLM-Based \\QA Agent}\\
\end{longtable}
\end{singlespace}
\vspace{-0.5cm}
\end{center}

\subsection{Assignment}

In modern triage systems, the assignment of issues to appropriate components and developers can be partially or fully automated through predefined rules or machine learning–based algorithms. These assignment mechanisms typically leverage a combination of factors, including the issue's associated component, the historical workload distribution among developers, and prior assignment patterns. Effective component assignment ensures that the issue is directed to the correct functional module or subsystem, thereby enabling precise fault localization. Subsequent developer assignment further refines the process by selecting the most suitable individual or team based on domain expertise, prior experience with similar issues, and current availability.

Once an issue has been assigned, the triage system provides the assignee with comprehensive report details to facilitate efficient resolution. This information package generally includes the issue description, steps to reproduce the problem, relevant logs, screenshots, code snippets, and any other contextual data collected during the classification phase. Providing complete and well-structured report details is critical for ensuring that developers or maintenance teams gain an immediate and accurate understanding of the problem, thereby reducing the need for repeated clarification, minimizing handoff delays, and ultimately accelerating defect resolution in large-scale software systems.

\subsubsection{Component Assignment}

When assigning issues to components, triage teams typically evaluate the functional scope, historical fault patterns, and recent change history of each component. Components with a documented history of similar faults or recent code modifications are often prioritized for investigation, as they present a higher likelihood of being the fault source. In addition, the stability profile and maintenance ownership of a component are critical considerations; assigning issues to components actively maintained by dedicated teams can facilitate faster diagnosis and resolution. By aligning issue assignment with component expertise, change context, and operational responsibility, triagers can improve fault localization accuracy, reduce unnecessary cross-component investigation, and enhance the overall efficiency of defect management in large-scale software systems.

\textbf{Text Classification.} Early studies on text-based component assignment applied traditional machine learning and knowledge-engineering approaches, where statistical models and expert-defined rules were used to map event or report attributes to predefined categories~\cite{gupta2009multi}.
With the rise of transformer architectures, research gradually shifted toward fine-tuning pre-trained language models, achieving consistent improvements over earlier neural or feature-based baselines while showing that traditional TF-IDF methods remain competitive in certain constrained settings~\cite{dipongkor2023comparative}.
More recent efforts emphasize industrial applicability. Borg et al.~\cite{borg2024adopting} propose a routing framework that assigns reports to software modules rather than individuals and integrates confidence-based human oversight, balancing automation and reliability in large-scale maintenance workflows.

\textbf{Information Retrieval.}
Beyond text classification, another common approach to component assignment is information retrieval, which locates historical cases or components relevant to a new report.

(1) \textit{Similarity-Based Retrieval:} 
Early approaches primarily relied on textual similarity and statistical correlation to locate related incidents.
SAS~\cite{lou2013sas} detects incident beacons through anomaly analysis and probabilistic correlation, and retrieves recurring failures via signature-based matching and log-likelihood evaluation.
Ding et al.\cite{ding2014mining} extend textual retrieval by constructing structured action triples $\langle$verb, target, location$\rangle$, combining semantic parsing of descriptions with log-based context extraction.
CRISTAL\cite{palomba2015user} further enriches similarity computation by integrating textual information from commit notes and code identifiers, and computes asymmetric Dice similarity between reviews and issue reports.
Together, these studies represent the foundation of retrieval-based triage, relying on handcrafted linguistic or statistical features to quantify incident similarity.

(2) \textit{Representation Learning for Retrieval:} 
With advances in deep learning, representation learning has become central to retrieval accuracy and scalability.
LiDAR~\cite{chen2020lidar} learns joint textual and component embeddings to generate unified vector representations, from which linkage confidence between incidents can be efficiently computed.
Zhang et al.\cite{zhang2020efficient} employ a deep neural model that projects bug descriptions into discriminative latent spaces, balancing component frequency bias through class-weighted loss and enabling near real-time assignment.
Beyond textual embeddings, Xu et al.\cite{xu2023method} incorporate structured and behavioral features by encoding call-stack frequency and positional information, allowing machine learning models to predict fault-prone components and reduce reassignment delay.
Collectively, these methods advance incident retrieval from surface-level similarity toward semantically grounded representation learning.

(3) \textit{Hybrid and LLM-Enhanced Retrieval:} 
Recent work integrates retrieval-based correlation mining with large language model reasoning to enhance interpretability and adaptability.
COLA~\cite{kuang2024cola} combines probabilistic and graph-based correlation mining, leveraging conditional probability, node2vec, and Skip-gram, with an LLM reasoning module based on two-round prompting, in-context learning, and parameter-efficient tuning for improved F1 performance.
Goel et al.~\cite{goel2024x} further demonstrate the potential of LLMs by incorporating X-lifecycle data into GPT-4–based reasoning pipelines for root cause recommendation and monitor classification, integrating both event semantics and dependency metadata.
These hybrid designs illustrate a shift toward intelligent retrieval systems that unify statistical correlation, learned representation, and natural language reasoning within a single framework.

Overall, the evolution of information retrieval methods reflects a clear trajectory, from lexical similarity to representation learning and finally to LLM-based reasoning, toward more adaptive, semantically rich, and context-aware incident triage.

\textbf{Social Network Modeling.} Following retrieval-based approaches, researchers have explored social and dependency structures to enhance triage effectiveness.
Su et al.~\cite{su2021reducing} leveraged a bug-tossing knowledge graph within a learning-to-rank framework, later extended by DEEPTRIAG~\cite{su2023still} through deep ensemble modeling.
At the service level, COT~\cite{wang2021fast} employed graph-based reasoning to capture interactions among incidents and components.
These efforts move triage modeling toward relational and graph-centric representations.

\textbf{Data Bias Modeling.} In parallel, some studies address the data bias and concept drift that emerge in continuously evolving bug report streams.
Chrupala et al.\cite{chrupala2012learning} applied online learning algorithms to adapt component assignment models in real time, mitigating distributional shifts.
Mandal et al.\cite{mandal2019improving} further tackled data imbalance by combining ensemble classification with information retrieval, enabling robust resolution prediction across both frequent and rare problem categories.
These approaches underscore growing attention to temporal dynamics and long-tail effects, establishing data bias modeling as a complement to content- and structure-based triage.

\begin{center}
\footnotesize
\begin{singlespace}
\begin{longtable}{ccccl}
\caption{
 Summary of component assignment methods categorized by their problem formulation. The abbreviations in the ``Category'' column stand for: \textbf{TC} = Text Classification, \textbf{IR} = Information Retrieval, \textbf{SN} = Social Network Modeling, and \textbf{DB} = Data Bias Modeling. The abbreviations in the ``Data'' column stand for: \textbf{L} = Logs, \textbf{M} = Metrics, \textbf{IT} = Incident Tickets, \textbf{A} = Alerts, \textbf{R} = Reviews, \textbf{BR} = Bug Reports, \textbf{HD} = Historical Documents, \textbf{CC} = Code Commits, and \textbf{DN} = Dependency Networks.
} 
\label{table:deduplication} \\

\toprule
\textbf{Category} & \textbf{Technique} & \textbf{Year} & \textbf{Data} & \textbf{Core Method} \\
\midrule
\endfirsthead

\multicolumn{5}{c}%
{\tablename\ \thetable\ -- \textit{Continued from previous page}} \\
\toprule
\textbf{Category} & \textbf{Technique} & \textbf{Year} & \textbf{Data} & \textbf{Core Method} \\
\midrule
\endhead

\midrule \multicolumn{5}{r}{\textit{Continued on next page}} \\
\endfoot

\bottomrule
\endlastfoot
        \multirow{3}{*}{\textbf{TC}} & Gupta et al.~\cite{gupta2009multi} & 2009 & IT, L, M & \makecell[l]{Machine-Learning + Naive Bayesian + Knowledge-Engineering Approach} \\
        & Dipongkor et al. ~\cite{dipongkor2023comparative} & 2023 & BR & Transformer-Based Models + DeBERTa \\
        & Borg et al.~\cite{borg2024adopting} & 2024 & BR, R & \makecell[l]{Logistic Regression + Confidence-Based Human-in-the-Loop Strategy}\\
        \midrule
        \multirow{10}{*}{\textbf{IR}} & SAS~\cite{lou2013sas} & 2013 & IT, L & FCA + DMI + Signature-Based Retrieval\\
        & Ding et al. ~\cite{ding2014mining} & 2014 & L, HD & Cosine Score + Generating Triple Structures\\
        & CRISTAL ~\cite{palomba2015user} & 2015 & IT, CC & Asymmetric Dice Similarity Coefficient\\
        & LiDAR~\cite{chen2020lidar} & 2020 & IT, DN & Calculate Linkage Confidence Score\\
        &Zhang et al.~\cite{zhang2020efficient} & 2020 & BR & \makecell[l]{Neural Network + Text-Projection Features with Class-Based Weighting}\\
        &Xu et al.~\cite{xu2023method} & 2023 & BR & \makecell[l]{Aggregating the Frequency and Position of Component Functions + Trained Machine \\Learning Model}\\
        & COLA~\cite{kuang2024cola} & 2024 & A, SOP & \makecell[l]{Conditional Probability + Node2Vec + Skip-Gram + Jaccard Similarity Denoising + \\Two-Round Prompting + ICL}\\
        & Goel~\cite{goel2024x} & 2024 & IT, M & Multi-Stage Data Processing + LLM Inference\\
        \midrule
        \multirow{3}{*}{\textbf{SN}} & Su et al. ~\cite{su2021reducing} & 2021 & BR & \makecell[l]{LambdaMART-Based Learning-to-Rank Framework + Bug Tossing Knowledge Graph}\\
        & DEEPTRIAG~\cite{su2023still} & 2023 & BR & Deep Ensemble Model\\
        & COT~\cite{wang2021fast} & 2021 & IT & Text Parsing + Graph Construction + SVM + Decision Tree\\
        \midrule
        \multirow{2}{*}{\textbf{DB}} & Chrupala et al.~\cite{chrupala2012learning} & 2012 & BR & \makecell[l]{Online Learning Algorithms}\\
        & Mandal et al.~\cite{mandal2019improving} & 2019 & IT & \makecell[l]{Linear SVM  +  MLP Ensemble Classifier +  Query Predefined Solutions + CORI Algorithm}\\
\end{longtable}
\end{singlespace}
\vspace{-0.5cm}
\end{center}


\subsubsection{Developer Assignment} 

During issue assignment, triage teams often consider developers' historical performance and prior experience in resolving similar problems. Developers who demonstrate a proven track record of successfully addressing comparable issues or consistently delivering high-quality fault fixes are frequently prioritized for assignment. Beyond individual expertise, effective collaboration and communication are essential for addressing complex, cross-cutting faults. Consequently, triagers may allocate issues to teams or individuals with a documented history of successful collaboration, thereby ensuring that established communication channels and coordination mechanisms are leveraged. Such assignment strategies not only increase the likelihood of timely resolution but also help reduce coordination overhead and improve overall fault management efficiency in large-scale software systems. 

From a methodological perspective, prior studies on automated developer assignment can be categorized based on how they model the assignment problem. Text Classification (TC) approaches treat the task as a supervised classification problem, where each developer corresponds to a potential class label and models are trained to predict the most suitable assignee based on the textual content of bug reports~\cite{lin2009empirical,sureka2015decision,florea2017parallel,wei2018enhancing,zhao2019unified,guo2020developer,wang2021effective}. In contrast, Information Retrieval (IR) methods frame assignment as a similarity search, ranking candidate developers according to the relevance of their historical records to the current issue, often using techniques such as TF-IDF, BM25, or topic modeling to compute textual similarity~\cite{peng2017improving}. Social or Collaboration Network (SN) approaches exploit the structural relationships among developers, files, and modules, capturing patterns of past collaboration and expertise propagation through graph-based models or network embeddings~\cite{wu2011drex,naguib2013bug}. Optimization or Decision-making (OPT) approaches formulate assignment as a constrained optimization or sequential decision problem, balancing multiple objectives such as expertise, workload, and cost, and often employing search algorithms or evolutionary strategies~\cite{rahman2009optimized,liu2016multi,gupta2021decentralized,jahanshahi2021dabt}. Finally, Other / Hybrid (OTH) approaches integrate multiple paradigms or techniques, such as combining rule-based, ML, embedding, and network features, to leverage complementary information sources for more robust and context-aware assignment~\cite{zhang2014novel,tian2016learning,yadav2019ranking,huang2019predicting,yu2021bug,devaiya2021castr,devaiya2021evaluating}.

\textbf{Text Classification.} 
Focusing on TC, prior work has progressed from traditional supervised models to deep learning and PLM-based approaches for automated bug assignment.

(1) \textit{Traditional Machine Learning Approaches:} 
As one of the earliest studies on automated bug assignment, Anvik et al.~\cite{anvik2006automating,anvik2006should} trained Naïve Bayes and SVM models on labeled bug reports, while Anvik et al.~\cite{anvik2011reducing} incorporated project metadata and historical reports to produce ranked developer lists. Ahsan et al.~\cite{ahsan2009automatic} combined feature selection with Latent Semantic Indexing to reduce TF-IDF matrices for classifier training. Costriage~\cite{park2011costriage} integrated per-developer SVM classifiers with collaborative filtering and cost estimation for effort-aware ranking.

(2) \textit{Deep Learning Approaches:} 
Deep learning methods capture semantic and sequential dependencies. Lee et al.~\cite{lee2017applying} used CNNs with Word2Vec embeddings and dynamic retraining. DeepTriage~\cite{mani2019deeptriage} employed bidirectional RNNs with attention, while iTriage~\cite{xi2019bug} integrated sequence-to-sequence textual modeling with metadata features.

(3) \textit{Pre-trained Language Models and Ensemble Strategies:} 
Following this wave of neural architectures, pre-trained language models (PLMs) and model ensembles have further advanced the state of the art in automated assignment. 
LBT-P~\cite{lee2022light} leverages multi-layer PLM embeddings, mitigating catastrophic forgetting, and producing efficient developer rankings. 
Wang et al.\cite{wang2024empirical} empirically studied combinations of embeddings and classifiers, and Dipongkor\cite{dipongkor2024ensemble} demonstrated that ensembles of transformer-based LLMs outperform individual models.

(4) \textit{Industrial Scale and Robust Deployment:} 
The latest studies focus on practical, large-scale deployment, robustness, and multi-source integration.
Park et al.~\cite{park2016cost} extended Costriage to non-textual reports using LDA and content-boosted collaborative filtering. Jonsson et al.~\cite{jonsson2016automated} applied ensemble-based stacked classifiers to assign bug reports to teams. Sarkar et al.~\cite{sarkar2019improving} combined logistic regression with incremental learning and high-confidence prediction for robust industrial triage.
More recent work, such as BTAL~\cite{zhang2025btal} integrates multiple bug report sources, encodes textual content with BERT and TextCNN, and applies adaptive loss functions to mitigate class imbalance. FLSCL~\cite{wang2025fixer} leverages supervised contrastive learning over fixers' reports, producing robust embeddings fed into Bi-LSTM or BERT classifiers.

Taken together, the evolution of TC-based bug triage reflects both methodological innovation and practical impact: advances in representation learning and model ensembling have enhanced predictive performance, while integration of multi-source data and robust industrial strategies ensures applicability in large-scale, real-world software development environments.

\textbf{Information Retrieval.} 
Early IR-based approaches primarily leveraged textual similarity, while later studies incorporated temporal, contextual, structural, and network information to capture evolving developer expertise.

(1) \textit{Textual Similarity and Fuzzy Expertise Modeling:} 
Bugzie~\cite{tamrawi2011bugzie} represents developer expertise using fuzzy sets combined with a cache-based mechanism, continuously updating membership scores from newly fixed bug reports. Incoming bugs are assigned by aggregating relevant fuzzy sets, improving accuracy and efficiency. Panda et al.~\cite{panda2022topic} introduced a framework combining LDA topic modeling with Intuitionistic Fuzzy Sets (IFS) to represent uncertainty in developer expertise. Similarity measures and fuzzy $\alpha$-cut selection enable the identification of developers likely to resolve new bugs.

(2) \textit{Contextual and Structural Enhancements:} 
Subsequent work extended textual similarity by modeling temporal and contextual dynamics. VTBA~\cite{sajedi2020vocabulary} ranks developers based on technical term matching filtered via Stack Overflow and weights historical fixes by recency, representing each developer as a ``document'' of past bug sub-documents. Goyal~\cite{goyal2017effective} presents Visheshagya for time-based assignment, W8Prioritizer for prioritization via AHP, and NRFixer for predicting fixability of non-reproducible bugs, forming a comprehensive recommender system for diverse bug types.
Structural and authorship information has also been leveraged: Linares-Vásquez et al.\cite{linares2012triaging} combine IR with source code authorship, ranking developers by file headers, while Shokripour et al.\cite{shokripour2013so} restrict candidate files to those within the reported component and compute each developer's expertise based on the number of past change activities on those files, weighted by the recency of the changes.

(3) \textit{Latent and Network-Based Representations:} 
Another line of work explored latent relationships through topic modeling and network-based representations of developer-bug interactions. 
Dretom~\cite{xie2012dretom} models developer expertise and interest using topic models derived from historical bug-resolving records, enabling the ranking of developers for new bug reports.
Building on network representations, BugFixer~\cite{hu2014effective} constructs a Developer–Component–Bug network, ranking developers based on both connectivity and similarity to historical bug-fix reports. Furthermore, $TopicMiner^{MTM}$~\cite{xia2017topicminer} extends LDA with multi-feature topic modeling that incorporates product and component information, enabling incremental developer assignment and significantly improving triage accuracy.

(4) \textit{Interactive Retrieval Systems:} 
Complementing automated retrieval, interactive systems have been designed to assist developers in managing and exploring bug reports.
PorchLight~\cite{bortis2013porchlight} introduces a specialized query language for tagging bug reports, enabling developers to organize and explore them in meaningful groups and thus mitigating the inefficiencies of one-by-one inspection in large-scale triaging scenarios. 

IR-based approaches have evolved from purely textual similarity and fuzzy expertise modeling to incorporate temporal, contextual, structural, and network information. The integration of interactive systems further enhances developer support, collectively enabling more accurate and scalable bug assignment.

\textbf{Social or Collaboration Network.} 
SN approaches exploit the structural relationships among developers, files, and modules to capture collaboration patterns and expertise propagation, enabling more informed bug assignment decisions.

(1) \textit{Early Graph-Based Methods:} 
Initial studies focused on modeling developer collaboration and bug flow using static or manually constructed networks.
Jeong et al.\cite{jeong2009improving} construct bug tossing graphs to represent developer collaboration and team structure, integrating historical assignment sequences to predict suitable developers and identify shorter paths to the fixer. Building on this idea, Bhattacharya et al.\cite{bhattacharya2012automated} leverage multi-feature tossing graphs to model developer activity and bug flow more comprehensively. They combine these graphs with incremental classifiers to provide textual support, effectively identifying potential developers while reducing tossing path lengths. Similarly, FixerCache~\cite{wang2014fixercache} introduces an unsupervised strategy that dynamically maintains developer caches for each component, ranking developers by activeness. This approach achieves high prediction accuracy and diversity while avoiding the training overhead associated with supervised models.

(2) \textit{Community-Level Extensions:} 
Subsequent approaches extended static networks to community-level structures, emphasizing collaborative expertise and developer clusters. 
DECOBA~\cite{banitaan2013decoba} constructs social networks of developers based on their bug-fixing contributions and detects communities. For a new bug report, it assigns a relevant developer community and ranks developers within that community by experience, ensuring that collaborative expertise is leveraged to identify the most suitable developers for triage and resolution.

(3) \textit{Graph Neural Networks and Contrastive Learning:} 
Recent advances leverage graph neural networks and contrastive learning to dynamically model bug–developer relationships in an end-to-end manner. 
Wu et al.~\cite{wu2022spatial} propose a spatial–temporal dynamic GNN for automated bug triaging. It models evolving developer collaboration networks using a joint random walk (JRWalk) mechanism, and learns node spatial-temporal features through a graph recurrent convolutional neural network (GRCNN), enabling the prediction of the most suitable bug fixers.
PCG\cite{dai2024pcg} and NCGBT~\cite{dong2024neighborhood} further enhance these models by applying contrastive learning to refine node representations, jointly considering structural and semantic relationships for candidate developer assignment.

Overall, the evolution of SN-based approaches highlights a progression from simple representations of developer interactions to more sophisticated models that integrate collaboration, community structures, and semantic context, enabling more accurate and adaptive bug triage.

\textbf{Optimization or Decision-making.} 
In OPT approaches, developer assignment is modeled as a constrained optimization problem, aiming to maximize overall project efficiency while minimizing manual triage effort. Early studies laid the theoretical foundation for expertise-based assignment.

(1) \textit{Foundational Optimization Models:} 
Baysal et al.\cite{baysal2009bug} proposed a theoretical framework that infers developer expertise from historical bug-fixing records, combining preference elicitation and expertise recommendation to allocate bugs optimally. While largely conceptual due to evaluation challenges, it laid the groundwork for optimization-based assignment. T-REC\cite{pahins2019t} extended this perspective by retrieving similar historical bugs using FastText and BM25F, fusing rankings probabilistically to recommend Top-K technical groups, thereby reducing manual triage and bug tossing. RAPTOR~\cite{kashiwa2019raptor} formalized developer assignment as a multi-knapsack problem, integrating bug priority, severity, developer experience, and activity to maximize overall project bug-fixing efficiency under developer time constraints.

(2) \textit{Multi-Objective and Dependency-Aware Optimization:} 
Building on optimization, subsequent approaches incorporate task dependencies and multi-objective search to better balance developer workload and bug handling order. 
Etemadi et al.\cite{etemadi2021scheduling} proposed a scheduling-driven assignment method that decomposes bugs into subtasks, models dependencies via a task dependency graph (TDG), and applies a multi-objective evolutionary algorithm to generate Pareto-optimal schedules minimizing fixing time and cost. Almhana et al.\cite{almhana2021considering} leveraged file-level dependencies to define bug report relations, applying NSGA-II search to produce Pareto-optimal sequences that balance bug priority and cognitive load for developers.

(3) \textit{Integration of Historical Context and Scheduling Constraints:} 
To integrate historical context and scheduling constraints, some methods combine optimization with dependency-aware models and dynamic decision-making. 
Jahanshahi et al.~\cite{jahanshahi2022wayback} leverage the Wayback Machine to reconstruct historical bug triage scenarios, dynamically updating bug dependency graphs and developer workloads. It enables integration of custom prioritization or assignment algorithms, including optimization-based methods like S-DABT~\cite{jahanshahi2022s}, allowing assignments to be scheduled and balanced while respecting dependencies and historical context. 
Building on this, they then apply S-DABT~\cite{jahanshahi2022s}, which integrates textual bug data, estimated fixing costs, bug dependencies, and developers' schedules into an integer programming framework, predicting suitability scores with SVM and LDA, and optimizing assignments to balance workload and respect dependencies.
Finally, they propose ADPTriage~\cite{jahanshahi2023adptriage} 

(4) \textit{Online Learning and Multi-Agent Strategies:} 
Recent works explore online learning and multi-agent systems to adaptively assign developers under uncertainty. 
Singh et al.\cite{singh2025navigating} propose Enhanced\_CMAB\_Triage, a contextual multi-armed bandit approach that integrates bug features, developer activity, and similarity-based expert selection to balance exploration and exploitation for cold-start bugs. 
Triangle~\cite{yu2021triangle} adopts a multi-agent architecture, including Analyser, Triage Decider, and Team Manager. The analyzer performs semantic distillation to extract key phrases, Triage Decider selects 5 team candidates by combining TF-IDF and LLM, and Team Manager queries monitor logs to generate enriched information. The negotiation mechanism uses team candidate voting, confirming the team if over half agree, otherwise reselecting with feedback of enriched information, up to 5 rounds.

OPT-based methods have transitioned from static formulations to adaptive, multi-objective, and dependency-aware frameworks, with recent studies incorporating online learning and multi-agent coordination to cope with real-world uncertainty.

\textbf{Other / Hybrid.} 
OTH approaches integrate multiple paradigms, such as rule-based reasoning, machine learning, feature embedding, and network modeling, to leverage complementary information sources for more robust and context-aware developer assignment.

(1) \textit{Integrating Historical and Log-Based Signals Beyond Textual Reports:} 
Some studies extend beyond traditional textual bug reports by incorporating historical execution or service data.
WHOSEFAULT~\cite{servant2012whosefault} integrates fault localization, history mining, and expertise mapping to directly assign developers to execution failures without relying on textual bug reports.
DeCaf~\cite{bansal2020decaf} employs machine learning and pattern mining over large-scale service logs to automatically diagnose and triage KPI performance regressions.
Lim et al.~\cite{lim2014identifying} employ a Hidden Markov Random Field (HMRF)-based clustering model to discretize performance metrics and identify recurrent or previously unseen performance issues.

(2) \textit{Multi-Feature and Knowledge-Enhanced Recommendation:} 
A second line of work enriches developer assignment by combining heterogeneous features and personalized expertise.
Yang et al.\cite{yang2014towards} extract latent topics from historical bug reports and compute multi-feature similarities (\eg component, product, severity, priority) to jointly recommend developers and predict bug severity on large-scale datasets.
Bhattacharya et al.\cite{bhattacharya2010fine} refine classification and tossing prediction by incorporating additional report attributes, intra-fold updates, and multi-feature tossing graphs.
KSAP~\cite{zhang2016ksap} retrieves similar resolved reports via K-nearest-neighbor search, ranks developers using heterogeneous proximity within a multi-entity collaboration network, and fuses textual and social features to generate a ranked developer list.

(3) \textit{Neural and Graph-Based Hybrid Models:} 
Recent hybrid approaches adopt neural architectures to capture complex patterns in bug reports and developer behavior.
DeepTriage~\cite{pham2020deeptriage} constructs an ensemble of FastTree binary classifiers with gradient boosting, supplemented by an inverted index to mitigate cold-start issues.
BRAIN~\cite{chen2020icmbrain} leverages GRU-based sequential modeling, attention masking, and CNN-based language encoding to classify incidents using textual, conversational, and environmental information.
GCBT~\cite{dai2023graph} builds a bipartite bug–developer graph where bug nodes are initialized through NLP pre-training and developer nodes via attribute encoding. Spatial–temporal graph convolutions model evolving expertise, and an IR-based classifier matches bugs to developers, enabling correlation-aware triaging.

(4) \textit{Rule-Based and Personalized Tossing Integration:} 
Hybrid strategies also combine rule-based reasoning with developer-specific behavior modeling.
EDR\_SI~\cite{sun2017enhancing} enhances developer recommendation by integrating expertise and coding habits through Collaborative Topic Modeling (CTM), providing not only ranked developers but also personalized contextual information such as code files and developer networks.
AutoAnalysis~\cite{wang2024comet} employs a rule-based decision tree to encode engineers' historical experience for root cause identification, generating interpretable incident summaries that guide downstream language models.
BPTRM~\cite{wei2025improving} introduces personalized tossing relationships by learning a developer transition matrix via attention over historical tossing paths, refining bug–developer matching through both textual and behavioral signals.

Overall, OTH approaches highlight a shift toward integrating heterogeneous information and intelligent reasoning mechanisms, aiming to enhance contextual awareness, personalization, and interpretability in developer assignment.

\begin{center}
\footnotesize
\begin{singlespace}
\begin{longtable}{cccc>{\raggedright\arraybackslash}p{7cm}}
\caption{
Summary of developer assignment approaches categorized by problem formulation. 
The abbreviations in the ``Category'' column stand for: \textbf{TC} = Text Classification, \textbf{IR} = Information Retrieval, \textbf{SN} = Social Network Modeling, \textbf{OPT} = Optimization / Decision-making, and \textbf{OTH} = Other / Hybrid. The abbreviations in the ``Data'' column stand for: \textbf{BR} = Bug Reports, \textbf{CC} = Code Commits, \textbf{MD} = Metadata (developer assignments, users, comments, milestones, tags), \textbf{TS} = Tossing Sequences, \textbf{L} = Logs, \textbf{M} = Metrics, \textbf{ST} = Stack Traces, and \textbf{IT} = Incident Tickets.
} 
\label{table:developer_assignment} \\

\toprule
\textbf{Category} & \textbf{Technique} & \textbf{Year} & \textbf{Data} & \textbf{Core Method} \\
\midrule
\endfirsthead

\multicolumn{5}{c}%
{\tablename\ \thetable\ -- \textit{Continued from previous page}} \\
\toprule
\textbf{Category} & \textbf{Technique} & \textbf{Year} & \textbf{Data} & \textbf{Core Method} \\
\midrule
\endhead

\midrule \multicolumn{5}{r}{\textit{Continued on next page}} \\
\endfoot

\bottomrule
\endlastfoot

\multirow{26}{*}{\textbf{TC}} 
    & Anvik et al.~\cite{anvik2006should} & 2006 & BR & Text Categorization  +  Supervised Learning  +  Naïve Bayes / SVM / C4.5  +  Developer Ranking \\
    & Anvik et al.~\cite{anvik2011reducing} & 2011 & BR, MD & Textual Feature Extraction  +  Supervised Learning Classifier  +  Developer Ranking \\
    & Ahsan et al.~\cite{ahsan2009automatic} & 2009 & BR & TF–IDF  +  LSI  +  SVM \\ 
    & Costriage~\cite{park2011costriage} & 2011 & BR, MD & SVM  +  Cost-Aware Adjustment  +  Collaborative Filtering \\
    & Lee et al.~\cite{lee2017applying} & 2017 & BR, MD & CNN  +  Word2Vec Embeddings  +  Developer Classification  +  Classifier Management \\
    & Deeptriage~\cite{mani2019deeptriage} & 2019 & BR, ST & Deep Bi-Directional RNN with Attention  +  LSTM  +  Text Classification  +  Probability Scoring \\
    & iTriage~\cite{xi2019bug} & 2019 & BR, TS & Seq2Seq Feature Learning  +  Classifier \\
    & LBT-P~\cite{lee2022light} & 2022 & BR, MD & Knowledge Distillation  +  PLM Layer-Wise Embedding  +  Multi-Layer Classifier  +  Weighted Output \\
    & Wang et al.~\cite{wang2024empirical} & 2024 & BR, MD & Word2Vec / GloVe / NextBug / ELMo / BERT  +  TextCNN / LSTM / Bi-LSTM / Attention  +  MLP / Naive Bayes  +  Embedding-Based Text Classification \\
    & Dipongkor et al.~\cite{dipongkor2024ensemble} & 2024 & BR & Fine-Tuned LLMs  +  Voting/Stacking Ensemble  +  Hinge Loss Sequence Classification \\
    & Park et al.~\cite{park2016cost} & 2016 & BR, MD, CC & SVM  +  LDA Topic Modeling  +  Content-Based Recommendation  +  Content-Boosted Collaborative Filtering +  Dynamic Developer Profiles \\
    & Jonsson et al.~\cite{jonsson2016automated} & 2016 & BR, MD & Stacked Generalization  +  Ensemble Classifiers  +  Text Feature Encoding  +  Team-level Classification \\ 
    & Sarkar et al.~\cite{sarkar2019improving} & 2019 & BR, MD, L & Textual Features  +  Categorical   Features  +  Logistic Regression  +  Incremental Learning  +  High-Confidence Filtering \\
    & BTAL~\cite{zhang2025btal} & 2025 & BR, MD & BERT Embeddings  +  TextCNN Local Features  +  Multi-Source Metadata Fusion  +  Adaptive Loss Function \\
    & FLSCL~\cite{wang2025fixer} & 2025 & BR, MD & ELMo Embeddings  +  BERT Embeddings  +  Bi-LSTM-Attention Classifier  +  Fixer-Level Supervised Contrastive Learning  +  Cross-Entropy Loss \\ 
    \midrule
\multirow{19}{*}{\textbf{IR}}
    & Bugzie~\cite{tamrawi2011bugzie} & 2011 & BR & Fuzzy Sets  +  Cache-Based Dynamic Scoring  +  Aggregation of Relevant Terms for Developer Ranking \\
    & Panda et al.~\cite{panda2022topic} & 2022 & BR, MD & LDA Topic Modeling  +  Intuitionistic Fuzzy Sets   +  IFSim Similarity Measures  +  Fuzzy $\alpha$-Cut Expert Selection \\
    & VTBA~\cite{sajedi2020vocabulary} & 2020 & BR, MD & Technical Terms Filtering  +  Stack Overflow Vocabulary  +  Developer-as-Document  +  Sub-Document Modeling  +  Time-Aware Weighting  +  IR-based Ranking \\
    & Goyal et al.~\cite{goyal2017effective} & 2017 & BR, MD & Time-based IR Ranking (Visheshagya)  +  Parameter Prioritization (W8Prioritizer)  +  NRFixer Model for Non-Reproducible Bugs \\
    & Linares-Vásquez et al.~\cite{linares2012triaging} & 2012 & BR, MD & Latent Semantic Indexing   +  Code Authorship Analysis for Developer Ranking \\
    & Shokripour et al.~\cite{shokripour2013so} & 2013 & BR, CC & Noun Filtering  +  Location-Based Developer Recommendation \\
    & Dretom~\cite{xie2012dretom} & 2012 & BR, MD & Topic Modeling  +  Developer Expertise Modeling  +  Interest-Aware Ranking \\
    & BugFixer~\cite{hu2014effective} & 2014 & BR, CC, MD & Vector Space Model  +  Developer-Component-Bug Network  +  Ranking \\
    & $TopicMiner^{MTM}$~\cite{xia2017topicminer} & 2011 & BR & Multi-Feature Topic Model  +  TopicMiner \\ 
    & PorchLight~\cite{bortis2013porchlight} & 2013 & BR, MD & Tagging via Bug Tagging Language  +  Query-Based Bug Set Creation  +  Interactive UI for Group Triaging \\
    \midrule
\multirow{9}{*}{\textbf{SN}}
    & Jeong et al.~\cite{jeong2009improving} & 2009 & BR, TS, MD & Bug Tossing Graphs  +  Markov Chain Modeling  +  Weighted Breadth-First Search  +  Developer Prediction Integration \\
    & Bhattacharya et al.~\cite{bhattacharya2012automated} & 2012 & BR, TS, MD & Multi-Feature Tossing Graphs  +  Incremental Classification  +  Developer Activity Labeling \\
    & FixerCache~\cite{wang2014fixercache} & 2014 & BR & Unsupervised Developer Cache  +  Activeness Scoring \\
    & DECOBA~\cite{banitaan2013decoba} & 2013 & BR, MD & Bug Term Matrix  +  Developer Collaboration Network  +  Community Detection  +  Community-Based Assignment  +  Expertise Ranking \\
    & Wu et al.~\cite{wu2022spatial} & 2022 & BR, MD, TS & Joint Random Walk   +  Graph Recurrent Convolutional Neural Network  +  Developer Prediction \\
    \multirow{6}{*}{\textbf{SN (continued)}}& PCG~\cite{dai2024pcg} & 2024 & BR, MD & Embedding Initialization  +  Prototype Clustering Augmentation  +  Graph Collaborative Filtering  +  Semantic Contrastive Learning  +  Multitask Joint Learning \\
    & NCGBT~\cite{dong2024neighborhood} & 2024 & BR, MD & Bipartite Graph Modeling  +  Pre-trained Language Model Initialization  +  Multi-Layer Graph Neural Network  +  Neighborhood Contrastive Learning  +  BPR Loss Optimization \\
    \midrule
\multirow{19}{*}{\textbf{OPT}}
    & Baysal et al.~\cite{baysal2009bug} & 2009 & BR, MD & Preference Elicitation  +  Expertise Recommendation  +  Task Allocation \\
    & IT-REC~\cite{pahins2019t} & 2019 & BR, MD & FastText Vector Space Model  +  Extended BM25F IR  +  Probabilistic Top-K Ranking \\
    & RAPTOR~\cite{kashiwa2019raptor} & 2019 & BR, MD & Multi-Knapsack Modeling  +  Suitability Score Calculation  +  Linear Programming Solver \\
    & Etemadi et al.~\cite{etemadi2021scheduling} & 2021 & BR, MD, CC & NSGA-II  +  Task Dependency Graph  +  Developer-Task Assignment Vector  +  Scheduling Vector  +  Greedy Local Search  +  Multi-Objective Evaluation \\
    & Almhana et al.~\cite{almhana2021considering} & 2021 & BR, MD, CC, TS & Bug File Localization  +  Dependency Calculation  +  Multi-Objective NSGA-II Search  +  Pareto-Optimal Bug Sequencing \\
    & Wayback~\cite{jahanshahi2022wayback} & 2022 & BR, MD, TS, L & Historical Event Replay  +  Bug Dependency Graph Update  +  Developer Load Tracking  +  Modular Algorithm Integration \\
    & S-DABT~\cite{jahanshahi2022s} & 2022 & BR, MD, TS & SVM Text Classification  +  LDA-Based Cost \& Dependency Estimation  +  Schedule-Aware Programming  +  Multi-Objective Optimization \\
    & ADPTriage~\cite{jahanshahi2023adptriage} & 2023 & BR, MD, TS, L & LDA Topic Modeling  +  Markov Decision Process  +  Approximate Dynamic Programming  +  Dynamic Bug Assignment \\
    & Singh et al.~\cite{singh2025navigating} & 2025 & BR, MD, TS & Developer Activity Level  +  Similarity-Based Candidate Filtering  +  Contextual Multi-Armed Bandits  +  Exploration-Exploitation Optimization \\
    & Triangle~\cite{yu2021triangle} & 2021 & IT, M & Multi-Agent Architecture  +  Semantic Distillation  +  TF-IDF + Negotiation Mechanism\\
    \midrule
\multirow{25}{*}{\textbf{OTH}}
    & WhoseFault~\cite{servant2012whosefault} & 2012 & CC & Fault Localization  +  History Mining (line-level)  +  Expertise Assignment for Ranked Developer List\\
    & DeCaf~\cite{bansal2020decaf} & 2020 & L & Random Forests  +  Pattern Mining  +  Custom Scoring \\
    & Lim et al.~\cite{lim2014identifying} & 2014 & M & HMRF-Based Clustering  +  EM Optimization \\
    & Yang et al.~\cite{yang2014towards} & 2014 & BR, MD & LDA Topic Modeling  +  Developer Ranking  +  KNN \\
    & Bhattacharya et al.~\cite{bhattacharya2010fine} & 2010 & BR, MD & Incremental Naïve Bayes / Bayesian Networks Multi-Feature Tossing Graphs  +  Ranking \\
    & KSAP~\cite{zhang2016ksap} & 2016 & BR, MD & K-Nearest-Neighbor Search  +  Heterogeneous Developer Collaboration Network  +  Meta-Path Proximity Ranking \\
    & DeepTriage~\cite{pham2020deeptriage} & 2020 & BR, MD, L, ST & MART  +  LightGBM  +  CNN  +  Inverted Index  +  Clustering \\
    & BRAIN~\cite{chen2020icmbrain} & 2020 & IT, MD, M, ST & GRU-Based Model  +  Attention Masking Strategy  +  CNN-Based Neural Language Model \\
    & GCBT~\cite{dai2023graph} & 2023 & BR, MD & Triaging Graph Construction  +  NLP-Based Node Initialization  +  Spatial Convolution  +  Temporal Convolution  +  Bug Embedding Augmentation  +  IR-Based Classifier \\
    & EDR\_SI~\cite{sun2017enhancing} & 2017 & BR, CC, MD & Historical Commits Preprocessing  +  Collaborative Topic Modeling  +  Personalized Developer–Code Ranking  +  Supplementary Information Construction \\
    & AutoAnalysis~\cite{wang2024comet} & 2024 & IT, L, MD, ST & Log filtering  +  Keyword Extraction  +  Incident Embedding  +  Similarity-Based Retrieval  +  LLM-Based Triage \\
    & BPTRM~\cite{wei2025improving} & 2025 & BR, MD, TS & BERT Embedding  +  Bug Attribute Encoding  +  SIMIR / SVM / DeepTriage  +  Attention-Based Ability Matching  +  Tossing Transition Probability Matrix  +  Top-K Recommendation \\
\end{longtable}
\end{singlespace}
\vspace{-0.5cm}
\end{center}

\subsection{Postmortem Process}

Following the resolution of an incident or bug, the triage process often includes a structured postmortem phase aimed at capturing actionable knowledge and improving future issue management. This phase typically involves labeling and categorizing the resolved issue with metadata such as root cause classification, impacted components, resolution type, and severity level. These labels enable efficient indexing, retrieval, and statistical analysis of historical cases. In addition, linking the issue to related incidents, commits, or configuration changes supports traceability and facilitates the identification of recurring fault patterns.

A well-executed postmortem process also includes documenting lessons learned, decision rationales, and any procedural improvements identified during resolution. Such documentation not only serves as a reference for preventing similar issues in the future but also strengthens automated triage systems by enriching training datasets with high-quality, structured information. Furthermore, tracking post-resolution follow-up actions, such as applying preventive patches, updating monitoring rules, or refining alert thresholds, ensures that the organization benefits from a continuous feedback loop between operational experience and fault management practices. By systematically embedding these postmortem practices into triage workflows, organizations can enhance knowledge reuse, improve fault detection accuracy, and reduce mean time to resolution in large-scale software and service environments.

\subsubsection{Continuous Triage}
In large-scale online service systems, incident triage aims to assign newly reported incidents to the most appropriate teams. However, due to incomplete information and evolving context, initial assignments are often inaccurate, requiring repeated reassessment and cross-team discussion. This iterative refinement process, referred to as continuous incident triage~\cite{chen2019continuous}, highlights the need for approaches capable of dynamically updating triage decisions as new information emerges. Existing research on continuous triage can be broadly grouped into three methodological paradigms.

(1) \textit{Learning-Driven Refinement: }Learning-driven methods, such as DeepCT~\cite{chen2019continuous}, formalize triage as an incremental learning process, where models progressively update assignment predictions based on ongoing interactions or discussion histories. Earlier probabilistic frameworks, such as Shao et al.'s work~\cite{shao2008efficient}, also reflect this idea, modeling team transitions as evolving state sequences to capture the dependencies among successive assignment decisions.

(2) \textit{Data-Driven Adaptation: }Data-driven approaches extend this paradigm by incorporating model uncertainty and distributional drift into the triage process. For instance, Scouts~\cite{gao2020scouts} detects shifts in incident patterns and dynamically adjusts routing decisions, while Ticket-BERT~\cite{liu2023ticket} employs an iterative fine-tuning cycle that continuously integrates newly labeled incidents to maintain model adaptability. These techniques emphasize automated model updating as a mechanism for sustaining performance under evolving data conditions.

(3) \textit{Human-in-the-Loop Collaboration: }In contrast, team-driven approaches integrate human expertise into the triage loop. Triangle~\cite{yu2021triangle} exemplifies this perspective by coupling semantic enrichment with consensus-based decision refinement, where automated recommendations are iteratively reviewed and adjusted through human feedback until convergence.

Overall, continuous triage methods share a common goal of enabling adaptive and iterative decision-making. They differ primarily in the source of feedback, but all aim to transform triage from a static one-shot prediction into a dynamic, continuously improving process.

\subsubsection{User Feedback Analysis}
Only a limited number of studies have investigated feedback issues arising after a single distribution failure. Considering the similarity between app user reviews and user submissions following triage failures, several lines of research have explored how user feedback can support post-release issue management, evolving from early integration with code artifacts to more advanced text-driven and learning-based analyses.

Early studies, such as Shokripour et al.~\cite{shokripour2013so}, incorporated user feedback into software repositories by linking review content with code commits and historical bug data to support defect localization and assignment. Palomba et al.~\cite{palomba2015user} extended this idea by constructing traceability links between user reviews and code changes, enabling release-level monitoring and assessment of quality variations.

Subsequent research shifted toward mining and modeling user feedback as an independent information source. Gao et al.~\cite{gao2018online} adopted topic modeling to identify emerging issues across app versions, dynamically combining prior and current topics to capture evolving user concerns. Later, Etaiwi et al.~\cite{etaiwi2020order} and Malgaonkar et al.~\cite{malgaonkar2022prioritizing} advanced the analysis by applying text preprocessing, clustering, and embedding-based similarity measures to prioritize user-reported issues and support developer decision-making.

Besides, Di et al.~\cite{di2021investigating} revisited the connection between user feedback and software quality by correlating feedback metrics, code characteristics, and app ratings. This line of work bridges earlier repository-based and review-based methods, highlighting the feedback loop between user experience and software maintenance activities.

\begin{table*}[ht]
    \footnotesize
    \centering
    \caption{Summary of postmortem processing based on their category, the year, and the data. The abbreviations in the ``Data'' column stand for: \textbf{M} = Metrics, \textbf{TR} = Triage Results, \textbf{IT} = Incident Tickets, \textbf{HTS} = Historical Ticket Sequences, \textbf{R} = Reviews, \textbf{RR} = Release Ranks, and \textbf{CC} = Code Commits.}
    \label{table:datasets}
    \begin{tabular}{ccccl}
        \toprule
        \textbf{Category} & \textbf{Technique}  & \textbf{Year} & \textbf{Data} & \textbf{Core Method}\\
        \midrule
        \multirow{5}{*}{Continuous Triage} & DeepCT~\cite{chen2019continuous} & 2019 & D, TR & GRU+Attention-Mask Strategy\\
        &Shao et al.~\cite{shao2008efficient}& 2008 & HTS, TR & First-Order Markov+Variable-Order Markov\\
        &Scouts ~\cite{gao2020scouts} & 2020 & IT & \makecell[l]{Random Forest+Improved Change Point Detection}\\
        &Ticket-BERT~\cite{liu2023ticket} & 2023 & IT & Fine-Tune\\
        & Triangle~\cite{yu2021triangle} & 2021 &IT, M &Semantic Distillation+Discussion Group Voting\\
        \midrule
        \multirow{7}{*}{User Feedback Analysis}
        &Shokripour et al.~\cite{shokripour2013so} & 2013 & R, CC & \makecell[l]{Noun Extraction Process+Simple Term Weighting Scheme}\\
        &CRISTAL~\cite{palomba2015user} & 2015 & R, RR & Review Coverage+Monitoring Component\\
        & Gao et al. ~\cite{gao2018online} & 2018 & R & \makecell[l]{IDEA to Automatically Identify Emerging Issues+AOLDA for Online\\ Review Analysis}\\
        &Etaiwi et al.~\cite{etaiwi2020order}& 2020 & R & \makecell[l]{CLAP Tool+ExactAlgorithm+BioConcert+KwikSort}\\
        &Malgaonkar et al.~\cite{malgaonkar2022prioritizing} & 2022 & R, IT & DistilBERT+Cosine Similarity\\
        &Di et al.~\cite{di2021investigating}& 2021 & R, M & Spearman's Rank Correlation Coefficient\\
        \bottomrule
    \end{tabular}
\vspace{-0.5cm}
\end{table*}

\begin{tcolorbox}[colback=gray!10,
	colframe=black,
	width=15cm,
	arc=2mm, auto outer arc,
	title={\textbf{Finding 1: Evolution of Triage Methods Across the Lifecycle}}]
	\textit{Over time, triage methods have evolved to encompass the entire lifecycle, from data preprocessing to postmortem analysis. Early research primarily focused on data deduplication and feature extraction to enhance input quality, followed by advances in prioritization techniques such as severity assessment and issue classification. Subsequent studies emphasized assignment optimization for components and developers, while recent efforts have introduced continuous triage and feedback-driven refinement mechanisms. This evolutionary trajectory signifies a transition from isolated task automation to adaptive, end-to-end triage systems capable of continuous learning and improvement in real-world settings.} 
\end{tcolorbox}
\section{Challenges in Triage Practice}\label{sec_5}

Triage, including bug triage in software engineering and incident triage in large-scale online or cloud services, is a multifaceted process that assigns reports, tickets, or alerts to the most appropriate resolver, whether an individual developer, a team, or an automated system. As modern software and service infrastructures evolve, triage practices encounter increasingly complex challenges arising from both technical barriers and operational demands. Drawing on recent literature and industrial practice, we summarize the key challenges as follows.

\subsection{Data Quality, Diversity, and Representation}

\subsubsection{Data Noise and Incomplete Information} 

Bug reports and incident tickets are often poorly structured, ambiguous, or irrelevant. Common issues include missing reproduction steps, vague or noisy descriptions, absent contextual information such as environment details, severity, or affected component, and inconsistent field usage across projects and systems.

\begin{itemize}
    \item \textbf{Multimodal Data Noise.} The integration of multimodal inputs such as text, code, logs, metrics, and configurations increases noise, especially in cloud service environments where heterogeneous data sources coexist~\cite{mani2019deeptriage, li2021revisiting}.  For text data, there is a large volume of logs and discussion texts with significant noise, making direct processing difficult. Existing methods struggle to extract high-quality key points from a large amount of text~\cite{wang2024comet}. Code snippets lack structural information in feature representation, making it difficult to learn effectively~\cite{aung2022multi}. For review of apps, vocabulary mismatches always happen between user reviews and source code or issues reported in issue trackers~\cite{palomba2015user}.

    \item \textbf{Human and System-induced Reporting Errors.} Reports are frequently affected by human errors, false alarms, or transient system anomalies, leading to non-reproducible or irrelevant cases such as alert storms or incidental incidents~\cite{guo2011not}. Besides, there may have been cases where this value was not a person with expertise in that area of the source code. For example, the person who fixed the bug may not have commit rights, and another project member may be required to commit the fix~\cite{shokripour2013so}. 

    \item \textbf{Heterogeneity Across Organizations and Platforms.} Triage systems operate across diverse organizations and platforms, where data structures, naming conventions, and semantic expressions vary widely~\cite{lee2017applying, liu2023ticket}. Monitoring alerts in production, customer support tickets, and automatically generated system events often exhibit distinct temporal, textual, and structural characteristics. Large-scale online service systems do not share much similarity: they are developed for different purposes by different teams. They possess various code logic, implementation languages, and architectures~\cite{zheng2019ifeedback}. Furthermore, update patterns can also differ across distinct features of the same application, as well as for the same feature at different time points. These differences thus require systematic analysis~\cite{wang2017app}.
\end{itemize}

\subsubsection{Class Imbalance and High Dimensionality}

Assignments typically involve hundreds or even thousands of potential targets, including developers, teams, or components. The majority of reports concentrate on a small subset of assignees, creating a long-tail distribution. Models tend to overfit frequent classes while underperforming on rare ones, reducing assignment accuracy and leading to overloaded teams or prolonged unassigned cases.
At the data collection level, a large number of possible teams not only challenge machine learning models in terms of complexity but also exhibit extreme data imbalance. A wide variety of data formats supported in modern incident management systems requires the routing system to use different approaches to parsing this content and provide useful features information to machine learning~\cite{pham2020deeptriage}. 
Besides, the disproportionate ratio of class labels in the training data affects the performance of the classification prediction model~\cite{aung2022multi}.
\subsection{Dynamic Ecosystem and Evolving Context}

\subsubsection{Concept Drift and Temporal Variability} 

Frequent organizational and system changes, such as team restructuring, personnel turnover, or component refactoring, rapidly invalidate historical assignment patterns.  This results in concept drift, where models must adapt to shifting data distributions in real time~\cite{chrupala2012learning, zhang2020efficient, borg2024adopting}. However, understanding the issue and providing appropriate healing action depend heavily on domain knowledge~\cite{ding2014mining}. The dependency structure is neither known nor fixed in large online service systems. Therefore, it is hard to use the dependency graph to identify linked incidents~\cite{chen2020lidar}. High-frequency updates in practical environments impose stringent requirements on both adaptability and stability.

\subsubsection{Cold-start Problem}

New developers, teams, or components lack sufficient historical data, making it difficult to assess their expertise or availability. Cold-start challenges are especially pronounced in system rollouts, team expansion, or platform migration, which are common in cloud monitoring and DevOps contexts~\cite{singh2025navigating}.

\subsubsection{Knowledge and Expertise Decay} 

Developer expertise diminishes over time, and inactive or departing experts leave behind outdated assignment traces~\cite{matter2009assigning}. The system components are mostly developed by different teams, and expert knowledge is limited. There is a lack of a systematic mechanism for sharing knowledge of historical events, resulting in incident resolution relying on a few experts and a longer MTTR~\cite{lou2017experience}. Effective triage systems must detect these changes to avoid misdirected assignments. Evolving code ownership and product lifecycles further require dynamic tracking of responsibility boundaries.

\subsection{Multi-factor and Multi-step Assignment Complexity}

\subsubsection{Multi-dimensional Decision Factors}

Accurate triage requires integrating multiple factors beyond text classification. These include current workload and availability of developers or teams, skills, and historical contributions such as authorship or activity records, individual or team preferences where pull-based assignment is adopted, and task-related constraints such as deadlines, severity, priority, or cost~\cite{park2016cost, zou2020how, etemadi2021scheduling}. It may be the case that the developer who actually fixed a bug according to the ``Line 10 Rule'' is not the optimal choice for that report, whereas we treat it as such~\cite{shokripour2013so}. Besides, the problem space is complex, with diverse causes for events involving hardware, network, resource competition, and other aspects, requiring a comprehensive analysis of various types of monitoring data~\cite{lou2013sas}. Consequently, the feature vectors tend to be extremely large, which renders it highly challenging for a single event router to integrate monitoring data from all teams. Furthermore, the events themselves are rare occurrences, making it difficult to address this challenge by increasing the number of training samples~\cite{gao2020scouts}.

\subsubsection{Complex Priorities and Dependency Structures}

Reports may contain explicit or implicit dependencies such as blocker, related, or duplicate relationships~\cite{ jahanshahi2023adptriage}. Failures often cascade across teams or components, requiring algorithms to jointly handle interdependent cases and update dependency queues dynamically. The large-scale online service system has numerous components developed by different teams, leading to a lack of comprehensive understanding among engineers about the system, as well as a lack of knowledge-sharing mechanisms, resulting in slow event resolution and long average recovery times~\cite{lou2013sas}. The virtualization and dynamic allocation of resources in cloud environments make it difficult to associate physical and virtual resources. The integration of scattered event tickets, CMDB, and monitoring data is needed to eliminate data silos~\cite{gupta2009multi}.

\subsubsection{Reassignment Cost}

Incorrect assignments frequently result in issue tossing, where reports are repeatedly reassigned among teams or individuals.  This significantly increases resolution time and reduces overall efficiency~\cite{shokripour2013so}.  Optimization must therefore consider not only predictive accuracy but also tossing path length and average time to resolution~\cite{jeong2009improving, su2021reducing}.

\subsection{Scalability, Real-world Integration, and Explainability}

\subsubsection{Scalability and Efficiency}

Deployed triage systems must scale to millions of reports or alerts, often in real time, as seen in Microsoft Azure or global banking alert systems~\cite{zhao2020understanding, lee2022light}. Besides, the agile development model in cloud services requires the development and deployment process to be flexible and meet certain production requirements in terms of execution time, or error tolerance to maintain high service availability~\cite{pham2020deeptriage}. This demands models with high inference speed and efficient resource utilization. Today, ticket routing is usually driven by expert decisions. It is not uncommon that, due to human error or inexperience, a ticket is mistakenly transferred to a group that cannot solve the problem, which might lead to a long and inefficient routing sequence. In such cases, not only are resources wasted, but it would also take a longer time to close tickets, causing customer dissatisfaction~\cite{shao2008efficient}. Some existing methods train developer recommendation and issue type assignment as independent tasks, resulting in task repetition and overlooking the correlating information between tasks~\cite{aung2022multi}. 

\subsubsection{Integration with Real-world Workflows}

Triage systems must integrate seamlessly into DevOps, agile, and team-specific processes without disrupting established practices~\cite{sajedi2020guidelines}. Outputs should be interpretable, allowing engineers to understand assignment rationales and confidence levels. Large-scale organizations further require mechanisms for cross-team permissions and inter-departmental communication. However, traditional hierarchical multi-label classification algorithms, when applied to ticket classification, fail to fully incorporate the domain experts' prior knowledge, so that it is unable to quickly adapt to the new system environment. The existing loss functions are ineffective in distinguishing between different types of misclassifications, making it difficult to accurately assess classification performance~\cite{zeng2017knowledge}.

\subsubsection{Lack of Standardized Evaluation and Reproducibility} 

Research on triage often lacks standardized task definitions, evaluation metrics, dataset preprocessing pipelines, and gold standards, making cross-study comparison difficult~\cite{sajedi2020guidelines}. Industrial data remains largely proprietary, limiting reproducibility and slowing technology transfer. The black-box models (such as DNNs) perform well in event classification, but their lack of interpretability makes it difficult for engineers to understand and trust the prediction results~\cite{remil2021interpretable}.


\begin{tcolorbox}[colback=gray!10,
	colframe=black,
	width=15cm,
	arc=2mm, auto outer arc,
	title={\textbf{Finding 2: Practical Challenges in Triage Practice}}]		
	\textit{In practice, triage processes face considerable challenges arising from data quality issues, dynamic project ecosystems, and operational complexity. Real-world datasets are often noisy, incomplete, and heterogeneous, while the continual evolution of project contexts results in concept drift and knowledge obsolescence. Furthermore, multifactor decision dependencies and the absence of standardized evaluation frameworks impede scalability and seamless integration into development workflows. Collectively, these limitations underscore a persistent gap between research-oriented prototypes and deployable triage solutions.} 
\end{tcolorbox}

\section{Evaluation and Benchmarking}\label{sec_6}

\subsection{Evaluation Metrics}
The evaluation of triage techniques has relied on a wide range of metrics that capture not only predictive accuracy but also ranking effectiveness, computational efficiency, and practical impact on software engineering workflows. For clarity, these metrics can be categorized into four groups.

\subsubsection{Accuracy Metrics}
For binary triage tasks, conventional classification metrics are commonly employed to evaluate whether the model correctly distinguishes between positive and negative cases. These metrics provide a foundational assessment of model discrimination and general predictive reliability. Representative metrics include:

\begin{itemize}
    \item \textbf{Accuracy:} The proportion of correctly predicted instances among all predictions. Accuracy has been employed to evaluate classifiers for team assignment in industrial contexts~\cite{jonsson2016automated} and to compare deep learning models against human triagers~\cite{lee2017applying}.
    
    \item \textbf{Precision, Recall, and $F$-score:} Precision quantifies the fraction of correctly predicted positive instances among all positive predictions, whereas recall measures the fraction of true positives correctly identified. The F-score provides a harmonic balance between the two, with variants such as $F1$ and $F2$ used depending on whether precision or recall is prioritized. These metrics have been foundational since early bug triage research~\cite{anvik2006should, ahsan2009automatic}, and remain widely adopted, with recent studies employing $F2$-scores to emphasize recall-oriented evaluation~\cite{aung2022multi}.  

    \item \textbf{AUC (Area Under the ROC Curve) and Area Under the Precision–Recall Curve:} These metrics assess the model's overall ranking capability across different classification thresholds. They are particularly useful in imbalanced datasets, such as those involving rare component assignments or low-severity bugs~\cite{catolino2019not, li2021revisiting}.  

    \item \textbf{Cohen's Kappa:} Occasionally adopted to measure the degree of agreement between predicted and actual classifications beyond random chance~\cite{ahsan2009automatic}.  
 
\end{itemize}

\subsubsection{Ranking and Top-k Metrics} 
While binary classification metrics focus on the correctness of a single predicted label, multi-class or recommendation-based triage tasks typically aim to rank multiple candidate labels (\eg developers, priorities, or solutions). In such cases, ranking-based and top-$k$ metrics are more appropriate to capture the quality of ordered recommendations. Key evaluation measures include:

\begin{itemize}
    \item \textbf{Top-k Accuracy (Hit@$k$, Acc@$k$):} Determines whether the correct label appears within the top-$k$ ranked predictions, commonly with $k \in \{1,3,5,10,20\}$~\cite{mani2019deeptriage, su2021reducing, lee2022light, dai2023graph, dai2024pcg}.
    
    \item \textbf{Mean Reciprocal Rank (MRR):} Evaluates the ranking position of the correct label by averaging the reciprocal rank across all instances. MRR has been widely used to assess the quality of ranked developer recommendations and other triage-related retrieval tasks, including those employing transformer-based or time-sensitive features~\cite{shokripour2015time, dipongkor2023comparative, su2023still}.  

    \item \textbf{Mean Average Precision (MAP):} The metric measures the mean of the Average Precision (AP) values computed across all queries, such as bug reports. It not only accounts for ranking quality but also integrates recall considerations, providing a comprehensive assessment of retrieval performance. Compared with Mean Reciprocal Rank (MRR), MAP offers a more holistic reflection of the overall ranking effectiveness, as it rewards systems that maintain high precision across varying recall levels.~\cite{zhang2016ksap,  panda2022topic}.

    \item \textbf{Normalized Discounted Cumulative Gain (NDCG@$k$):} Measures ranking quality by assigning higher importance to correct predictions appearing earlier in the ranked list, thereby capturing both relevance and position sensitivity~\cite{su2021reducing}.  

    \item \textbf{Precision@$k$, Recall@$k$, and $F1@k$:} Extensions of the traditional classification metrics adapted to top-$k$ recommendation scenarios. These are often employed to evaluate developer recommendation models by measuring the proportion of correct labels among the top-$k$ candidates~\cite{xia2013accurate, panda2022topic}.  

    \item \textbf{Diversity Metrics:} Used to assess the variety of recommended items, ensuring that suggestions (\eg developers or solutions) are not redundant or overly concentrated. For instance, diversity metrics have been used in caching-based triage systems to promote balanced and diverse developer recommendations~\cite{wang2014fixercache}.  

    \item \textbf{Nearest False Negatives (NFN) and Nearest False Positives (NFP):} These two metrics are used to evaluate the performance of priority classification tasks by quantifying the proximity between the predicted and actual priority levels of misclassified bug reports. Specifically, NFN measures how close an incorrectly predicted lower-priority report is to its true higher-priority label, while NFP assesses the reverse case. Together, they capture the severity of misclassification and provide a nuanced understanding of how prediction errors deviate from true priority rankings~\cite{kanwal2012bug}.

\end{itemize}

\subsubsection{Efficiency Metrics}
Since triage is often performed in time-critical operational environments, efficiency-oriented evaluation plays an essential role in assessing both the responsiveness and the practical utility of triage systems. Commonly used measures include:

\begin{itemize}
    \item \textbf{Assignment Accuracy and Fixing Time:} These metrics jointly evaluate predictive correctness and temporal efficiency. The total time saved by an automated triage system can be quantified using two complementary measures: \textit{gain-in}, representing the time saved through correct routing, and \textit{gain-out}, denoting the time saved by preventing incorrect assignments~\cite{gao2020scouts}.
    
    \item \textbf{Recommendation and Query Response Time:} Captures the latency required to generate triage recommendations or respond to queries. This metric directly reflects the system's computational efficiency and its suitability for deployment in real-time settings~\cite{tamrawi2011bugzie}.  

    \item \textbf{Time to Mitigation (TTM), First Triage Time (FTT), and Mean Steps to Resolve (MSTR):} Measure the timeliness and process efficiency of triage workflows, reflecting how quickly a reported issue progresses from identification to actionable resolution~\cite{shao2008efficient, wang2024comet}.  

    \item \textbf{Reduction in Tossing Path Lengths:} Evaluates the degree to which a triage approach reduces unnecessary reassignments before reaching the correct developer or team. Shorter tossing paths indicate improved routing precision and decreased communication overhead~\cite{jeong2009improving}.  

    \item \textbf{Distribution of Due Dates and Convergence Analysis:} Analyze the temporal dynamics of triage decisions, including how assignment timing aligns with project deadlines or optimization objectives. These advanced metrics are used to evaluate the balance between assignment delay and developer suitability, as well as to assess model convergence and stability over time~\cite{jahanshahi2023adptriage}.  

\end{itemize}

\subsubsection{User-Centered and Qualitative Evaluation}
Beyond quantitative performance metrics, several studies emphasize user-centered and qualitative assessments to evaluate the perceived usability, interpretability, and adoption potential of triage systems.

\begin{itemize}
    \item \textbf{Likert-Scale Surveys:} Often implemented as five-point questionnaires designed to assess dimensions such as ease of use, perceived usefulness, and the likelihood of adoption by practitioners~\cite{bortis2013porchlight}.
    
    \item \textbf{Qualitative Feedback and Interviews:} Provide insights into practitioners' perceptions regarding the interpretability, reliability, and practical value of triage systems. For example, structured interviews have been conducted to capture developers' experiences with new triage tools~\cite{bortis2013porchlight}, while broader surveys have explored how practitioners perceive automated bug management techniques more generally~\cite{zou2020how}.  

\end{itemize}

\subsubsection{Structural Classification Metrics}
\begin{itemize}

    \item \textbf{Hierarchical Classification Accuracy:} In hierarchical or multi-label triage scenarios, such as categorizing issue types or ticket hierarchies, accuracy is often assessed using specialized indicators including Hamming Loss, HMC-Loss, H-Loss, and Parent–Child Error. These metrics evaluate not only label correctness but also the consistency of predictions within hierarchical structures~\cite{zeng2017knowledge}. 

\end{itemize}

\subsection{Datasets}

High-quality, large-scale datasets provide indispensable experimental foundations and evaluation benchmarks for algorithmic innovation and technological advancement.  They play a pivotal role in fostering knowledge integration and accelerating innovation across academia and industry.  Similar to other data-driven fields, datasets serve as a cornerstone in the triage domain.  However, while industrial-grade software systems naturally generate heterogeneous and large-scale data, academic research often lacks access to these real-world datasets.  Consequently, many empirical studies rely on incomplete, small-scale, or simulated data, limiting the industrial applicability of their findings.  This challenge is particularly acute in the context of software fault diagnosis, where the absence of high-quality datasets poses a significant barrier to validation and reproducibility.

Advancing triage research requires joint efforts from both industry and academia.  Industrial stakeholders are uniquely positioned to provide production-level event logs, bug reports, and observability data, whereas academia can contribute by curating labeled datasets, developing advanced triage solutions, and facilitating open sharing of resources.  The construction of standardized datasets and the establishment of unified evaluation metrics demand collaborative engagement across multiple parties.  Based on a survey of existing efforts, we summarize the current landscape of publicly available triage-related datasets as follows.

\subsubsection{Bug Triage Datasets} 

In the domain of bug triage, the open culture of large-scale open-source projects has significantly facilitated research by providing extensive repositories of publicly accessible defect data. Such datasets not only enable reproducibility of experimental results but also constitute valuable resources for advancing subsequent studies.

\textbf{Data Sources and Characteristics.} The most widely adopted datasets originate from public defect tracking systems (\eg Bugzilla, JIRA) maintained by prominent open-source communities such as Mozilla (Firefox), Eclipse, Apache, and GCC. These datasets primarily consist of structured bug reports, which typically include metadata such as titles, detailed descriptions, submitters, associated products and components, operating systems, priorities, and severity levels. Additionally, the ``tossing history'' of bug reports, records of developer reassignments, has emerged as a critical feature for modeling collaboration networks and analyzing assignment processes. Based on a comprehensive literature and repository review, we assemble publicly accessible triage datasets, summarized in Table~\ref{table:datasets}.

\begin{center}
\footnotesize
\begin{singlespace}
\begin{longtable}{p{3cm}p{5cm}p{6cm}}
\caption{
 Summary of publicly available datasets.} 
\label{table:toolkits} \\
\toprule
\textbf{Name} & \textbf{Data} & \textbf{Details}\\
\midrule
\endfirsthead

\multicolumn{3}{c}%
{\tablename\ \thetable\ -- \textit{Continued from previous page}} \\
\toprule
\textbf{Name} & \textbf{Data}  & \textbf{Details}\\
\midrule
\endhead

\midrule \multicolumn{3}{r}{\textit{Continued on next page}} \\
\endfoot

\bottomrule
\endlastfoot
        \multirow{6}{*}{MultiTriage~\cite{aung2022multi}} & Bug Reports from Eclipse \& Github OSS projects & \url{https://github.com/thazin31086/MultiTriage/tree/master/Project/Data}\\
        & Original Issue Report of the Open-Source ASP.NET Core Project & \url{https://github.com/dotnet/aspnetcore}\\
        & The Preprocessed Partial Dataset Used for MultiTriage &\url{http://dx.doi.org/10.5281/zenodo.5532458}\\
        Wu et al.~\cite{wu2022spatial} & Fixed bug reports from two OSS projects, namely Eclipse and Mozilla. & \url{https://github.com/ssea-lab/BugTriage/tree/master/raw%20data}\\
        ADPTriage~\cite{jahanshahi2023adptriage} & Bug reports from \text{EclipseJDT} \& GCC \& Mozilla OSS projects & \url{https://github.com/HadiJahanshahi/ADPTriage/tree/main/Toy_Example}\\
        Zhang et al.~\cite{zhang2016towards} & Bug reports from Eclipse \& GCC \& Mozilla \& Netbeans \& OpenOffice OSS projects & \url{https://github.com/ProgrammerCJC/SPFR} \\
        VTBA~\cite{sajedi2020vocabulary} & Bug reports from 13 popular GitHub projects (\eg Angular.js, Rails, Elasticsearch) & \url{https://github.com/TaskAssignment/VTBA} \\
        Li et al.~\cite{li2021revisiting} & Bug reports from 6 OSS projects (\eg Cassandra, Flex, Hbase) & \url{https://github.com/lizx2017/textMyth} \\
        S-DABT~\cite{jahanshahi2022s} & Bug reports from EclipseJDT \& GCC \& Mozilla \& OpenOffice OSS projects &\url{https://github.com/HadiJahanshahi/SDABT/tree/main/dat} \\
        Wang et al.~\cite{wang2024empirical} & Bug reports from EclipseJDT \& GCC \& Firefox OSS projects & \url{https://github.com/AI4BA/dl4ba} \\
        Gao et al. ~\cite{gao2018online}& User reviews of 6 popular apps & \url{https://github.com/ReMine-Lab/IDEA}\\
        Di et al.~\cite{di2021investigating}& Reviews and versions of Android apps\& code quality metrics &\url{https://github.com/sealuzh/user-satisfaction}\\
        ART~\cite{sun2024art}& Microservice datasets \& failure cases & \url{https://github.com/bbyldebb/ART}\\
        Sadlek et al.~\cite{sadlek2025severity} & Testbed data with Windows/Ubuntu hosts \& alerts \& logs from 5 attack scenarios & \url{http://dx.doi.org/10.5281/zenodo.14547668}\\
        Gao et al.~\cite{gao2022listening}& Reviews from 5 apps (\eg eBay, Viber) & \url{https://github.com/monsterLee599/SOLAR}\\
        Haering et al. \cite{haering2021automatically} & Problem reviews and bug reports from 4 open-source apps (\eg Firefox, VLC)& \url{https://mast.informatik.uni-hamburg.de/replication-packages/}\\
        Phetrungnapha et al.~\cite{phetrungnapha2019classification} & Mobile app user review dataset with titles, descriptions, ratings, labeled as feature requests, problem discoveries, etc & \url{https://mast.informatik.uni-hamburg.de/app-review-analysis}\\
        Linares et al.~\cite{linares2012triaging} & Issue \& Commit Comment from ArgoUML, JEdit, MuCommander &\url{http://www.cs.wm.edu/semeru/data/icsm2012-authorship/}\\
        Lim et al.~\cite{lim2014identifying}& Performance Metric Data Records & \url{http://www.dropbox.com/s/pj1miqu00ryoj9a/HMRF-kMedoid-EM.zip}\\
        Nagwani et al.~\cite{nagwani2023artificial} & GitHub bug repository & \url{https://github.com/orgs/github/repositories}\\
        Bushehrian et al.~\cite{bushehrian2023code} & One-phase and Two-phase method from Camel, CloudStack, Geode and Hbase & \url{https://github.com/Ziba-Ghane/CQC.git}\\
        Dipongkor~\cite{dipongkor2023fusing} & Sun Firefox, JDT, Netbeans, GUO Firefox, GCC datasets & \url{https://github.com/farhan-93/bugtriage}\\
        \multirow{2}{*}{$Triage_{Expert-Recency}$~\cite{goyal2024integrated}} & Mozilla Firefox project bug reports & \url{https://bugzilla.mozilla.org/describecomponents.cgi?product=Firefox}\\
        & SeaMonkey project bug reports & \url{https://bugzilla.mozilla.org/describecomponents.cgi?product=SeaMonkey}\\
        Jalal~\cite{jalal2024enhancement}&Eclipse project's Bugzilla bug report dataset with 48 features & \url{https://data.mendeley.com/datasets/t6d9y7yt54/1} \\
        Sajedi et al.~\cite{sajedi2016crowdsourced} & Bug reports from 20 large GitHub projects & \url{https://github.com/alisajedi/BugTriaging} \\
        Bernardi et al.~\cite{BERNARDI2018111} & Bug and communication data from Apache httpd, GNU GCC, Mozilla Firefox, and Xorg Xserver projects & \url{https://github.com/mlbresearch/talking-data} \\
        GitBugs~\cite{patil2025gitbugs} & Bug report dataset from 9 open-source projects & \url{https://github.com/av9ash/gitbugs/}\\
        Krasniqi~\cite{krasniqi2022automatically} & Bug reports from six OSS projects (Bugzilla \& Jira) & \url{https://zenodo.org/record/6412840} \\
        Bugayenko et al.~\cite{bugayenko2022automatically} & Software development tasks (``puzzles'') extracted from industrial code repositories & \url{https://github.com/cqfn/pdd-data-analysis} \\
        James et al.~\cite{james2022separating} & crash reports from large-scale OSS bug repositories & \url{https://github.com/kedjames/crashsearch-triage} \\
        MSR2013~\cite{LamkanfiMSR13,MOHSIN2022107711} & Reported bugs extracted from the Eclipse and Mozilla projects. & \url{https://github.com/ansymo/msr2013-bug_dataset} \\
        Bug Triaging~\cite{MOHSIN2022107711} & Bugs tagged in the Eclipse dataset. & \url{https://www.kaggle.com/datasets/monika11/bug-triagingbug-assignment/data} \\
        \makecell[l]{Bugzilla-Mozilla~\cite{singh2023empirical}\\~\cite{qian2023survey, yadav2024developer, lyakajigari2024review}} & System Bug report &\url{https://bugzilla.mozilla.org/home}\\
        \makecell[l]{Eclipse~\cite{singh2023empirical,qian2023survey,samir2023improving}\\~\cite{park2011costriage,tamrawi2011bugzie,murphy2004automatic,xi2019bug}\\~\cite{yang2014towards,nagwani2023artificial,yadav2024developer,lyakajigari2024review} }& Bugs reported \& Bugs changed & \url{https://bugs.eclipse.org/bugs/}\\
        \makecell[l]{NetBeans~\cite{nagwani2023artificial,singh2023empirical,qian2023survey}\\~\cite{yadav2024developer,lyakajigari2024review}} & NetBeans bug repository & \url{https://issues.apache.org/jira/projects/NETBEANS}\\
        Apache~\cite{tamrawi2011bugzie,park2011costriage} & Bugs reported \& Bugs changed & \url{https://issues.apache.org/jira/}\\
        GCC~\cite{nagwani2023artificial,tamrawi2011bugzie} & Bugs reported \& Bugs changed & \url{http://gcc.gnu.org/bugzilla/}\\
        Linux kernel~\cite{park2011costriage} & Bugs reported \& Bugs changed & \url{https://bugzilla.kernel.org/}\\
        Gentoo~\cite{xi2019bug}
        & Bugs reported \& Bugs changed & \url{https://bugs.gentoo.org)}\\

\end{longtable}
\end{singlespace}
\vspace{-0.5cm}
\end{center}

\textbf{Data Accessibility.} Most datasets used in defect classification research are fully open. Numerous studies explicitly acknowledge that their data was collected from publicly available platforms, with some works providing processed datasets or reproducible packages~\cite{su2021reducing, aung2022multi}. 

\textbf{Data Scale and Type.} These datasets are generally large-scale, ranging from tens of thousands to several hundred thousand reports. For instance, Mozilla and Eclipse datasets often include between 100,000 and 500,000 defect reports. While the core data is textual (titles and descriptions), structured metadata provides supplementary classification features.

\textbf{Industrial Proprietary Data.} Beyond open-source repositories, certain studies have leveraged proprietary defect datasets from companies such as LG Electronics, IBM (Jazz), and Microsoft (Windows Vista). These datasets offer insights into defect characteristics within real industrial environments. However, due to commercial confidentiality, they are rarely made publicly available, limiting reproducibility and wider academic adoption.

\subsubsection{Incident Triage Datasets}

In contrast to defect classification, acquiring datasets for incident management, classification, and prioritization in large-scale online service systems remains significantly more challenging.

\textbf{Data Sources and Characteristics.} Research in this area largely depends on proprietary datasets collected from production environments of technology companies such as Microsoft (particularly Azure), IBM, and Google. Unlike defect reports, incident datasets are inherently multimodal, encompassing user-submitted natural language incident tickets alongside diverse machine-generated data, including system logs, performance metrics, service dependency graphs, configuration change logs, and alerts. The heterogeneity and complexity of this data underscore the core challenges of incident triage.

\textbf{Data Accessibility.} Due to concerns over trade secrets, system security, and user privacy, nearly all incident triage datasets remain proprietary. Consequently, most published studies describe only high-level data characteristics (\eg ``six months of incident data totaling 90 GB'' or ``terabytes of daily service logs''), without releasing raw or anonymized datasets. This lack of accessibility severely impedes reproducibility and prevents fair cross-study comparisons.

\textbf{Research Challenges and Outlook.} The field of incident triage is heavily constrained by its reliance on proprietary industrial data. To overcome this impasse and facilitate technological transfer from research to practice, collaborative initiatives are urgently needed to build and release benchmark datasets that faithfully represent real-world industrial scenarios. Establishing standardized datasets and unified evaluation metrics would be a critical enabler for accelerating innovation, enhancing comparability of research outcomes, and fostering the development of advanced and practically deployable triage solutions.

\subsection{Toolkits}
In this section, we present a curated list of open-source code repositories and toolkits derived from the surveyed papers on bug triage, incident triage, and related tasks. These resources are based on works that explicitly indicate code availability in the provided document.  We focus on those with accessible links or confirmed releases, as they enable researchers, developers, and practitioners to replicate experiments, extend models, or integrate them into real-world systems. Table~\ref{table:toolkits} summarizes publicly available toolkits, which serve as essential baselines for benchmarking new approaches.

\begin{center}
\footnotesize
\begin{singlespace}
\begin{longtable}{ccp{4cm}p{7cm}}
\caption{
 Summary of publicly available triage toolkits.} 
\label{table:toolkits} \\
\toprule
\textbf{Name} & \textbf{Year} & \textbf{Data}  & \textbf{Details}\\
\midrule
\endfirsthead

\multicolumn{4}{c}%
{\tablename\ \thetable\ -- \textit{Continued from previous page}} \\
\toprule
\textbf{Name} & \textbf{Year} & \textbf{Data}  & \textbf{Details}\\
\midrule
\endhead

\midrule \multicolumn{4}{r}{\textit{Continued on next page}} \\
\endfoot

\bottomrule
\endlastfoot
        MultiTriage~\cite{aung2022multi} & 2022 & Bug Reports from Eclipse \& Github OSS projects & A neural network based bug triage learning model to recommend the list of developers and issue types most relevant to a new issue report. \url{https://github.com/thazin31086/MultiTriage} \\
        LR-BKG~\cite{su2021reducing} & 2021 & Bug Reports from Mozilla OSS projects & A learning-to-rank framework that learns to distinguish correct, erroneous and irrelevant bugcomponent assignments, based on a rich set of features derived from bug tossing knowledge graph. \url{https://github.com/SuYanqi/LR-BKG} \\
        Wu et al.~\cite{wu2022spatial} & 2022 & Fixed bug reports from two OSS projects, namely Eclipse and Mozilla. & A spatial–temporal dynamic graph neural network (ST-DGNN) framework to improve automated bug triaging by modeling developer collaboration networks over time and predicting the most suitable bug fixers. \url{https://github.com/ssea-lab/BugTriage/tree/master/GRCNN} \\
        \text{ADPTriage}~\cite{jahanshahi2023adptriage} & 2023 & Bug reports from \text{EclipseJDT} \& GCC \& Mozilla OSS projects & A triage model for ITS accounts for the uncertainties, which not only assigns the bugs to the most appropriate developers or postpones them to the future but also determines the assignment timing according to the likelihood of having a particular bug type in the system and possible changes in developers' schedules in the future. \url{https://github.com/HadiJahanshahi/ADPTriage/tree/main/Toy_Example} \\
        Zhang et al.~\cite{zhang2016towards} & 2016 & Bug reports from Eclipse \& GCC \& Mozilla \& Netbeans \& OpenOffice OSS projects & An automatic approach to perform severity prediction and fixer recommendation Based on the features (\eg textural similarity and developers' experience) extracted from top-K nearest neighbours of the new bug report. \url{https://github.com/ProgrammerCJC/SPFR} \\
        VTBA~\cite{sajedi2020vocabulary} & 2020 & Bug reports from 13 popular GitHub projects (\eg Angular.js, Rails, Elasticsearch) & An vocabulary and time-aware bug-assignment approach by matching technical terms filtered via Stack Overflow and weighting historical fixes based on recency. \url{https://github.com/TaskAssignment/VTBA} \\
        Wayback~\cite{jahanshahi2022wayback} & 2022 & Bug reports from EclipseJDT \&  Mozilla \& OpenOffice OSS projects & An event-replay-based approach to reconstructing historical bug triage scenarios, enabling dependency-aware and workload-balanced assignment through dynamic bug dependency graph updates. \url{https://github.com/HadiJahanshahi/WaybackMachine} \\
        S-DABT~\cite{jahanshahi2022s} & 2022 & Bug reports from EclipseJDT \&  Mozilla \& OpenOffice OSS projects & An schedule and dependency-aware bug triage approach, which utilizes integer programming and machine learning techniques to assign bugs to suitable developers.
        \url{https://github.com/HadiJahanshahi/SDABT} \\
        SevPredict~\cite{arshad2024sevpredict} & 2024 & Bug reports from 8 OSS projects (\eg Mozilla Core, Mozilla Firefox) & A GPT--2-based framework for automated bug severity prediction, which preprocesses bug report text, extracts sentiment features, and inputs these into a fine-tuned transformer model, capturing semantic and contextual patterns to generate real-time severity predictions for integration with bug tracking systems. 
        \url{https://huggingface.co/spaces/AliArshad/SeverityPrediction}\\
        Wang et al.~\cite{wang2024empirical} & 2024 & Bug reports from EclipseJDT \& GCC \& Firefox OSS projects & An empirical approach to evaluating word embedding and deep learning combinations for automated bug assignment. \url{https://github.com/AI4BA/dl4ba} \\
        AutoAnalysis~\cite{wang2024comet}& 2024 & ERP Incident tickets &SplitSD4X groups incidents via subgroup discovery to summarize black box explanations.\url{https://github.com/RemilYoucef/split-sd4x}\\
        Gao et al. ~\cite{gao2018online} & 2018 & User reviews of 6 popular apps & Proposes IDEA framework, uses AOLDA to track version-sensitive topic distribution, detects emerging app issues, and labels topics with semantics and sentiment.\url{https://github.com/ReMine-Lab/IDEA}\\
        Di et al.~\cite{di2021investigating}& 2021 & Reviews and versions of Android apps\& code quality metrics &Classifies user reviews into PD/FR via URM, correlates app ratings with code quality metrics to identify critical user feedback.\url{https://github.com/sealuzh/user-satisfaction}\\
        ART~\cite{sun2024art} & 2024 & Microservice datasets \& failure cases & Proposes ART unsupervised framework, uses Transformer/GRU/GraphSAGE to model multi-dependencies, unifying AD, FT, and RCL.\url{https://github.com/bbyldebb/ART}\\
        Sadlek et al.~\cite{sadlek2025severity} & 2025 & Testbed data with Windows/Ubuntu hosts \& alerts \& logs from 5 attack scenarios & Proposes severity-based cyber incident triage via kill chain attack graphs, uses MulVAL to generate graphs and match alert sequences.\url{http://dx.doi.org/10.5281/zenodo.14547668}\\
        Gao et al.~\cite{gao2022listening}& 2022 & Reviews from 5 apps (\eg eBay, Viber) & Proposes SOLAR framework with review helpfulness prediction, topic-sentiment modeling, and multi-factor ranking to summarize useful app reviews.\url{https://github.com/monsterLee599/SOLAR}\\
        Haering et al. \cite{haering2021automatically} & 2021 & Problem reviews and bug reports from 4 open-source apps (\eg Firefox, VLC)& Proposes DeepMatcher, uses DistilBERT for text embedding and cosine similarity to match app reviews with bug reports.\url{https://mast.informatik.uni-hamburg.de/replication-packages/}\\
        Sajedi~\cite{sajedi2016crowdsourced} & 2016 & Bug reports from 20 large GitHub projects & An expertise-aware bug triaging approach leveraging developers' Stack Overflow activities to identify suitable assignees. \url{https://github.com/alisajedi/BugTriaging} \\
        He et al.~\cite{he2021automatic} & 2021 & Bug reports from Chromium, Mozilla Core \& Firefox, NetBeans, and Eclipse OSS projects & An end-to-end hierarchical attention network approach for automatic bug triage. \url{https://github.com/username1017/Bug-triage-with-HAN} \\
        Bhuyan et al.~\cite{bhuyan2021evaluating} & 2021 & Bug reports from Mozilla projects & A web browser plug-in for Google Chrome which recommends developer expertise based on the bug report. \url{https://chromewebstore.google.com/detail/recommend-expertise/clpcpddhohohhfcnkiknfopaeikbngid?pli=1} \\
        Krasniqi~\cite{krasniqi2022automatically} & 2022 & Bug reports from six OSS projects (Bugzilla \& Jira) & An automated quality-based bug classification approach leveraging feature selection and machine learning algorithms. \url{https://zenodo.org/record/6412840} \\
        James et al.~\cite{james2022separating} & 2022 & Crash reports from large-scale OSS bug repositories & An LSH- and sequential-pattern-mining–based approach for fingerprinting and clustering crash reports to identify duplicate and related bugs. \url{https://github.com/kedjames/crashsearch-triage} \\
\end{longtable}
\end{singlespace}
\vspace{-0.5cm}
\end{center}

\begin{tcolorbox}[colback=gray!10,
	colframe=black,
	width=15cm,
	arc=2mm, auto outer arc,
	title={\textbf{Finding 3: Evaluation and Benchmarking of Triage Approaches}}]		
	\textit{In evaluation, triage approaches are examined using a range of accuracy, ranking, efficiency, and user-centered metrics. Traditional metrics such as accuracy, precision, recall, and Top-k performance remain predominant, while efficiency metrics and qualitative user feedback offer complementary insights into real-world applicability. Most existing studies utilize open-source bug repositories such as Mozilla and Eclipse; however, research on incident triage remains constrained by the scarcity of industrial datasets and the lack of standardized benchmarks. These limitations underscore the necessity for unified datasets, consistent evaluation protocols, and reproducible toolkits to facilitate fair and practical comparisons across triage methods.} 
\end{tcolorbox}
\vspace{-0.5cm}
\section{Future Trends and Opportunities}\label{sec_7}

\subsection{Fusion of Heterogeneous Data Sources}
Triage in modern systems is inherently multimodal, drawing on diverse data sources such as logs, alerts, engineer discussions, postmortem analyses, and system documentation. Each of these sources provides a distinct perspective: logs capture fine-grained system behaviors, discussions reflect human reasoning and contextualization~\cite{wang2024comet}, while documentation encodes structured definitions and domain-specific knowledge. Existing approaches, however, often treat these modalities in isolation or perform naive concatenation at the representation level, leading to information loss and reduced interpretability.

Beyond textual and structured data, triage increasingly involves visual artifacts such as defect screenshots~\cite{sun2025tixfusion}. These visual elements are often indispensable for root cause analysis, as they encode operational signals that are difficult to fully convey in text. For instance, error screenshots and short video captures uploaded by beta users provide direct visual insights into the state of the operating system, offering contextual cues that are otherwise lost in textual descriptions. 

However, the reliance on such visual information places a heavy burden on human triage engineers. Each submitted ticket must be carefully examined, including both text and visual attachments, before being assigned to the appropriate development team. This process is both time-intensive and expertise-dependent: experienced engineers with domain-specific knowledge of mobile operating systems and related components can typically process only around forty tickets per day~\cite{sun2025tixfusion}. The scarcity of such skilled engineers, coupled with their central role in the pipeline, often results in significant delays and bottlenecks, ultimately reducing the overall efficiency of incident management. This tension highlights a critical gap: while visual artifacts greatly enrich interpretability, they simultaneously exacerbate the challenges of scalability and automation in current triage practices.

A promising research direction lies in the systematic fusion of heterogeneous data, where the complementary strengths of different modalities are jointly leveraged. For instance, temporal correlations between log bursts and discussion timelines may yield stronger signals for incident categorization, while aligning system documentation with log templates could enable automatic mapping of raw anomalies to meaningful fault categories. Designing architectures that dynamically weight diverse inputs, adapt to noisy signals, and resolve conflicts across modalities will be essential for developing robust and interpretable triage systems.

\subsection{Integration of Domain Knowledge}
Despite the growing success of data-driven techniques, effective triage in real-world settings continues to rely heavily on domain expertise. Practitioners regularly employ service-specific terminology, heuristics, and taxonomies to make timely and accurate decisions. However, most existing triage models fail to explicitly encode such structured knowledge, resulting in limited interpretability and poor generalization across projects or services.

Recent studies have begun to bridge this gap by integrating knowledge graphs, ontologies, and rule-based constraints to enhance data-driven triage. For example, Jahanshahi et al.~\cite{jahanshahi2023adptriage} propose policy learning constrained by project-specific knowledge, while Wu et al.~\cite{wu2022spatial} incorporate developer collaboration graphs to capture implicit expertise and team dynamics.

Future research should focus on knowledge-enhanced triage frameworks, where structured representations, such as failure taxonomies, component hierarchies, and domain ontologies, are integrated with representation learning and LLM-based reasoning. This trend aligns with the broader movement in AIOps toward knowledge-grounded automation, where decisions are both data-driven and semantically interpretable. Knowledge-guided models inherently promote explainability, as their reasoning aligns with concepts familiar to engineers, thereby improving trust, transparency, and adoption in industrial environments.

\subsection{Human-in-the-Loop and Continuous Learning}
While automation continues to advance, human expertise remains indispensable in high-stakes triage scenarios. Prior research has consistently shown that practitioners' feedback is critical for refining model predictions, resolving ambiguous cases, and maintaining operational trust~\cite{chen2019continuous}. However, existing pipelines often treat human input as a post hoc verification step rather than a core learning signal.

Future triage frameworks should embrace human-in-the-loop (HITL) paradigms, where engineers' corrections, rationales, and annotations are continuously incorporated into model retraining processes, enabling mutual adaptation between humans and AI systems. Semi-supervised and reinforcement learning techniques could exploit weak supervision signals, such as corrective labels, ranking feedback, or textual annotations, to update models online and improve performance over time.

Developing transparent feedback interfaces and explainable reasoning mechanisms will be crucial to facilitate effective collaboration between AI and human experts. Such mechanisms not only enhance model accountability and interpretability but also foster long-term trust and continuous learning in production triage environments.

\subsection{Generalizability and Model Adaptability in Software Evolution}
A persistent limitation of conventional triage systems lies in their dependence on static rules or narrowly trained models, which struggle to cope with the diversity and dynamism of large-scale service environments~\cite{dipongkor2023comparative}. As incident patterns evolve and architectures change, these static approaches degrade in accuracy, scalability, and adaptability~\cite{aung2022multi}. While heuristic and rule-based automation has reduced manual effort to some extent, it remains inadequate for capturing the complex and multifaceted nature of real-world failure contexts.

Recent studies advocate for continuous model adaptation and cross-project generalization, leveraging techniques such as domain adaptation, transfer learning, and incremental fine-tuning~\cite{su2021reducing}. Future research should emphasize rigorous benchmarking across heterogeneous environments and issue types, accompanied by standardized evaluation frameworks that reflect evolving operational realities. Moreover, integrating adaptive feedback loops and hybrid designs, combining LLM reasoning with domain-specific knowledge graphs or causal inference models, holds promise for creating triage systems that are intelligent, resilient, and scalable.

\section{Conclusion}

With the increasing scale and complexity of software and service systems, as well as the continuous advancement of deployment technologies, triage has emerged as a critical research frontier in software engineering. This survey provides a comprehensive review of 234 studies on software triage published between 2004 and 2025, summarizing existing system architectures and technical approaches. Through an extensive analysis of prior research, we outlined the major progress achieved in triage automation and examined the empirical differences between academic research and industrial practice. To promote further research and practical adoption, we consolidated widely used datasets and evaluation metrics that establish a consistent framework for assessing triage effectiveness. The survey also identified several promising directions for future work, emphasizing the importance of closer collaboration between academia and industry. Our goal is to contribute to the advancement of modern software systems by offering researchers a comprehensive reference and valuable insights that may inspire future exploration in triage technologies.

\bibliographystyle{ACM-Reference-Format}
\bibliography{ref}


\begin{thebibliography}{196}


\ifx \showCODEN    \undefined \def \showCODEN     #1{\unskip}     \fi
\ifx \showISBNx    \undefined \def \showISBNx     #1{\unskip}     \fi
\ifx \showISBNxiii \undefined \def \showISBNxiii  #1{\unskip}     \fi
\ifx \showISSN     \undefined \def \showISSN      #1{\unskip}     \fi
\ifx \showLCCN     \undefined \def \showLCCN      #1{\unskip}     \fi
\ifx \shownote     \undefined \def \shownote      #1{#1}          \fi
\ifx \showarticletitle \undefined \def \showarticletitle #1{#1}   \fi
\ifx \showURL      \undefined \def \showURL       {\relax}        \fi
\providecommand\bibfield[2]{#2}
\providecommand\bibinfo[2]{#2}
\providecommand\natexlab[1]{#1}
\providecommand\showeprint[2][]{arXiv:#2}

\bibitem[dbl(2025)]%
        {dblp}
 \bibinfo{year}{2025}\natexlab{}.
\newblock \bibinfo{title}{DBLP}.
\newblock
\urldef\tempurl%
\url{https://dblp.org/}
\showURL{%
\tempurl}


\bibitem[Ahsan et~al\mbox{.}(2009)]%
        {ahsan2009automatic}
\bibfield{author}{\bibinfo{person}{Syed~Nadeem Ahsan}, \bibinfo{person}{Javed Ferzund}, {and} \bibinfo{person}{Franz Wotawa}.} \bibinfo{year}{2009}\natexlab{}.
\newblock \showarticletitle{Automatic software bug triage system (bts) based on latent semantic indexing and support vector machine}. In \bibinfo{booktitle}{\emph{2009 Fourth International Conference on Software Engineering Advances}}. IEEE, \bibinfo{pages}{216--221}.
\newblock


\bibitem[Akila et~al\mbox{.}(2014)]%
        {akila2014survey}
\bibfield{author}{\bibinfo{person}{V Akila}, \bibinfo{person}{G Zayaraz}, {and} \bibinfo{person}{V Govindasamy}.} \bibinfo{year}{2014}\natexlab{}.
\newblock \showarticletitle{Bug triage in open source systems: a review}.
\newblock \bibinfo{journal}{\emph{International Journal of Collaborative Enterprise}} \bibinfo{volume}{4}, \bibinfo{number}{4} (\bibinfo{year}{2014}), \bibinfo{pages}{299--319}.
\newblock


\bibitem[Al-Bayati et~al\mbox{.}(2024)]%
        {jalal2024enhancement}
\bibfield{author}{\bibinfo{person}{Jalal Sadoon~Hameed Al-Bayati}, \bibinfo{person}{Mohammed Al-Shamma}, {and} \bibinfo{person}{Furat~Nidhal Tawfeeq}.} \bibinfo{year}{2024}\natexlab{}.
\newblock \showarticletitle{Enhancement of Recommendation Engine Technique for Bug System Fixes}.
\newblock \bibinfo{journal}{\emph{Journal of Advances in Information Technology}} \bibinfo{volume}{15}, \bibinfo{number}{4} (\bibinfo{year}{2024}).
\newblock


\bibitem[Alazzam et~al\mbox{.}(2020)]%
        {alazzam2020automatic}
\bibfield{author}{\bibinfo{person}{Iyad Alazzam}, \bibinfo{person}{Ahmed Aleroud}, \bibinfo{person}{Zainab Al~Latifah}, {and} \bibinfo{person}{George Karabatis}.} \bibinfo{year}{2020}\natexlab{}.
\newblock \showarticletitle{Automatic bug triage in software systems using graph neighborhood relations for feature augmentation}.
\newblock \bibinfo{journal}{\emph{IEEE Transactions on Computational Social Systems}} \bibinfo{volume}{7}, \bibinfo{number}{5} (\bibinfo{year}{2020}), \bibinfo{pages}{1288--1303}.
\newblock


\bibitem[Alenezi et~al\mbox{.}(2013)]%
        {alenezi2013efficient}
\bibfield{author}{\bibinfo{person}{Mamdouh Alenezi}, \bibinfo{person}{Kenneth Magel}, {and} \bibinfo{person}{Shadi Banitaan}.} \bibinfo{year}{2013}\natexlab{}.
\newblock \showarticletitle{Efficient Bug Triaging Using Text Mining.}
\newblock \bibinfo{journal}{\emph{J. Softw.}} \bibinfo{volume}{8}, \bibinfo{number}{9} (\bibinfo{year}{2013}), \bibinfo{pages}{2185--2190}.
\newblock


\bibitem[Almhana and Kessentini(2021)]%
        {almhana2021considering}
\bibfield{author}{\bibinfo{person}{Rafi Almhana} {and} \bibinfo{person}{Marouane Kessentini}.} \bibinfo{year}{2021}\natexlab{}.
\newblock \showarticletitle{Considering dependencies between bug reports to improve bugs triage}.
\newblock \bibinfo{journal}{\emph{Automated Software Engineering}} \bibinfo{volume}{28}, \bibinfo{number}{1} (\bibinfo{year}{2021}), \bibinfo{pages}{1}.
\newblock


\bibitem[Anvik(2006)]%
        {anvik2006automating}
\bibfield{author}{\bibinfo{person}{John Anvik}.} \bibinfo{year}{2006}\natexlab{}.
\newblock \showarticletitle{Automating bug report assignment}. In \bibinfo{booktitle}{\emph{Proceedings of the 28th international conference on Software engineering}}. \bibinfo{pages}{937--940}.
\newblock


\bibitem[Anvik et~al\mbox{.}(2006)]%
        {anvik2006should}
\bibfield{author}{\bibinfo{person}{John Anvik}, \bibinfo{person}{Lyndon Hiew}, {and} \bibinfo{person}{Gail~C Murphy}.} \bibinfo{year}{2006}\natexlab{}.
\newblock \showarticletitle{Who should fix this bug?}. In \bibinfo{booktitle}{\emph{Proceedings of the 28th international conference on Software engineering}}. \bibinfo{pages}{361--370}.
\newblock


\bibitem[Anvik and Murphy(2011)]%
        {anvik2011reducing}
\bibfield{author}{\bibinfo{person}{John Anvik} {and} \bibinfo{person}{Gail~C Murphy}.} \bibinfo{year}{2011}\natexlab{}.
\newblock \showarticletitle{Reducing the effort of bug report triage: Recommenders for development-oriented decisions}.
\newblock \bibinfo{journal}{\emph{ACM Transactions on Software Engineering and Methodology (TOSEM)}} \bibinfo{volume}{20}, \bibinfo{number}{3} (\bibinfo{year}{2011}), \bibinfo{pages}{1--35}.
\newblock


\bibitem[Arshad et~al\mbox{.}(2024)]%
        {arshad2024sevpredict}
\bibfield{author}{\bibinfo{person}{Muhammad~Ali Arshad}, \bibinfo{person}{Adnan Riaz}, \bibinfo{person}{Rubia Fatima}, {and} \bibinfo{person}{Affan Yasin}.} \bibinfo{year}{2024}\natexlab{}.
\newblock \showarticletitle{SevPredict: Exploring the Potential of Large Language Models in Software Maintenance}.
\newblock \bibinfo{journal}{\emph{AI}} \bibinfo{volume}{5}, \bibinfo{number}{4} (\bibinfo{year}{2024}), \bibinfo{pages}{2739--2760}.
\newblock


\bibitem[Aung et~al\mbox{.}(2022)]%
        {aung2022multi}
\bibfield{author}{\bibinfo{person}{Thazin Win~Win Aung}, \bibinfo{person}{Yao Wan}, \bibinfo{person}{Huan Huo}, {and} \bibinfo{person}{Yulei Sui}.} \bibinfo{year}{2022}\natexlab{}.
\newblock \showarticletitle{Multi-triage: A multi-task learning framework for bug triage}.
\newblock \bibinfo{journal}{\emph{Journal of Systems and Software}}  \bibinfo{volume}{184} (\bibinfo{year}{2022}), \bibinfo{pages}{111133}.
\newblock


\bibitem[Banitaan and Alenezi(2013)]%
        {banitaan2013decoba}
\bibfield{author}{\bibinfo{person}{Shadi Banitaan} {and} \bibinfo{person}{Mamdouh Alenezi}.} \bibinfo{year}{2013}\natexlab{}.
\newblock \showarticletitle{Decoba: Utilizing developers communities in bug assignment}. In \bibinfo{booktitle}{\emph{2013 12th International Conference on Machine Learning and Applications}}, Vol.~\bibinfo{volume}{2}. IEEE, \bibinfo{pages}{66--71}.
\newblock


\bibitem[Bansal et~al\mbox{.}(2020)]%
        {bansal2020decaf}
\bibfield{author}{\bibinfo{person}{Chetan Bansal}, \bibinfo{person}{Sundararajan Renganathan}, \bibinfo{person}{Ashima Asudani}, \bibinfo{person}{Olivier Midy}, {and} \bibinfo{person}{Mathru Janakiraman}.} \bibinfo{year}{2020}\natexlab{}.
\newblock \showarticletitle{DeCaf: diagnosing and triaging performance issues in large-scale cloud services}. In \bibinfo{booktitle}{\emph{Proceedings of the ACM/IEEE 42nd International Conference on Software Engineering: Software Engineering in Practice}} (Seoul, South Korea) \emph{(\bibinfo{series}{ICSE-SEIP '20})}. \bibinfo{publisher}{Association for Computing Machinery}, \bibinfo{address}{New York, NY, USA}, \bibinfo{pages}{201–210}.
\newblock
\showISBNx{9781450371230}
\href{https://doi.org/10.1145/3377813.3381353}{doi:\nolinkurl{10.1145/3377813.3381353}}


\bibitem[Baysal et~al\mbox{.}(2009)]%
        {baysal2009bug}
\bibfield{author}{\bibinfo{person}{Olga Baysal}, \bibinfo{person}{Michael~W Godfrey}, {and} \bibinfo{person}{Robin Cohen}.} \bibinfo{year}{2009}\natexlab{}.
\newblock \showarticletitle{A bug you like: A framework for automated assignment of bugs}. In \bibinfo{booktitle}{\emph{2009 IEEE 17th International Conference on Program Comprehension}}. IEEE, \bibinfo{pages}{297--298}.
\newblock


\bibitem[Bernardi et~al\mbox{.}(2018)]%
        {BERNARDI2018111}
\bibfield{author}{\bibinfo{person}{Mario~Luca Bernardi}, \bibinfo{person}{Gerardo Canfora}, \bibinfo{person}{Giuseppe~A. {Di Lucca}}, \bibinfo{person}{Massimiliano {Di Penta}}, {and} \bibinfo{person}{Damiano Distante}.} \bibinfo{year}{2018}\natexlab{}.
\newblock \showarticletitle{The relation between developers’ communication and fix-Inducing changes: An empirical study}.
\newblock \bibinfo{journal}{\emph{Journal of Systems and Software}}  \bibinfo{volume}{140} (\bibinfo{year}{2018}), \bibinfo{pages}{111--125}.
\newblock
\showISSN{0164-1212}
\href{https://doi.org/10.1016/j.jss.2018.02.065}{doi:\nolinkurl{10.1016/j.jss.2018.02.065}}


\bibitem[Bhattacharya and Neamtiu(2010)]%
        {bhattacharya2010fine}
\bibfield{author}{\bibinfo{person}{Pamela Bhattacharya} {and} \bibinfo{person}{Iulian Neamtiu}.} \bibinfo{year}{2010}\natexlab{}.
\newblock \showarticletitle{Fine-grained incremental learning and multi-feature tossing graphs to improve bug triaging}. In \bibinfo{booktitle}{\emph{2010 IEEE International Conference on Software Maintenance}}. \bibinfo{pages}{1--10}.
\newblock
\href{https://doi.org/10.1109/ICSM.2010.5609736}{doi:\nolinkurl{10.1109/ICSM.2010.5609736}}


\bibitem[Bhattacharya et~al\mbox{.}(2012)]%
        {bhattacharya2012automated}
\bibfield{author}{\bibinfo{person}{Pamela Bhattacharya}, \bibinfo{person}{Iulian Neamtiu}, {and} \bibinfo{person}{Christian~R Shelton}.} \bibinfo{year}{2012}\natexlab{}.
\newblock \showarticletitle{Automated, highly-accurate, bug assignment using machine learning and tossing graphs}.
\newblock \bibinfo{journal}{\emph{Journal of Systems and Software}} \bibinfo{volume}{85}, \bibinfo{number}{10} (\bibinfo{year}{2012}), \bibinfo{pages}{2275--2292}.
\newblock


\bibitem[Bhuyan and Anvik(2021)]%
        {bhuyan2021evaluating}
\bibfield{author}{\bibinfo{person}{Shayla~Azad Bhuyan} {and} \bibinfo{person}{John Anvik}.} \bibinfo{year}{2021}\natexlab{}.
\newblock \showarticletitle{Evaluating Visual Explanation of Bug Report Assignment Recommendations (S).}. In \bibinfo{booktitle}{\emph{SEKE}}. \bibinfo{pages}{334--339}.
\newblock


\bibitem[Bocu et~al\mbox{.}(2023)]%
        {bocu2023survey}
\bibfield{author}{\bibinfo{person}{Razvan Bocu}, \bibinfo{person}{Alexandra Baicoianu}, {and} \bibinfo{person}{Arpad Kerestely}.} \bibinfo{year}{2023}\natexlab{}.
\newblock \showarticletitle{An Extended Survey Concerning the Significance of Artificial Intelligence and Machine Learning Techniques for Bug Triage and Management}.
\newblock \bibinfo{journal}{\emph{IEEE Access}}  \bibinfo{volume}{11} (\bibinfo{year}{2023}), \bibinfo{pages}{123924--123937}.
\newblock
\href{https://doi.org/10.1109/ACCESS.2023.3329732}{doi:\nolinkurl{10.1109/ACCESS.2023.3329732}}


\bibitem[Borg et~al\mbox{.}(2024)]%
        {borg2024adopting}
\bibfield{author}{\bibinfo{person}{Markus Borg}, \bibinfo{person}{Leif Jonsson}, \bibinfo{person}{Emelie Engstr{\"o}m}, \bibinfo{person}{B{\'e}la Bartalos}, {and} \bibinfo{person}{Attila Szab{\'o}}.} \bibinfo{year}{2024}\natexlab{}.
\newblock \showarticletitle{Adopting automated bug assignment in practice—a longitudinal case study at Ericsson}.
\newblock \bibinfo{journal}{\emph{Empirical Software Engineering}} \bibinfo{volume}{29}, \bibinfo{number}{5} (\bibinfo{year}{2024}), \bibinfo{pages}{126}.
\newblock


\bibitem[Bortis and van~der Hoek(2013)]%
        {bortis2013porchlight}
\bibfield{author}{\bibinfo{person}{Gerald Bortis} {and} \bibinfo{person}{André van~der Hoek}.} \bibinfo{year}{2013}\natexlab{}.
\newblock \showarticletitle{PorchLight: A tag-based approach to bug triaging}. In \bibinfo{booktitle}{\emph{2013 35th International Conference on Software Engineering (ICSE)}}. \bibinfo{pages}{342--351}.
\newblock
\href{https://doi.org/10.1109/ICSE.2013.6606580}{doi:\nolinkurl{10.1109/ICSE.2013.6606580}}


\bibitem[Bugayenko et~al\mbox{.}(2022)]%
        {bugayenko2022automatically}
\bibfield{author}{\bibinfo{person}{Yegor Bugayenko}, \bibinfo{person}{Ayomide Bakare}, \bibinfo{person}{Arina Cheverda}, \bibinfo{person}{Mirko Farina}, \bibinfo{person}{Artem Kruglov}, \bibinfo{person}{Yaroslav Plaksin}, \bibinfo{person}{Giancarlo Succi}, {and} \bibinfo{person}{Witold Pedrycz}.} \bibinfo{year}{2022}\natexlab{}.
\newblock \showarticletitle{Automatically Prioritizing and Assigning Tasks from Code Repositories in Puzzle Driven Development}. In \bibinfo{booktitle}{\emph{2022 IEEE/ACM 19th International Conference on Mining Software Repositories (MSR)}}. \bibinfo{pages}{722--723}.
\newblock
\href{https://doi.org/10.1145/3524842.3528512}{doi:\nolinkurl{10.1145/3524842.3528512}}


\bibitem[Bushehrian and Ghane(2023)]%
        {bushehrian2023code}
\bibfield{author}{\bibinfo{person}{Omid Bushehrian} {and} \bibinfo{person}{Ziba Ghane}.} \bibinfo{year}{2023}\natexlab{}.
\newblock \showarticletitle{Code quality control by bug report classification}.
\newblock \bibinfo{journal}{\emph{Software Quality Journal}} \bibinfo{volume}{31}, \bibinfo{number}{3} (\bibinfo{year}{2023}), \bibinfo{pages}{991--1007}.
\newblock


\bibitem[Catolino et~al\mbox{.}(2019)]%
        {catolino2019not}
\bibfield{author}{\bibinfo{person}{Gemma Catolino}, \bibinfo{person}{Fabio Palomba}, \bibinfo{person}{Andy Zaidman}, {and} \bibinfo{person}{Filomena Ferrucci}.} \bibinfo{year}{2019}\natexlab{}.
\newblock \showarticletitle{Not all bugs are the same: Understanding, characterizing, and classifying bug types}.
\newblock \bibinfo{journal}{\emph{Journal of Systems and Software}}  \bibinfo{volume}{152} (\bibinfo{year}{2019}), \bibinfo{pages}{165--181}.
\newblock


\bibitem[Chen et~al\mbox{.}(2019a)]%
        {chen2019empirical}
\bibfield{author}{\bibinfo{person}{Junjie Chen}, \bibinfo{person}{Xiaoting He}, \bibinfo{person}{Qingwei Lin}, \bibinfo{person}{Yong Xu}, \bibinfo{person}{Hongyu Zhang}, \bibinfo{person}{Dan Hao}, \bibinfo{person}{Feng Gao}, \bibinfo{person}{Zhangwei Xu}, \bibinfo{person}{Yingnong Dang}, {and} \bibinfo{person}{Dongmei Zhang}.} \bibinfo{year}{2019}\natexlab{a}.
\newblock \showarticletitle{An empirical investigation of incident triage for online service systems}. In \bibinfo{booktitle}{\emph{2019 IEEE/ACM 41st International Conference on Software Engineering: Software Engineering in Practice (ICSE-SEIP)}}. IEEE, \bibinfo{pages}{111--120}.
\newblock


\bibitem[Chen et~al\mbox{.}(2019b)]%
        {chen2019continuous}
\bibfield{author}{\bibinfo{person}{Junjie Chen}, \bibinfo{person}{Xiaoting He}, \bibinfo{person}{Qingwei Lin}, \bibinfo{person}{Hongyu Zhang}, \bibinfo{person}{Dan Hao}, \bibinfo{person}{Feng Gao}, \bibinfo{person}{Zhangwei Xu}, \bibinfo{person}{Yingnong Dang}, {and} \bibinfo{person}{Dongmei Zhang}.} \bibinfo{year}{2019}\natexlab{b}.
\newblock \showarticletitle{Continuous Incident Triage for Large-Scale Online Service Systems}. In \bibinfo{booktitle}{\emph{2019 34th IEEE/ACM International Conference on Automated Software Engineering (ASE)}}. \bibinfo{pages}{364--375}.
\newblock
\href{https://doi.org/10.1109/ASE.2019.00042}{doi:\nolinkurl{10.1109/ASE.2019.00042}}


\bibitem[Chen et~al\mbox{.}(2020b)]%
        {chen2020survey}
\bibfield{author}{\bibinfo{person}{Junjie Chen}, \bibinfo{person}{Jibesh Patra}, \bibinfo{person}{Michael Pradel}, \bibinfo{person}{Yingfei Xiong}, \bibinfo{person}{Hongyu Zhang}, \bibinfo{person}{Dan Hao}, {and} \bibinfo{person}{Lu Zhang}.} \bibinfo{year}{2020}\natexlab{b}.
\newblock \showarticletitle{A survey of compiler testing}.
\newblock \bibinfo{journal}{\emph{ACM Computing Surveys (CSUR)}} \bibinfo{volume}{53}, \bibinfo{number}{1} (\bibinfo{year}{2020}), \bibinfo{pages}{1--36}.
\newblock


\bibitem[Chen et~al\mbox{.}(2022)]%
        {chen2022oas}
\bibfield{author}{\bibinfo{person}{Jia Chen}, \bibinfo{person}{Peng Wang}, {and} \bibinfo{person}{Wei Wang}.} \bibinfo{year}{2022}\natexlab{}.
\newblock \showarticletitle{Online summarizing alerts through semantic and behavior information}. In \bibinfo{booktitle}{\emph{Proceedings of the 44th International Conference on Software Engineering}} (Pittsburgh, Pennsylvania) \emph{(\bibinfo{series}{ICSE '22})}. \bibinfo{publisher}{Association for Computing Machinery}, \bibinfo{address}{New York, NY, USA}, \bibinfo{pages}{1646–1657}.
\newblock
\showISBNx{9781450392211}
\href{https://doi.org/10.1145/3510003.3510055}{doi:\nolinkurl{10.1145/3510003.3510055}}


\bibitem[Chen et~al\mbox{.}(2021b)]%
        {chen2021deepip}
\bibfield{author}{\bibinfo{person}{Junjie Chen}, \bibinfo{person}{Shu Zhang}, \bibinfo{person}{Xiaoting He}, \bibinfo{person}{Qingwei Lin}, \bibinfo{person}{Hongyu Zhang}, \bibinfo{person}{Dan Hao}, \bibinfo{person}{Yu Kang}, \bibinfo{person}{Feng Gao}, \bibinfo{person}{Zhangwei Xu}, \bibinfo{person}{Yingnong Dang}, {and} \bibinfo{person}{Dongmei Zhang}.} \bibinfo{year}{2021}\natexlab{b}.
\newblock \showarticletitle{How incidental are the incidents? characterizing and prioritizing incidents for large-scale online service systems}. In \bibinfo{booktitle}{\emph{Proceedings of the 35th IEEE/ACM International Conference on Automated Software Engineering}} (Virtual Event, Australia) \emph{(\bibinfo{series}{ASE '20})}. \bibinfo{publisher}{Association for Computing Machinery}, \bibinfo{address}{New York, NY, USA}, \bibinfo{pages}{373–384}.
\newblock
\showISBNx{9781450367684}
\href{https://doi.org/10.1145/3324884.3416624}{doi:\nolinkurl{10.1145/3324884.3416624}}


\bibitem[Chen et~al\mbox{.}(2014)]%
        {chen2014ar}
\bibfield{author}{\bibinfo{person}{Ning Chen}, \bibinfo{person}{Jialiu Lin}, \bibinfo{person}{Steven~CH Hoi}, \bibinfo{person}{Xiaokui Xiao}, {and} \bibinfo{person}{Boshen Zhang}.} \bibinfo{year}{2014}\natexlab{}.
\newblock \showarticletitle{AR-miner: mining informative reviews for developers from mobile app marketplace}. In \bibinfo{booktitle}{\emph{Proceedings of the 36th international conference on software engineering}}. \bibinfo{pages}{767--778}.
\newblock


\bibitem[Chen et~al\mbox{.}(2020c)]%
        {chen2020lidar}
\bibfield{author}{\bibinfo{person}{Yujun Chen}, \bibinfo{person}{Xian Yang}, \bibinfo{person}{Hang Dong}, \bibinfo{person}{Xiaoting He}, \bibinfo{person}{Hongyu Zhang}, \bibinfo{person}{Qingwei Lin}, \bibinfo{person}{Junjie Chen}, \bibinfo{person}{Pu Zhao}, \bibinfo{person}{Yu Kang}, \bibinfo{person}{Feng Gao}, \bibinfo{person}{Zhangwei Xu}, {and} \bibinfo{person}{Dongmei Zhang}.} \bibinfo{year}{2020}\natexlab{c}.
\newblock \showarticletitle{Identifying linked incidents in large-scale online service systems}. In \bibinfo{booktitle}{\emph{Proceedings of the 28th ACM Joint Meeting on European Software Engineering Conference and Symposium on the Foundations of Software Engineering}} (Virtual Event, USA) \emph{(\bibinfo{series}{ESEC/FSE 2020})}. \bibinfo{publisher}{Association for Computing Machinery}, \bibinfo{address}{New York, NY, USA}, \bibinfo{pages}{304–314}.
\newblock
\showISBNx{9781450370431}
\href{https://doi.org/10.1145/3368089.3409768}{doi:\nolinkurl{10.1145/3368089.3409768}}


\bibitem[Chen et~al\mbox{.}(2020a)]%
        {chen2020icmbrain}
\bibfield{author}{\bibinfo{person}{Zhuangbin Chen}, \bibinfo{person}{Yu Kang}, \bibinfo{person}{Liqun Li}, \bibinfo{person}{Xu Zhang}, \bibinfo{person}{Hongyu Zhang}, \bibinfo{person}{Hui Xu}, \bibinfo{person}{Yangfan Zhou}, \bibinfo{person}{Li Yang}, \bibinfo{person}{Jeffrey Sun}, \bibinfo{person}{Zhangwei Xu}, \bibinfo{person}{Yingnong Dang}, \bibinfo{person}{Feng Gao}, \bibinfo{person}{Pu Zhao}, \bibinfo{person}{Bo Qiao}, \bibinfo{person}{Qingwei Lin}, \bibinfo{person}{Dongmei Zhang}, {and} \bibinfo{person}{Michael~R. Lyu}.} \bibinfo{year}{2020}\natexlab{a}.
\newblock \showarticletitle{Towards intelligent incident management: why we need it and how we make it}. In \bibinfo{booktitle}{\emph{Proceedings of the 28th ACM Joint Meeting on European Software Engineering Conference and Symposium on the Foundations of Software Engineering}} (Virtual Event, USA) \emph{(\bibinfo{series}{ESEC/FSE 2020})}. \bibinfo{publisher}{Association for Computing Machinery}, \bibinfo{address}{New York, NY, USA}, \bibinfo{pages}{1487–1497}.
\newblock
\showISBNx{9781450370431}
\href{https://doi.org/10.1145/3368089.3417055}{doi:\nolinkurl{10.1145/3368089.3417055}}


\bibitem[Chen et~al\mbox{.}(2021a)]%
        {chen2021grlia}
\bibfield{author}{\bibinfo{person}{Zhuangbin Chen}, \bibinfo{person}{Jinyang Liu}, \bibinfo{person}{Yuxin Su}, \bibinfo{person}{Hongyu Zhang}, \bibinfo{person}{Xuemin Wen}, \bibinfo{person}{Xiao Ling}, \bibinfo{person}{Yongqiang Yang}, {and} \bibinfo{person}{Michael~R. Lyu}.} \bibinfo{year}{2021}\natexlab{a}.
\newblock \showarticletitle{Graph-based Incident Aggregation for Large-Scale Online Service Systems}. In \bibinfo{booktitle}{\emph{2021 36th IEEE/ACM International Conference on Automated Software Engineering (ASE)}}. \bibinfo{pages}{430--442}.
\newblock
\href{https://doi.org/10.1109/ASE51524.2021.9678746}{doi:\nolinkurl{10.1109/ASE51524.2021.9678746}}


\bibitem[Christ et~al\mbox{.}(2010)]%
        {christ2010modern}
\bibfield{author}{\bibinfo{person}{Michael Christ}, \bibinfo{person}{Florian Grossmann}, \bibinfo{person}{Daniela Winter}, \bibinfo{person}{Roland Bingisser}, {and} \bibinfo{person}{Elke Platz}.} \bibinfo{year}{2010}\natexlab{}.
\newblock \showarticletitle{Modern triage in the emergency department}.
\newblock \bibinfo{journal}{\emph{Deutsches {\"A}rzteblatt International}} \bibinfo{volume}{107}, \bibinfo{number}{50} (\bibinfo{year}{2010}), \bibinfo{pages}{892}.
\newblock


\bibitem[Chrupa{\l}a(2012)]%
        {chrupala2012learning}
\bibfield{author}{\bibinfo{person}{Grzegorz Chrupa{\l}a}.} \bibinfo{year}{2012}\natexlab{}.
\newblock \showarticletitle{Learning from evolving data streams: online triage of bug reports}. In \bibinfo{booktitle}{\emph{Proceedings of the 13th Conference of the European Chapter of the Association for Computational Linguistics}}. \bibinfo{pages}{613--622}.
\newblock


\bibitem[Cook et~al\mbox{.}(2020)]%
        {cook2020using}
\bibfield{author}{\bibinfo{person}{Byron Cook}, \bibinfo{person}{Bj{\"o}rn D{\"o}bel}, \bibinfo{person}{Daniel Kroening}, \bibinfo{person}{Norbert Manthey}, \bibinfo{person}{Martin Pohlack}, \bibinfo{person}{Elizabeth Polgreen}, \bibinfo{person}{Michael Tautschnig}, {and} \bibinfo{person}{Pawel Wieczorkiewicz}.} \bibinfo{year}{2020}\natexlab{}.
\newblock \showarticletitle{Using model checking tools to triage the severity of security bugs in the Xen hypervisor}. In \bibinfo{booktitle}{\emph{\# PLACEHOLDER\_PARENT\_METADATA\_VALUE\#}}, Vol.~\bibinfo{volume}{1}. TU Wien Academic Press, \bibinfo{pages}{185--193}.
\newblock


\bibitem[Dai et~al\mbox{.}(2024)]%
        {dai2024pcg}
\bibfield{author}{\bibinfo{person}{Jie Dai}, \bibinfo{person}{Qingshan Li}, \bibinfo{person}{Shenglong Xie}, \bibinfo{person}{Daizhen Li}, {and} \bibinfo{person}{Hua Chu}.} \bibinfo{year}{2024}\natexlab{}.
\newblock \showarticletitle{PCG: A joint framework of graph collaborative filtering for bug triaging}.
\newblock \bibinfo{journal}{\emph{Journal of Software: Evolution and Process}} \bibinfo{volume}{36}, \bibinfo{number}{9} (\bibinfo{year}{2024}), \bibinfo{pages}{e2673}.
\newblock


\bibitem[Dai et~al\mbox{.}(2023)]%
        {dai2023graph}
\bibfield{author}{\bibinfo{person}{Jie Dai}, \bibinfo{person}{Qingshan Li}, \bibinfo{person}{Hui Xue}, \bibinfo{person}{Zhao Luo}, \bibinfo{person}{Yinglin Wang}, {and} \bibinfo{person}{Siyuan Zhan}.} \bibinfo{year}{2023}\natexlab{}.
\newblock \showarticletitle{Graph collaborative filtering-based bug triaging}.
\newblock \bibinfo{journal}{\emph{Journal of Systems and Software}}  \bibinfo{volume}{200} (\bibinfo{year}{2023}), \bibinfo{pages}{111667}.
\newblock


\bibitem[Dao and Yang(2023)]%
        {dao2023automated}
\bibfield{author}{\bibinfo{person}{Anh-Hien Dao} {and} \bibinfo{person}{Cheng-Zen Yang}.} \bibinfo{year}{2023}\natexlab{}.
\newblock \showarticletitle{Automated Priority Prediction for Bug Reports Using Comment Intensiveness Features and SMOTE Data Balancing}.
\newblock \bibinfo{journal}{\emph{International Journal of Software Engineering and Knowledge Engineering}} \bibinfo{volume}{33}, \bibinfo{number}{03} (\bibinfo{year}{2023}), \bibinfo{pages}{415--433}.
\newblock
\href{https://doi.org/10.1142/S021819402350002X}{doi:\nolinkurl{10.1142/S021819402350002X}}


\bibitem[Devaiya et~al\mbox{.}(2021a)]%
        {devaiya2021evaluating}
\bibfield{author}{\bibinfo{person}{Disha Devaiya}, \bibinfo{person}{John Anvik}, \bibinfo{person}{Meher Bheree}, {and} \bibinfo{person}{Farjana~Yeasmin Omee}.} \bibinfo{year}{2021}\natexlab{a}.
\newblock \showarticletitle{Evaluating a Tool for Creating Bug Report Assignment Recommenders (S).}. In \bibinfo{booktitle}{\emph{SEKE}}. \bibinfo{pages}{271--274}.
\newblock


\bibitem[Devaiya et~al\mbox{.}(2021b)]%
        {devaiya2021castr}
\bibfield{author}{\bibinfo{person}{Disha~Thakarshibhai Devaiya}, \bibinfo{person}{John Anvik}, \bibinfo{person}{Farjana~Yeasmin Omee}, {and} \bibinfo{person}{Meher Bheree}.} \bibinfo{year}{2021}\natexlab{b}.
\newblock \showarticletitle{CASTR: Assisting Bug Report Assignment Recommender Creation.}. In \bibinfo{booktitle}{\emph{SEKE}}. \bibinfo{pages}{639}.
\newblock


\bibitem[Di~Sorbo et~al\mbox{.}(2021)]%
        {di2021investigating}
\bibfield{author}{\bibinfo{person}{Andrea Di~Sorbo}, \bibinfo{person}{Giovanni Grano}, \bibinfo{person}{Corrado Aaron~Visaggio}, {and} \bibinfo{person}{Sebastiano Panichella}.} \bibinfo{year}{2021}\natexlab{}.
\newblock \showarticletitle{Investigating the criticality of user-reported issues through their relations with app rating}.
\newblock \bibinfo{journal}{\emph{Journal of Software: Evolution and Process}} \bibinfo{volume}{33}, \bibinfo{number}{3} (\bibinfo{year}{2021}), \bibinfo{pages}{e2316}.
\newblock


\bibitem[Ding et~al\mbox{.}(2014)]%
        {ding2014mining}
\bibfield{author}{\bibinfo{person}{Rui Ding}, \bibinfo{person}{Qiang Fu}, \bibinfo{person}{Jian~Guang Lou}, \bibinfo{person}{Qingwei Lin}, \bibinfo{person}{Dongmei Zhang}, {and} \bibinfo{person}{Tao Xie}.} \bibinfo{year}{2014}\natexlab{}.
\newblock \showarticletitle{Mining historical issue repositories to heal large-scale online service systems}. In \bibinfo{booktitle}{\emph{2014 44th Annual IEEE/IFIP International Conference on Dependable Systems and Networks}}. IEEE, \bibinfo{pages}{311--322}.
\newblock


\bibitem[Dipongkor et~al\mbox{.}(2023)]%
        {dipongkor2023fusing}
\bibfield{author}{\bibinfo{person}{Atish~Kumar Dipongkor}, \bibinfo{person}{Md~Saiful Islam}, \bibinfo{person}{Ishtiaque Hussain}, \bibinfo{person}{Sira Yongchareon}, {and} \bibinfo{person}{Sajib Mistry}.} \bibinfo{year}{2023}\natexlab{}.
\newblock \showarticletitle{On fusing artificial and convolutional neural network features for automatic bug assignments}.
\newblock \bibinfo{journal}{\emph{IEEE Access}}  \bibinfo{volume}{11} (\bibinfo{year}{2023}), \bibinfo{pages}{49493--49508}.
\newblock


\bibitem[Dipongkor and Moran(2023)]%
        {dipongkor2023comparative}
\bibfield{author}{\bibinfo{person}{Atish~Kumar Dipongkor} {and} \bibinfo{person}{Kevin Moran}.} \bibinfo{year}{2023}\natexlab{}.
\newblock \showarticletitle{A comparative study of transformer-based neural text representation techniques on bug triaging}. In \bibinfo{booktitle}{\emph{2023 38th IEEE/ACM International Conference on Automated Software Engineering (ASE)}}. IEEE, \bibinfo{pages}{1012--1023}.
\newblock


\bibitem[Dong et~al\mbox{.}(2024)]%
        {dong2024neighborhood}
\bibfield{author}{\bibinfo{person}{Haozhen Dong}, \bibinfo{person}{Hongmin Ren}, \bibinfo{person}{Jialiang Shi}, \bibinfo{person}{Yichen Xie}, {and} \bibinfo{person}{Xudong Hu}.} \bibinfo{year}{2024}\natexlab{}.
\newblock \showarticletitle{Neighborhood contrastive learning-based graph neural network for bug triaging}.
\newblock \bibinfo{journal}{\emph{Science of Computer Programming}}  \bibinfo{volume}{235} (\bibinfo{year}{2024}), \bibinfo{pages}{103093}.
\newblock


\bibitem[Etaiwi et~al\mbox{.}(2020)]%
        {etaiwi2020order}
\bibfield{author}{\bibinfo{person}{Layan Etaiwi}, \bibinfo{person}{Sylvie Hamel}, \bibinfo{person}{Yann-Ga{\"e}l Gu{\'e}h{\'e}neuc}, \bibinfo{person}{William Flageol}, {and} \bibinfo{person}{Rodrigo Morales}.} \bibinfo{year}{2020}\natexlab{}.
\newblock \showarticletitle{Order in chaos: prioritizing mobile app reviews using consensus algorithms}. In \bibinfo{booktitle}{\emph{2020 IEEE 44th Annual Computers, Software, and Applications Conference (COMPSAC)}}. IEEE, \bibinfo{pages}{912--920}.
\newblock


\bibitem[Etemadi et~al\mbox{.}(2021)]%
        {etemadi2021scheduling}
\bibfield{author}{\bibinfo{person}{Vahid Etemadi}, \bibinfo{person}{Omid Bushehrian}, \bibinfo{person}{Reza Akbari}, {and} \bibinfo{person}{Gregorio Robles}.} \bibinfo{year}{2021}\natexlab{}.
\newblock \showarticletitle{A scheduling-driven approach to efficiently assign bug fixing tasks to developers}.
\newblock \bibinfo{journal}{\emph{Journal of Systems and Software}}  \bibinfo{volume}{178} (\bibinfo{year}{2021}), \bibinfo{pages}{110967}.
\newblock


\bibitem[FitzGerald et~al\mbox{.}(2010)]%
        {fitzgerald2010emergency}
\bibfield{author}{\bibinfo{person}{Gerard FitzGerald}, \bibinfo{person}{George~A Jelinek}, \bibinfo{person}{Deborah Scott}, {and} \bibinfo{person}{Marie~Frances Gerdtz}.} \bibinfo{year}{2010}\natexlab{}.
\newblock \showarticletitle{Emergency department triage revisited}.
\newblock \bibinfo{journal}{\emph{Emergency Medicine Journal}} \bibinfo{volume}{27}, \bibinfo{number}{2} (\bibinfo{year}{2010}), \bibinfo{pages}{86--92}.
\newblock


\bibitem[Florea et~al\mbox{.}(2017)]%
        {florea2017parallel}
\bibfield{author}{\bibinfo{person}{Adrian-C{\u{a}}t{\u{a}}lin Florea}, \bibinfo{person}{John Anvik}, {and} \bibinfo{person}{R{\u{a}}zvan Andonie}.} \bibinfo{year}{2017}\natexlab{}.
\newblock \showarticletitle{Parallel implementation of a bug report assignment recommender using deep learning}. In \bibinfo{booktitle}{\emph{International Conference on artificial neural networks}}. Springer, \bibinfo{pages}{64--71}.
\newblock


\bibitem[Gao et~al\mbox{.}(2022)]%
        {gao2022listening}
\bibfield{author}{\bibinfo{person}{Cuiyun Gao}, \bibinfo{person}{Yaoxian Li}, \bibinfo{person}{Shuhan Qi}, \bibinfo{person}{Yang Liu}, \bibinfo{person}{Xuan Wang}, \bibinfo{person}{Zibin Zheng}, {and} \bibinfo{person}{Qing Liao}.} \bibinfo{year}{2022}\natexlab{}.
\newblock \showarticletitle{Listening to users' voice: Automatic summarization of helpful app reviews}.
\newblock \bibinfo{journal}{\emph{IEEE Transactions on Reliability}} \bibinfo{volume}{72}, \bibinfo{number}{4} (\bibinfo{year}{2022}), \bibinfo{pages}{1619--1631}.
\newblock


\bibitem[Gao et~al\mbox{.}(2015)]%
        {gao2015paid}
\bibfield{author}{\bibinfo{person}{Cuiyun Gao}, \bibinfo{person}{Baoxiang Wang}, \bibinfo{person}{Pinjia He}, \bibinfo{person}{Jieming Zhu}, \bibinfo{person}{Yangfan Zhou}, {and} \bibinfo{person}{Michael~R Lyu}.} \bibinfo{year}{2015}\natexlab{}.
\newblock \showarticletitle{PAID: Prioritizing app issues for developers by tracking user reviews over versions}. In \bibinfo{booktitle}{\emph{2015 IEEE 26th international symposium on software reliability engineering (ISSRE)}}. IEEE, \bibinfo{pages}{35--45}.
\newblock


\bibitem[Gao et~al\mbox{.}(2018)]%
        {gao2018online}
\bibfield{author}{\bibinfo{person}{Cuiyun Gao}, \bibinfo{person}{Jichuan Zeng}, \bibinfo{person}{Michael~R Lyu}, {and} \bibinfo{person}{Irwin King}.} \bibinfo{year}{2018}\natexlab{}.
\newblock \showarticletitle{Online app review analysis for identifying emerging issues}. In \bibinfo{booktitle}{\emph{Proceedings of the 40th international conference on software engineering}}. \bibinfo{pages}{48--58}.
\newblock


\bibitem[Gao et~al\mbox{.}(2020)]%
        {gao2020scouts}
\bibfield{author}{\bibinfo{person}{Jiaqi Gao}, \bibinfo{person}{Nofel Yaseen}, \bibinfo{person}{Robert MacDavid}, \bibinfo{person}{Felipe~Vieira Frujeri}, \bibinfo{person}{Vincent Liu}, \bibinfo{person}{Ricardo Bianchini}, \bibinfo{person}{Ramaswamy Aditya}, \bibinfo{person}{Xiaohang Wang}, \bibinfo{person}{Henry Lee}, \bibinfo{person}{David Maltz}, \bibinfo{person}{Yu Minlan}, {and} \bibinfo{person}{Arzani Behnaz}.} \bibinfo{year}{2020}\natexlab{}.
\newblock \showarticletitle{Scouts: Improving the diagnosis process through domain-customized incident routing}. In \bibinfo{booktitle}{\emph{Proceedings of the Annual conference of the ACM Special Interest Group on Data Communication on the applications, technologies, architectures, and protocols for computer communication}}. \bibinfo{pages}{253--269}.
\newblock


\bibitem[Ghosh et~al\mbox{.}(2024)]%
        {ghosh2024dilink}
\bibfield{author}{\bibinfo{person}{Supriyo Ghosh}, \bibinfo{person}{Karish Grover}, \bibinfo{person}{Jimmy Wong}, \bibinfo{person}{Chetan Bansal}, \bibinfo{person}{Rakesh Namineni}, \bibinfo{person}{Mohit Verma}, {and} \bibinfo{person}{Saravan Rajmohan}.} \bibinfo{year}{2024}\natexlab{}.
\newblock \showarticletitle{Dependency Aware Incident Linking in Large Cloud Systems}. In \bibinfo{booktitle}{\emph{Companion Proceedings of the ACM Web Conference 2024}} (Singapore, Singapore) \emph{(\bibinfo{series}{WWW '24})}. \bibinfo{publisher}{Association for Computing Machinery}, \bibinfo{address}{New York, NY, USA}, \bibinfo{pages}{141–150}.
\newblock
\showISBNx{9798400701726}
\href{https://doi.org/10.1145/3589335.3648311}{doi:\nolinkurl{10.1145/3589335.3648311}}


\bibitem[Goel et~al\mbox{.}(2024)]%
        {goel2024x}
\bibfield{author}{\bibinfo{person}{Drishti Goel}, \bibinfo{person}{Fiza Husain}, \bibinfo{person}{Aditya Singh}, \bibinfo{person}{Supriyo Ghosh}, \bibinfo{person}{Anjaly Parayil}, \bibinfo{person}{Chetan Bansal}, \bibinfo{person}{Xuchao Zhang}, {and} \bibinfo{person}{Saravan Rajmohan}.} \bibinfo{year}{2024}\natexlab{}.
\newblock \showarticletitle{X-lifecycle learning for cloud incident management using llms}. In \bibinfo{booktitle}{\emph{Companion Proceedings of the 32nd ACM International Conference on the Foundations of Software Engineering}}. \bibinfo{pages}{417--428}.
\newblock


\bibitem[Goyal(2017)]%
        {goyal2017effective}
\bibfield{author}{\bibinfo{person}{Anjali Goyal}.} \bibinfo{year}{2017}\natexlab{}.
\newblock \showarticletitle{Effective Bug Triage for Non-Reproducible Bugs}. In \bibinfo{booktitle}{\emph{2017 IEEE/ACM 39th International Conference on Software Engineering Companion (ICSE-C)}}. \bibinfo{pages}{487--488}.
\newblock
\href{https://doi.org/10.1109/ICSE-C.2017.41}{doi:\nolinkurl{10.1109/ICSE-C.2017.41}}


\bibitem[Goyal and Sardana(2024)]%
        {goyal2024integrated}
\bibfield{author}{\bibinfo{person}{Anjali Goyal} {and} \bibinfo{person}{Neetu Sardana}.} \bibinfo{year}{2024}\natexlab{}.
\newblock \showarticletitle{An Integrated Approach using Developer Profiles with Temporal Dynamics for Assignee Recommendation in Non-Reproducible Bugs}.
\newblock \bibinfo{journal}{\emph{Procedia Computer Science}}  \bibinfo{volume}{235} (\bibinfo{year}{2024}), \bibinfo{pages}{2833--2842}.
\newblock


\bibitem[Gu et~al\mbox{.}(2020)]%
        {gu2020linkcm}
\bibfield{author}{\bibinfo{person}{Jiazhen Gu}, \bibinfo{person}{Jiaqi Wen}, \bibinfo{person}{Zijian Wang}, \bibinfo{person}{Pu Zhao}, \bibinfo{person}{Chuan Luo}, \bibinfo{person}{Yu Kang}, \bibinfo{person}{Yangfan Zhou}, \bibinfo{person}{Li Yang}, \bibinfo{person}{Jeffrey Sun}, \bibinfo{person}{Zhangwei Xu}, \bibinfo{person}{Bo Qiao}, \bibinfo{person}{Liqun Li}, \bibinfo{person}{Qingwei Lin}, {and} \bibinfo{person}{Dongmei Zhang}.} \bibinfo{year}{2020}\natexlab{}.
\newblock \showarticletitle{Efficient customer incident triage via linking with system incidents}. In \bibinfo{booktitle}{\emph{Proceedings of the 28th ACM Joint Meeting on European Software Engineering Conference and Symposium on the Foundations of Software Engineering}} (Virtual Event, USA) \emph{(\bibinfo{series}{ESEC/FSE 2020})}. \bibinfo{publisher}{Association for Computing Machinery}, \bibinfo{address}{New York, NY, USA}, \bibinfo{pages}{1296–1307}.
\newblock
\showISBNx{9781450370431}
\href{https://doi.org/10.1145/3368089.3417061}{doi:\nolinkurl{10.1145/3368089.3417061}}


\bibitem[Guo et~al\mbox{.}(2011)]%
        {guo2011not}
\bibfield{author}{\bibinfo{person}{Philip~J. Guo}, \bibinfo{person}{Thomas Zimmermann}, \bibinfo{person}{Nachiappan Nagappan}, {and} \bibinfo{person}{Brendan Murphy}.} \bibinfo{year}{2011}\natexlab{}.
\newblock \showarticletitle{"Not my bug!" and other reasons for software bug report reassignments}. In \bibinfo{booktitle}{\emph{Proceedings of the ACM 2011 Conference on Computer Supported Cooperative Work}} (Hangzhou, China) \emph{(\bibinfo{series}{CSCW '11})}. \bibinfo{publisher}{Association for Computing Machinery}, \bibinfo{address}{New York, NY, USA}, \bibinfo{pages}{395–404}.
\newblock
\showISBNx{9781450305563}
\href{https://doi.org/10.1145/1958824.1958887}{doi:\nolinkurl{10.1145/1958824.1958887}}


\bibitem[Guo et~al\mbox{.}(2020)]%
        {guo2020developer}
\bibfield{author}{\bibinfo{person}{Shikai Guo}, \bibinfo{person}{Xinyi Zhang}, \bibinfo{person}{Xi Yang}, \bibinfo{person}{Rong Chen}, \bibinfo{person}{Chen Guo}, \bibinfo{person}{Hui Li}, {and} \bibinfo{person}{Tingting Li}.} \bibinfo{year}{2020}\natexlab{}.
\newblock \showarticletitle{Developer activity motivated bug triaging: via convolutional neural network}.
\newblock \bibinfo{journal}{\emph{Neural Processing Letters}} \bibinfo{volume}{51}, \bibinfo{number}{3} (\bibinfo{year}{2020}), \bibinfo{pages}{2589--2606}.
\newblock


\bibitem[Gupta and Freire(2021)]%
        {gupta2021decentralized}
\bibfield{author}{\bibinfo{person}{Chetna Gupta} {and} \bibinfo{person}{M{\'a}rio~M Freire}.} \bibinfo{year}{2021}\natexlab{}.
\newblock \showarticletitle{A decentralized blockchain oriented framework for automated bug assignment}.
\newblock \bibinfo{journal}{\emph{Information and Software Technology}}  \bibinfo{volume}{134} (\bibinfo{year}{2021}), \bibinfo{pages}{106540}.
\newblock


\bibitem[Gupta et~al\mbox{.}(2009)]%
        {gupta2009multi}
\bibfield{author}{\bibinfo{person}{Rajeev Gupta}, \bibinfo{person}{K~Hima Prasad}, \bibinfo{person}{Laura Luan}, \bibinfo{person}{Daniela Rosu}, {and} \bibinfo{person}{Chris Ward}.} \bibinfo{year}{2009}\natexlab{}.
\newblock \showarticletitle{Multi-dimensional knowledge integration for efficient incident management in a services cloud}. In \bibinfo{booktitle}{\emph{2009 IEEE International Conference on Services Computing}}. IEEE, \bibinfo{pages}{57--64}.
\newblock


\bibitem[Haering et~al\mbox{.}(2021)]%
        {haering2021automatically}
\bibfield{author}{\bibinfo{person}{Marlo Haering}, \bibinfo{person}{Christoph Stanik}, {and} \bibinfo{person}{Walid Maalej}.} \bibinfo{year}{2021}\natexlab{}.
\newblock \showarticletitle{Automatically matching bug reports with related app reviews}. In \bibinfo{booktitle}{\emph{2021 IEEE/ACM 43rd international conference on software engineering (ICSE)}}. IEEE, \bibinfo{pages}{970--981}.
\newblock


\bibitem[Hall et~al\mbox{.}(2011)]%
        {hall2011systematic}
\bibfield{author}{\bibinfo{person}{Tracy Hall}, \bibinfo{person}{Sarah Beecham}, \bibinfo{person}{David Bowes}, \bibinfo{person}{David Gray}, {and} \bibinfo{person}{Steve Counsell}.} \bibinfo{year}{2011}\natexlab{}.
\newblock \showarticletitle{A systematic literature review on fault prediction performance in software engineering}.
\newblock \bibinfo{journal}{\emph{IEEE Transactions on Software Engineering}} \bibinfo{volume}{38}, \bibinfo{number}{6} (\bibinfo{year}{2011}), \bibinfo{pages}{1276--1304}.
\newblock


\bibitem[Hassan et~al\mbox{.}(2019)]%
        {nodoze2019}
\bibfield{author}{\bibinfo{person}{Wajih~Ul Hassan}, \bibinfo{person}{Shengjian Guo}, \bibinfo{person}{Ding Li}, \bibinfo{person}{Zhengzhang Chen}, \bibinfo{person}{Kangkook Jee}, \bibinfo{person}{Zhichun Li}, {and} \bibinfo{person}{Adam Bates}.} \bibinfo{year}{2019}\natexlab{}.
\newblock \showarticletitle{NoDoze: Combatting Threat Alert Fatigue with Automated Provenance Triage}.
\newblock \bibinfo{journal}{\emph{Network and Distributed Systems Security Symposium}} (\bibinfo{year}{2019}).
\newblock
\urldef\tempurl%
\url{https://par.nsf.gov/biblio/10085663}
\showURL{%
\tempurl}


\bibitem[He and Yang(2021)]%
        {he2021automatic}
\bibfield{author}{\bibinfo{person}{Huoliang He} {and} \bibinfo{person}{ShunKun Yang}.} \bibinfo{year}{2021}\natexlab{}.
\newblock \showarticletitle{Automatic Bug Triage Using Hierarchical Attention Networks}. In \bibinfo{booktitle}{\emph{2021 IEEE 21st International Conference on Software Quality, Reliability and Security Companion (QRS-C)}}. \bibinfo{pages}{1043--1049}.
\newblock
\href{https://doi.org/10.1109/QRS-C55045.2021.00158}{doi:\nolinkurl{10.1109/QRS-C55045.2021.00158}}


\bibitem[Hosseini et~al\mbox{.}(2017)]%
        {hosseini2017systematic}
\bibfield{author}{\bibinfo{person}{Seyedrebvar Hosseini}, \bibinfo{person}{Burak Turhan}, {and} \bibinfo{person}{Dimuthu Gunarathna}.} \bibinfo{year}{2017}\natexlab{}.
\newblock \showarticletitle{A systematic literature review and meta-analysis on cross project defect prediction}.
\newblock \bibinfo{journal}{\emph{IEEE Transactions on Software Engineering}} \bibinfo{volume}{45}, \bibinfo{number}{2} (\bibinfo{year}{2017}), \bibinfo{pages}{111--147}.
\newblock


\bibitem[Hu et~al\mbox{.}(2014)]%
        {hu2014effective}
\bibfield{author}{\bibinfo{person}{Hao Hu}, \bibinfo{person}{Hongyu Zhang}, \bibinfo{person}{Jifeng Xuan}, {and} \bibinfo{person}{Weigang Sun}.} \bibinfo{year}{2014}\natexlab{}.
\newblock \showarticletitle{Effective bug triage based on historical bug-fix information}. In \bibinfo{booktitle}{\emph{2014 IEEE 25th international symposium on software reliability engineering}}. IEEE, \bibinfo{pages}{122--132}.
\newblock


\bibitem[Huang et~al\mbox{.}(2024)]%
        {huang2024faultprofit}
\bibfield{author}{\bibinfo{person}{Junjie Huang}, \bibinfo{person}{Jinyang Liu}, \bibinfo{person}{Zhuangbin Chen}, \bibinfo{person}{Zhihan Jiang}, \bibinfo{person}{Yichen Li}, \bibinfo{person}{Jiazhen Gu}, \bibinfo{person}{Cong Feng}, \bibinfo{person}{Zengyin Yang}, \bibinfo{person}{Yongqiang Yang}, {and} \bibinfo{person}{Michael~R Lyu}.} \bibinfo{year}{2024}\natexlab{}.
\newblock \showarticletitle{Faultprofit: Hierarchical fault profiling of incident tickets in large-scale cloud systems}. In \bibinfo{booktitle}{\emph{Proceedings of the 46th International Conference on Software Engineering: Software Engineering in Practice}}. \bibinfo{pages}{392--404}.
\newblock


\bibitem[Huang and Ma(2019)]%
        {huang2019predicting}
\bibfield{author}{\bibinfo{person}{Jinxiao Huang} {and} \bibinfo{person}{Yutao Ma}.} \bibinfo{year}{2019}\natexlab{}.
\newblock \showarticletitle{Predicting the fixer of software bugs via a collaborative multiplex network: Two case studies}. In \bibinfo{booktitle}{\emph{International Conference on Collaborative Computing: Networking, Applications and Worksharing}}. Springer, \bibinfo{pages}{469--488}.
\newblock


\bibitem[Jahanshahi and Cevik(2022)]%
        {jahanshahi2022s}
\bibfield{author}{\bibinfo{person}{Hadi Jahanshahi} {and} \bibinfo{person}{Mucahit Cevik}.} \bibinfo{year}{2022}\natexlab{}.
\newblock \showarticletitle{S-DABT: Schedule and dependency-aware bug triage in open-source bug tracking systems}.
\newblock \bibinfo{journal}{\emph{Information and Software Technology}}  \bibinfo{volume}{151} (\bibinfo{year}{2022}), \bibinfo{pages}{107025}.
\newblock


\bibitem[Jahanshahi et~al\mbox{.}(2023)]%
        {jahanshahi2023adptriage}
\bibfield{author}{\bibinfo{person}{Hadi Jahanshahi}, \bibinfo{person}{Mucahit Cevik}, \bibinfo{person}{Kianoush Mousavi}, {and} \bibinfo{person}{Ay{\c{s}}e Ba{\c{s}}ar}.} \bibinfo{year}{2023}\natexlab{}.
\newblock \showarticletitle{ADPTriage: Approximate dynamic programming for bug triage}.
\newblock \bibinfo{journal}{\emph{IEEE Transactions on Software Engineering}} \bibinfo{volume}{49}, \bibinfo{number}{10} (\bibinfo{year}{2023}), \bibinfo{pages}{4594--4609}.
\newblock


\bibitem[Jahanshahi et~al\mbox{.}(2022)]%
        {jahanshahi2022wayback}
\bibfield{author}{\bibinfo{person}{Hadi Jahanshahi}, \bibinfo{person}{Mucahit Cevik}, \bibinfo{person}{Jos{\'e} Navas-S{\'u}}, \bibinfo{person}{Ay{\c{s}}e Ba{\c{s}}ar}, {and} \bibinfo{person}{Antonio Gonz{\'a}lez-Torres}.} \bibinfo{year}{2022}\natexlab{}.
\newblock \showarticletitle{Wayback Machine: A tool to capture the evolutionary behavior of the bug reports and their triage process in open-source software systems}.
\newblock \bibinfo{journal}{\emph{Journal of Systems and Software}}  \bibinfo{volume}{189} (\bibinfo{year}{2022}), \bibinfo{pages}{111308}.
\newblock


\bibitem[Jahanshahi et~al\mbox{.}(2021)]%
        {jahanshahi2021dabt}
\bibfield{author}{\bibinfo{person}{Hadi Jahanshahi}, \bibinfo{person}{Kritika Chhabra}, \bibinfo{person}{Mucahit Cevik}, {and} \bibinfo{person}{Ay{\th}e Ba{\th}ar}.} \bibinfo{year}{2021}\natexlab{}.
\newblock \showarticletitle{DABT: A dependency-aware bug triaging method}. In \bibinfo{booktitle}{\emph{Proceedings of the 25th International Conference on Evaluation and Assessment in Software Engineering}}. \bibinfo{pages}{221--230}.
\newblock


\bibitem[James et~al\mbox{.}(2022)]%
        {james2022separating}
\bibfield{author}{\bibinfo{person}{Kedrian James}, \bibinfo{person}{Yufei Du}, \bibinfo{person}{Sanjeev Das}, {and} \bibinfo{person}{Fabian Monrose}.} \bibinfo{year}{2022}\natexlab{}.
\newblock \showarticletitle{Separating the Wheat from the Chaff: Using Indexing and Sub-Sequence Mining Techniques to Identify Related Crashes During Bug Triage}. In \bibinfo{booktitle}{\emph{2022 IEEE 22nd International Conference on Software Quality, Reliability and Security (QRS)}}. \bibinfo{pages}{31--42}.
\newblock
\href{https://doi.org/10.1109/QRS57517.2022.00014}{doi:\nolinkurl{10.1109/QRS57517.2022.00014}}


\bibitem[Jeong et~al\mbox{.}(2009)]%
        {jeong2009improving}
\bibfield{author}{\bibinfo{person}{Gaeul Jeong}, \bibinfo{person}{Sunghun Kim}, {and} \bibinfo{person}{Thomas Zimmermann}.} \bibinfo{year}{2009}\natexlab{}.
\newblock \showarticletitle{Improving bug triage with bug tossing graphs}. In \bibinfo{booktitle}{\emph{Proceedings of the 7th joint meeting of the European software engineering conference and the ACM SIGSOFT symposium on The foundations of software engineering}}. \bibinfo{pages}{111--120}.
\newblock


\bibitem[Jonsson et~al\mbox{.}(2016)]%
        {jonsson2016automated}
\bibfield{author}{\bibinfo{person}{Leif Jonsson}, \bibinfo{person}{Markus Borg}, \bibinfo{person}{David Broman}, \bibinfo{person}{Kristian Sandahl}, \bibinfo{person}{Sigrid Eldh}, {and} \bibinfo{person}{Per Runeson}.} \bibinfo{year}{2016}\natexlab{}.
\newblock \showarticletitle{Automated bug assignment: Ensemble-based machine learning in large scale industrial contexts}.
\newblock \bibinfo{journal}{\emph{Empirical Software Engineering}} \bibinfo{volume}{21}, \bibinfo{number}{4} (\bibinfo{year}{2016}), \bibinfo{pages}{1533--1578}.
\newblock


\bibitem[Kanwal and Maqbool(2012)]%
        {kanwal2012bug}
\bibfield{author}{\bibinfo{person}{Jaweria Kanwal} {and} \bibinfo{person}{Onaiza Maqbool}.} \bibinfo{year}{2012}\natexlab{}.
\newblock \showarticletitle{Bug prioritization to facilitate bug report triage}.
\newblock \bibinfo{journal}{\emph{Journal of Computer Science and Technology}} \bibinfo{volume}{27}, \bibinfo{number}{2} (\bibinfo{year}{2012}), \bibinfo{pages}{397--412}.
\newblock


\bibitem[Kashiwa(2019)]%
        {kashiwa2019raptor}
\bibfield{author}{\bibinfo{person}{Yutaro Kashiwa}.} \bibinfo{year}{2019}\natexlab{}.
\newblock \showarticletitle{RAPTOR: Release-aware and prioritized bug-fixing task assignment optimization}. In \bibinfo{booktitle}{\emph{2019 IEEE International Conference on Software Maintenance and Evolution (ICSME)}}. IEEE, \bibinfo{pages}{629--633}.
\newblock


\bibitem[Kitchenham(2004)]%
        {kitchenham2004procedures}
\bibfield{author}{\bibinfo{person}{Barbara Kitchenham}.} \bibinfo{year}{2004}\natexlab{}.
\newblock \showarticletitle{Procedures for performing systematic reviews}.
\newblock \bibinfo{journal}{\emph{Keele, UK, Keele University}} \bibinfo{volume}{33}, \bibinfo{number}{2004} (\bibinfo{year}{2004}), \bibinfo{pages}{1--26}.
\newblock


\bibitem[Krasniqi and Do(2022)]%
        {krasniqi2022automatically}
\bibfield{author}{\bibinfo{person}{Rrezarta Krasniqi} {and} \bibinfo{person}{Hyunsook Do}.} \bibinfo{year}{2022}\natexlab{}.
\newblock \showarticletitle{Automatically Capturing Quality-Related Concerns in Bug Report Descriptions for Efficient Bug Triaging}. In \bibinfo{booktitle}{\emph{Proceedings of the 26th International Conference on Evaluation and Assessment in Software Engineering}} (Gothenburg, Sweden) \emph{(\bibinfo{series}{EASE '22})}. \bibinfo{publisher}{Association for Computing Machinery}, \bibinfo{address}{New York, NY, USA}, \bibinfo{pages}{10–19}.
\newblock
\showISBNx{9781450396134}
\href{https://doi.org/10.1145/3530019.3530021}{doi:\nolinkurl{10.1145/3530019.3530021}}


\bibitem[Kuang et~al\mbox{.}(2024)]%
        {kuang2024cola}
\bibfield{author}{\bibinfo{person}{Jinxi Kuang}, \bibinfo{person}{Jinyang Liu}, \bibinfo{person}{Junjie Huang}, \bibinfo{person}{Renyi Zhong}, \bibinfo{person}{Jiazhen Gu}, \bibinfo{person}{Lan Yu}, \bibinfo{person}{Rui Tan}, \bibinfo{person}{Zengyin Yang}, {and} \bibinfo{person}{Michael~R. Lyu}.} \bibinfo{year}{2024}\natexlab{}.
\newblock \showarticletitle{Knowledge-aware Alert Aggregation in Large-scale Cloud Systems: a Hybrid Approach}. In \bibinfo{booktitle}{\emph{Proceedings of the 46th International Conference on Software Engineering: Software Engineering in Practice}} (Lisbon, Portugal) \emph{(\bibinfo{series}{ICSE-SEIP '24})}. \bibinfo{publisher}{Association for Computing Machinery}, \bibinfo{address}{New York, NY, USA}, \bibinfo{pages}{369–380}.
\newblock
\showISBNx{9798400705014}
\href{https://doi.org/10.1145/3639477.3639745}{doi:\nolinkurl{10.1145/3639477.3639745}}


\bibitem[Kumar~Dipongkor(2024)]%
        {dipongkor2024ensemble}
\bibfield{author}{\bibinfo{person}{Atish Kumar~Dipongkor}.} \bibinfo{year}{2024}\natexlab{}.
\newblock \showarticletitle{An Ensemble Method for Bug Triaging using Large Language Models}. In \bibinfo{booktitle}{\emph{Proceedings of the 2024 IEEE/ACM 46th International Conference on Software Engineering: Companion Proceedings}} (Lisbon, Portugal) \emph{(\bibinfo{series}{ICSE-Companion '24})}. \bibinfo{publisher}{Association for Computing Machinery}, \bibinfo{address}{New York, NY, USA}, \bibinfo{pages}{438–440}.
\newblock
\showISBNx{9798400705021}
\href{https://doi.org/10.1145/3639478.3641228}{doi:\nolinkurl{10.1145/3639478.3641228}}


\bibitem[Lamkanfi et~al\mbox{.}(2013)]%
        {LamkanfiMSR13}
\bibfield{author}{\bibinfo{person}{Ahmed Lamkanfi}, \bibinfo{person}{Javier Perez}, {and} \bibinfo{person}{Serge Demeyer}.} \bibinfo{year}{2013}\natexlab{}.
\newblock \showarticletitle{The Eclipse and Mozilla Defect Tracking Dataset: a Genuine Dataset for Mining Bug Information}. In \bibinfo{booktitle}{\emph{MSR '13: Proceedings of the 10th Working Conference on Mining Software Repositories, May 18-–19, 2013. San Francisco, California, USA}}.
\newblock


\bibitem[Lee et~al\mbox{.}(2022)]%
        {lee2022light}
\bibfield{author}{\bibinfo{person}{Jaehyung Lee}, \bibinfo{person}{Kisun Han}, {and} \bibinfo{person}{Hwanjo Yu}.} \bibinfo{year}{2022}\natexlab{}.
\newblock \showarticletitle{A light bug triage framework for applying large pre-trained language model}. In \bibinfo{booktitle}{\emph{Proceedings of the 37th IEEE/ACM international conference on automated software engineering}}. \bibinfo{pages}{1--11}.
\newblock


\bibitem[Lee et~al\mbox{.}(2017)]%
        {lee2017applying}
\bibfield{author}{\bibinfo{person}{Sun-Ro Lee}, \bibinfo{person}{Min-Jae Heo}, \bibinfo{person}{Chan-Gun Lee}, \bibinfo{person}{Milhan Kim}, {and} \bibinfo{person}{Gaeul Jeong}.} \bibinfo{year}{2017}\natexlab{}.
\newblock \showarticletitle{Applying deep learning based automatic bug triager to industrial projects}. In \bibinfo{booktitle}{\emph{Proceedings of the 2017 11th Joint Meeting on foundations of software engineering}}. \bibinfo{pages}{926--931}.
\newblock


\bibitem[Li et~al\mbox{.}(2021)]%
        {li2021warden}
\bibfield{author}{\bibinfo{person}{Liqun Li}, \bibinfo{person}{Xu Zhang}, \bibinfo{person}{Xin Zhao}, \bibinfo{person}{Hongyu Zhang}, \bibinfo{person}{Yu Kang}, \bibinfo{person}{Pu Zhao}, \bibinfo{person}{Bo Qiao}, \bibinfo{person}{Shilin He}, \bibinfo{person}{Pochian Lee}, \bibinfo{person}{Jeffrey Sun}, \bibinfo{person}{Feng Gao}, \bibinfo{person}{Li Yang}, \bibinfo{person}{Qingwei Lin}, \bibinfo{person}{Saravanakumar Rajmohan}, \bibinfo{person}{Zhangwei Xu}, {and} \bibinfo{person}{Dongmei Zhang}.} \bibinfo{year}{2021}\natexlab{}.
\newblock \showarticletitle{Fighting the Fog of War: Automated Incident Detection for Cloud Systems}. In \bibinfo{booktitle}{\emph{2021 USENIX Annual Technical Conference (USENIX ATC 21)}}. \bibinfo{publisher}{USENIX Association}, \bibinfo{pages}{131--146}.
\newblock
\showISBNx{978-1-939133-23-6}
\urldef\tempurl%
\url{https://www.usenix.org/conference/atc21/presentation/li-liqun}
\showURL{%
\tempurl}


\bibitem[Li and Huang(2024)]%
        {li2024automatic}
\bibfield{author}{\bibinfo{person}{Zexuan Li} {and} \bibinfo{person}{Kaixin Huang}.} \bibinfo{year}{2024}\natexlab{}.
\newblock \showarticletitle{Automatic bug assignments without texts: a study}.
\newblock \bibinfo{journal}{\emph{Frontiers of Computer Science}} \bibinfo{volume}{18}, \bibinfo{number}{4} (\bibinfo{year}{2024}), \bibinfo{pages}{184210}.
\newblock


\bibitem[Li and Zhong(2021)]%
        {li2021revisiting}
\bibfield{author}{\bibinfo{person}{Zexuan Li} {and} \bibinfo{person}{Hao Zhong}.} \bibinfo{year}{2021}\natexlab{}.
\newblock \showarticletitle{Revisiting textual feature of bug-triage approach}. In \bibinfo{booktitle}{\emph{2021 36th IEEE/ACM International Conference on Automated Software Engineering (ASE)}}. IEEE, \bibinfo{pages}{1183--1185}.
\newblock


\bibitem[Lim et~al\mbox{.}(2014)]%
        {lim2014identifying}
\bibfield{author}{\bibinfo{person}{Meng-Hui Lim}, \bibinfo{person}{Jian-Guang Lou}, \bibinfo{person}{Hongyu Zhang}, \bibinfo{person}{Qiang Fu}, \bibinfo{person}{Andrew Beng~Jin Teoh}, \bibinfo{person}{Qingwei Lin}, \bibinfo{person}{Rui Ding}, {and} \bibinfo{person}{Dongmei Zhang}.} \bibinfo{year}{2014}\natexlab{}.
\newblock \showarticletitle{Identifying Recurrent and Unknown Performance Issues}. In \bibinfo{booktitle}{\emph{2014 IEEE International Conference on Data Mining}}. \bibinfo{pages}{320--329}.
\newblock
\showISSN{2374-8486}
\href{https://doi.org/10.1109/ICDM.2014.96}{doi:\nolinkurl{10.1109/ICDM.2014.96}}


\bibitem[Lin et~al\mbox{.}(2014)]%
        {lin2014unveiling}
\bibfield{author}{\bibinfo{person}{Derek Lin}, \bibinfo{person}{Rashmi Raghu}, \bibinfo{person}{Vivek Ramamurthy}, \bibinfo{person}{Jin Yu}, \bibinfo{person}{Regunathan Radhakrishnan}, {and} \bibinfo{person}{Joseph Fernandez}.} \bibinfo{year}{2014}\natexlab{}.
\newblock \showarticletitle{Unveiling clusters of events for alert and incident management in large-scale enterprise it}. In \bibinfo{booktitle}{\emph{Proceedings of the 20th ACM SIGKDD International Conference on Knowledge Discovery and Data Mining}} (New York, New York, USA) \emph{(\bibinfo{series}{KDD '14})}. \bibinfo{publisher}{Association for Computing Machinery}, \bibinfo{address}{New York, NY, USA}, \bibinfo{pages}{1630–1639}.
\newblock
\showISBNx{9781450329569}
\href{https://doi.org/10.1145/2623330.2623360}{doi:\nolinkurl{10.1145/2623330.2623360}}


\bibitem[Lin(2018)]%
        {lin2018data}
\bibfield{author}{\bibinfo{person}{Tao Lin}.} \bibinfo{year}{2018}\natexlab{}.
\newblock \showarticletitle{A data triage retrieval system for cyber security operations center}.
\newblock  (\bibinfo{year}{2018}).
\newblock


\bibitem[Lin et~al\mbox{.}(2018)]%
        {lin2018car}
\bibfield{author}{\bibinfo{person}{Ying Lin}, \bibinfo{person}{Zhengzhang Chen}, \bibinfo{person}{Cheng Cao}, \bibinfo{person}{Lu-An Tang}, \bibinfo{person}{Kai Zhang}, \bibinfo{person}{Wei Cheng}, {and} \bibinfo{person}{Zhichun Li}.} \bibinfo{year}{2018}\natexlab{}.
\newblock \showarticletitle{Collaborative Alert Ranking for Anomaly Detection}. In \bibinfo{booktitle}{\emph{Proceedings of the 27th ACM International Conference on Information and Knowledge Management}} (Torino, Italy) \emph{(\bibinfo{series}{CIKM '18})}. \bibinfo{publisher}{Association for Computing Machinery}, \bibinfo{address}{New York, NY, USA}, \bibinfo{pages}{1987–1995}.
\newblock
\showISBNx{9781450360142}
\href{https://doi.org/10.1145/3269206.3272013}{doi:\nolinkurl{10.1145/3269206.3272013}}


\bibitem[Lin et~al\mbox{.}(2009)]%
        {lin2009empirical}
\bibfield{author}{\bibinfo{person}{Zhongpeng Lin}, \bibinfo{person}{Fengdi Shu}, \bibinfo{person}{Ye Yang}, \bibinfo{person}{Chenyong Hu}, {and} \bibinfo{person}{Qing Wang}.} \bibinfo{year}{2009}\natexlab{}.
\newblock \showarticletitle{An empirical study on bug assignment automation using chinese bug data}. In \bibinfo{booktitle}{\emph{2009 3rd International Symposium on Empirical Software Engineering and Measurement}}. IEEE, \bibinfo{pages}{451--455}.
\newblock


\bibitem[Linares-Vásquez et~al\mbox{.}(2012)]%
        {linares2012triaging}
\bibfield{author}{\bibinfo{person}{Mario Linares-Vásquez}, \bibinfo{person}{Kamal Hossen}, \bibinfo{person}{Hoang Dang}, \bibinfo{person}{Huzefa Kagdi}, \bibinfo{person}{Malcom Gethers}, {and} \bibinfo{person}{Denys Poshyvanyk}.} \bibinfo{year}{2012}\natexlab{}.
\newblock \showarticletitle{Triaging incoming change requests: Bug or commit history, or code authorship?}. In \bibinfo{booktitle}{\emph{2012 28th IEEE International Conference on Software Maintenance (ICSM)}}. \bibinfo{pages}{451--460}.
\newblock
\href{https://doi.org/10.1109/ICSM.2012.6405306}{doi:\nolinkurl{10.1109/ICSM.2012.6405306}}


\bibitem[Liu et~al\mbox{.}(2023b)]%
        {liu2023ipack}
\bibfield{author}{\bibinfo{person}{Jinyang Liu}, \bibinfo{person}{Shilin He}, \bibinfo{person}{Zhuangbin Chen}, \bibinfo{person}{Liqun Li}, \bibinfo{person}{Yu Kang}, \bibinfo{person}{Xu Zhang}, \bibinfo{person}{Pinjia He}, \bibinfo{person}{Hongyu Zhang}, \bibinfo{person}{Qingwei Lin}, \bibinfo{person}{Zhangwei Xu}, \bibinfo{person}{Saravan Rajmohan}, \bibinfo{person}{Dongmei Zhang}, {and} \bibinfo{person}{Michael~R. Lyu}.} \bibinfo{year}{2023}\natexlab{b}.
\newblock \showarticletitle{Incident-aware Duplicate Ticket Aggregation for Cloud Systems}. In \bibinfo{booktitle}{\emph{2023 IEEE/ACM 45th International Conference on Software Engineering (ICSE)}}. \bibinfo{pages}{2299--2311}.
\newblock
\href{https://doi.org/10.1109/ICSE48619.2023.00193}{doi:\nolinkurl{10.1109/ICSE48619.2023.00193}}


\bibitem[Liu et~al\mbox{.}(2016)]%
        {liu2016multi}
\bibfield{author}{\bibinfo{person}{Jin Liu}, \bibinfo{person}{Yiqiuzi Tian}, \bibinfo{person}{Xiao Yu}, \bibinfo{person}{Zhijiang Yang}, \bibinfo{person}{Xiangyang Jia}, \bibinfo{person}{Chuanxiang Ma}, {and} \bibinfo{person}{Zheng Xu}.} \bibinfo{year}{2016}\natexlab{}.
\newblock \showarticletitle{A multi-source approach for bug triage}.
\newblock \bibinfo{journal}{\emph{International Journal of Software Engineering and Knowledge Engineering}} \bibinfo{volume}{26}, \bibinfo{number}{09n10} (\bibinfo{year}{2016}), \bibinfo{pages}{1593--1604}.
\newblock


\bibitem[Liu et~al\mbox{.}(2023a)]%
        {liu2023ticket}
\bibfield{author}{\bibinfo{person}{Zhexiong Liu}, \bibinfo{person}{Cris Benge}, {and} \bibinfo{person}{Siduo Jiang}.} \bibinfo{year}{2023}\natexlab{a}.
\newblock \showarticletitle{Ticket-bert: Labeling incident management tickets with language models}.
\newblock \bibinfo{journal}{\emph{arXiv preprint arXiv:2307.00108}} (\bibinfo{year}{2023}).
\newblock


\bibitem[Lou et~al\mbox{.}(2013)]%
        {lou2013sas}
\bibfield{author}{\bibinfo{person}{Jian-Guang Lou}, \bibinfo{person}{Qingwei Lin}, \bibinfo{person}{Rui Ding}, \bibinfo{person}{Qiang Fu}, \bibinfo{person}{Dongmei Zhang}, {and} \bibinfo{person}{Tao Xie}.} \bibinfo{year}{2013}\natexlab{}.
\newblock \showarticletitle{Software analytics for incident management of online services: An experience report}. In \bibinfo{booktitle}{\emph{2013 28th IEEE/ACM International Conference on Automated Software Engineering (ASE)}}. \bibinfo{pages}{475--485}.
\newblock
\href{https://doi.org/10.1109/ASE.2013.6693105}{doi:\nolinkurl{10.1109/ASE.2013.6693105}}


\bibitem[Lou et~al\mbox{.}(2017)]%
        {lou2017experience}
\bibfield{author}{\bibinfo{person}{Jian-Guang Lou}, \bibinfo{person}{Qingwei Lin}, \bibinfo{person}{Rui Ding}, \bibinfo{person}{Qiang Fu}, \bibinfo{person}{Dongmei Zhang}, {and} \bibinfo{person}{Tao Xie}.} \bibinfo{year}{2017}\natexlab{}.
\newblock \showarticletitle{Experience report on applying software analytics in incident management of online service}.
\newblock \bibinfo{journal}{\emph{automated software engineering}} \bibinfo{volume}{24}, \bibinfo{number}{4} (\bibinfo{year}{2017}), \bibinfo{pages}{905--941}.
\newblock


\bibitem[Lyakajigari et~al\mbox{.}(2024)]%
        {lyakajigari2024review}
\bibfield{author}{\bibinfo{person}{Shiva~Theja Lyakajigari}, \bibinfo{person}{Sindhu Lakkavtri}, \bibinfo{person}{Sai~Mehar Ankathi}, \bibinfo{person}{Ramu Kuchipudi}, {and} \bibinfo{person}{Ramakrishna Kolikipogu}.} \bibinfo{year}{2024}\natexlab{}.
\newblock \showarticletitle{review on Automated software Bug Triage system}. In \bibinfo{booktitle}{\emph{2024 International Conference on Healthcare Innovations, Software and Engineering Technologies (HISET)}}. IEEE, \bibinfo{pages}{346--349}.
\newblock


\bibitem[Malgaonkar et~al\mbox{.}(2022)]%
        {malgaonkar2022prioritizing}
\bibfield{author}{\bibinfo{person}{Saurabh Malgaonkar}, \bibinfo{person}{Sherlock~A Licorish}, {and} \bibinfo{person}{Bastin Tony~Roy Savarimuthu}.} \bibinfo{year}{2022}\natexlab{}.
\newblock \showarticletitle{Prioritizing user concerns in app reviews--A study of requests for new features, enhancements and bug fixes}.
\newblock \bibinfo{journal}{\emph{Information and Software Technology}}  \bibinfo{volume}{144} (\bibinfo{year}{2022}), \bibinfo{pages}{106798}.
\newblock


\bibitem[Mandal et~al\mbox{.}(2019)]%
        {mandal2019improving}
\bibfield{author}{\bibinfo{person}{Atri Mandal}, \bibinfo{person}{Shivali Agarwal}, \bibinfo{person}{Nikhil Malhotra}, \bibinfo{person}{Giriprasad Sridhara}, \bibinfo{person}{Anupama Ray}, {and} \bibinfo{person}{Daivik Swarup}.} \bibinfo{year}{2019}\natexlab{}.
\newblock \showarticletitle{Improving it support by enhancing incident management process with multi-modal analysis}. In \bibinfo{booktitle}{\emph{International Conference on Service-Oriented Computing}}. Springer, \bibinfo{pages}{431--446}.
\newblock


\bibitem[Mani et~al\mbox{.}(2019)]%
        {mani2019deeptriage}
\bibfield{author}{\bibinfo{person}{Senthil Mani}, \bibinfo{person}{Anush Sankaran}, {and} \bibinfo{person}{Rahul Aralikatte}.} \bibinfo{year}{2019}\natexlab{}.
\newblock \showarticletitle{Deeptriage: Exploring the effectiveness of deep learning for bug triaging}. In \bibinfo{booktitle}{\emph{Proceedings of the ACM India joint international conference on data science and management of data}}. \bibinfo{pages}{171--179}.
\newblock


\bibitem[Matter et~al\mbox{.}(2009)]%
        {matter2009assigning}
\bibfield{author}{\bibinfo{person}{Dominique Matter}, \bibinfo{person}{Adrian Kuhn}, {and} \bibinfo{person}{Oscar Nierstrasz}.} \bibinfo{year}{2009}\natexlab{}.
\newblock \showarticletitle{Assigning bug reports using a vocabulary-based expertise model of developers}. In \bibinfo{booktitle}{\emph{2009 6th IEEE International Working Conference on Mining Software Repositories}}. \bibinfo{pages}{131--140}.
\newblock
\href{https://doi.org/10.1109/MSR.2009.5069491}{doi:\nolinkurl{10.1109/MSR.2009.5069491}}


\bibitem[Meher et~al\mbox{.}(2024)]%
        {meher2024deep}
\bibfield{author}{\bibinfo{person}{Jyoti~Prakash Meher}, \bibinfo{person}{Sourav Biswas}, {and} \bibinfo{person}{Rajib Mall}.} \bibinfo{year}{2024}\natexlab{}.
\newblock \showarticletitle{Deep learning-based software bug classification}.
\newblock \bibinfo{journal}{\emph{Information and Software Technology}}  \bibinfo{volume}{166} (\bibinfo{year}{2024}), \bibinfo{pages}{107350}.
\newblock


\bibitem[Mohsin et~al\mbox{.}(2022)]%
        {MOHSIN2022107711}
\bibfield{author}{\bibinfo{person}{Hufsa Mohsin}, \bibinfo{person}{Chongyang Shi}, \bibinfo{person}{Shufeng Hao}, {and} \bibinfo{person}{He Jiang}.} \bibinfo{year}{2022}\natexlab{}.
\newblock \showarticletitle{SPAN: A self-paced association augmentation and node embedding-based model for software bug classification and assignment}.
\newblock \bibinfo{journal}{\emph{Knowledge-Based Systems}}  \bibinfo{volume}{236} (\bibinfo{year}{2022}), \bibinfo{pages}{107711}.
\newblock
\showISSN{0950-7051}
\href{https://doi.org/10.1016/j.knosys.2021.107711}{doi:\nolinkurl{10.1016/j.knosys.2021.107711}}


\bibitem[Murphy and Cubranic(2004)]%
        {murphy2004automatic}
\bibfield{author}{\bibinfo{person}{G Murphy} {and} \bibinfo{person}{Davor Cubranic}.} \bibinfo{year}{2004}\natexlab{}.
\newblock \showarticletitle{Automatic bug triage using text categorization}. In \bibinfo{booktitle}{\emph{Proceedings of the sixteenth international conference on software engineering \& knowledge engineering}}. Citeseer, \bibinfo{pages}{1--6}.
\newblock


\bibitem[Naguib et~al\mbox{.}(2013)]%
        {naguib2013bug}
\bibfield{author}{\bibinfo{person}{Hoda Naguib}, \bibinfo{person}{Nitesh Narayan}, \bibinfo{person}{Bernd Br{\"u}gge}, {and} \bibinfo{person}{Dina Helal}.} \bibinfo{year}{2013}\natexlab{}.
\newblock \showarticletitle{Bug report assignee recommendation using activity profiles}. In \bibinfo{booktitle}{\emph{2013 10th Working Conference on Mining Software Repositories (MSR)}}. IEEE, \bibinfo{pages}{22--30}.
\newblock


\bibitem[Nagwani and Suri(2023)]%
        {nagwani2023artificial}
\bibfield{author}{\bibinfo{person}{Naresh~Kumar Nagwani} {and} \bibinfo{person}{Jasjit~S. Suri}.} \bibinfo{year}{2023}\natexlab{}.
\newblock \showarticletitle{An artificial intelligence framework on software bug triaging, technological evolution, and future challenges: A review}.
\newblock \bibinfo{journal}{\emph{International Journal of Information Management Data Insights}} \bibinfo{volume}{3}, \bibinfo{number}{1} (\bibinfo{year}{2023}), \bibinfo{pages}{100153}.
\newblock
\showISSN{2667-0968}
\href{https://doi.org/10.1016/j.jjimei.2022.100153}{doi:\nolinkurl{10.1016/j.jjimei.2022.100153}}


\bibitem[Nath et~al\mbox{.}(2021)]%
        {nath2021principal}
\bibfield{author}{\bibinfo{person}{Vaskar Nath}, \bibinfo{person}{David Sheldon}, {and} \bibinfo{person}{John Alphonso-Gibbs}.} \bibinfo{year}{2021}\natexlab{}.
\newblock \showarticletitle{Principal component analysis and entropy-based selection for the improvement of bug triage}. In \bibinfo{booktitle}{\emph{2021 20th IEEE International Conference on Machine Learning and Applications (ICMLA)}}. IEEE, \bibinfo{pages}{541--546}.
\newblock


\bibitem[Neysiani et~al\mbox{.}(2020)]%
        {neysiani2020efficient}
\bibfield{author}{\bibinfo{person}{Behzad~Soleimani Neysiani}, \bibinfo{person}{Seyed~Morteza Babamir}, {and} \bibinfo{person}{Masayoshi Aritsugi}.} \bibinfo{year}{2020}\natexlab{}.
\newblock \showarticletitle{Efficient feature extraction model for validation performance improvement of duplicate bug report detection in software bug triage systems}.
\newblock \bibinfo{journal}{\emph{Information and Software Technology}}  \bibinfo{volume}{126} (\bibinfo{year}{2020}), \bibinfo{pages}{106344}.
\newblock


\bibitem[Notaro et~al\mbox{.}(2021)]%
        {notaro2021survey}
\bibfield{author}{\bibinfo{person}{Paolo Notaro}, \bibinfo{person}{Jorge Cardoso}, {and} \bibinfo{person}{Michael Gerndt}.} \bibinfo{year}{2021}\natexlab{}.
\newblock \showarticletitle{A Survey of AIOps Methods for Failure Management}.
\newblock \bibinfo{journal}{\emph{ACM Trans. Intell. Syst. Technol.}} \bibinfo{volume}{12}, \bibinfo{number}{6}, Article \bibinfo{articleno}{81} (\bibinfo{date}{Nov.} \bibinfo{year}{2021}), \bibinfo{numpages}{45}~pages.
\newblock
\showISSN{2157-6904}
\href{https://doi.org/10.1145/3483424}{doi:\nolinkurl{10.1145/3483424}}


\bibitem[Pahins et~al\mbox{.}(2019)]%
        {pahins2019t}
\bibfield{author}{\bibinfo{person}{C{\'\i}cero Augusto De~Lara Pahins}, \bibinfo{person}{Fabr{\'\i}cio D'Morison}, \bibinfo{person}{Thiago~M Rocha}, \bibinfo{person}{Larissa~M Almeida}, \bibinfo{person}{Arthur~F Batista}, {and} \bibinfo{person}{Diego~F Souza}.} \bibinfo{year}{2019}\natexlab{}.
\newblock \showarticletitle{T-REC: Towards accurate bug triage for technical groups}. In \bibinfo{booktitle}{\emph{2019 18th IEEE International Conference on Machine Learning and Applications (ICMLA)}}. IEEE, \bibinfo{pages}{889--895}.
\newblock


\bibitem[Palomba et~al\mbox{.}(2015)]%
        {palomba2015user}
\bibfield{author}{\bibinfo{person}{Fabio Palomba}, \bibinfo{person}{Mario Linares-V{\'a}squez}, \bibinfo{person}{Gabriele Bavota}, \bibinfo{person}{Rocco Oliveto}, \bibinfo{person}{Massimiliano Di~Penta}, \bibinfo{person}{Denys Poshyvanyk}, {and} \bibinfo{person}{Andrea De~Lucia}.} \bibinfo{year}{2015}\natexlab{}.
\newblock \showarticletitle{User reviews matter! tracking crowdsourced reviews to support evolution of successful apps}. In \bibinfo{booktitle}{\emph{2015 IEEE international conference on software maintenance and evolution (ICSME)}}. IEEE, \bibinfo{pages}{291--300}.
\newblock


\bibitem[Panda and Nagwani(2022)]%
        {panda2022topic}
\bibfield{author}{\bibinfo{person}{Rama~Ranjan Panda} {and} \bibinfo{person}{Naresh~Kumar Nagwani}.} \bibinfo{year}{2022}\natexlab{}.
\newblock \showarticletitle{Topic modeling and intuitionistic fuzzy set-based approach for efficient software bug triaging}.
\newblock \bibinfo{journal}{\emph{Knowledge and Information Systems}} \bibinfo{volume}{64}, \bibinfo{number}{11} (\bibinfo{year}{2022}), \bibinfo{pages}{3081--3111}.
\newblock


\bibitem[Panda and Nagwani(2024)]%
        {panda2024software}
\bibfield{author}{\bibinfo{person}{Rama~Ranjan Panda} {and} \bibinfo{person}{Naresh~Kumar Nagwani}.} \bibinfo{year}{2024}\natexlab{}.
\newblock \showarticletitle{Software bug priority prediction technique based on intuitionistic fuzzy representation and class imbalance learning}.
\newblock \bibinfo{journal}{\emph{Knowledge and Information Systems}} \bibinfo{volume}{66}, \bibinfo{number}{3} (\bibinfo{year}{2024}), \bibinfo{pages}{2135--2164}.
\newblock


\bibitem[Park et~al\mbox{.}(2011)]%
        {park2011costriage}
\bibfield{author}{\bibinfo{person}{Jin-woo Park}, \bibinfo{person}{Mu-Woong Lee}, \bibinfo{person}{Jinhan Kim}, \bibinfo{person}{Seung-won Hwang}, {and} \bibinfo{person}{Sunghun Kim}.} \bibinfo{year}{2011}\natexlab{}.
\newblock \showarticletitle{Costriage: A cost-aware triage algorithm for bug reporting systems}. In \bibinfo{booktitle}{\emph{Proceedings of the AAAI conference on artificial intelligence}}, Vol.~\bibinfo{volume}{25}. \bibinfo{pages}{139--144}.
\newblock


\bibitem[Park et~al\mbox{.}(2016)]%
        {park2016cost}
\bibfield{author}{\bibinfo{person}{Jin-woo Park}, \bibinfo{person}{Mu-Woong Lee}, \bibinfo{person}{Jinhan Kim}, \bibinfo{person}{Seung-won Hwang}, {and} \bibinfo{person}{Sunghun Kim}.} \bibinfo{year}{2016}\natexlab{}.
\newblock \showarticletitle{Cost-aware triage ranking algorithms for bug reporting systems}.
\newblock \bibinfo{journal}{\emph{Knowledge and Information Systems}} \bibinfo{volume}{48}, \bibinfo{number}{3} (\bibinfo{year}{2016}), \bibinfo{pages}{679--705}.
\newblock


\bibitem[Patil(2025)]%
        {patil2025gitbugs}
\bibfield{author}{\bibinfo{person}{Avinash Patil}.} \bibinfo{year}{2025}\natexlab{}.
\newblock \showarticletitle{GitBugs: Bug Reports for Duplicate Detection, Retrieval Augmented Generation, Triage, and More}.
\newblock \bibinfo{journal}{\emph{arXiv preprint arXiv:2504.09651}} (\bibinfo{year}{2025}).
\newblock


\bibitem[Peng et~al\mbox{.}(2017)]%
        {peng2017improving}
\bibfield{author}{\bibinfo{person}{Xinyu Peng}, \bibinfo{person}{Pingyi Zhou}, \bibinfo{person}{Jin Liu}, {and} \bibinfo{person}{Xu Chen}.} \bibinfo{year}{2017}\natexlab{}.
\newblock \showarticletitle{Improving Bug Triage with Relevant Search.}. In \bibinfo{booktitle}{\emph{SEKE}}. \bibinfo{pages}{123--128}.
\newblock


\bibitem[Pham et~al\mbox{.}(2020)]%
        {pham2020deeptriage}
\bibfield{author}{\bibinfo{person}{Phuong Pham}, \bibinfo{person}{Vivek Jain}, \bibinfo{person}{Lukas Dauterman}, \bibinfo{person}{Justin Ormont}, {and} \bibinfo{person}{Navendu Jain}.} \bibinfo{year}{2020}\natexlab{}.
\newblock \showarticletitle{Deeptriage: Automated transfer assistance for incidents in cloud services}. In \bibinfo{booktitle}{\emph{Proceedings of the 26th ACM SIGKDD International Conference on Knowledge Discovery \& Data Mining}}. \bibinfo{pages}{3281--3289}.
\newblock


\bibitem[Phetrungnapha and Senivongse(2019)]%
        {phetrungnapha2019classification}
\bibfield{author}{\bibinfo{person}{Kittisak Phetrungnapha} {and} \bibinfo{person}{Twittie Senivongse}.} \bibinfo{year}{2019}\natexlab{}.
\newblock \showarticletitle{Classification of mobile application user reviews for generating tickets on issue tracking system}. In \bibinfo{booktitle}{\emph{2019 12th International Conference on Information \& Communication Technology and System (ICTS)}}. IEEE, \bibinfo{pages}{229--234}.
\newblock


\bibitem[Qian et~al\mbox{.}(2023)]%
        {qian2023survey}
\bibfield{author}{\bibinfo{person}{Cheng Qian}, \bibinfo{person}{Ming Zhang}, \bibinfo{person}{Yuanping Nie}, \bibinfo{person}{Shuaibing Lu}, {and} \bibinfo{person}{Huayang Cao}.} \bibinfo{year}{2023}\natexlab{}.
\newblock \showarticletitle{A survey on bug deduplication and triage methods from multiple points of view}.
\newblock \bibinfo{journal}{\emph{Applied Sciences}} \bibinfo{volume}{13}, \bibinfo{number}{15} (\bibinfo{year}{2023}), \bibinfo{pages}{8788}.
\newblock


\bibitem[Rahman et~al\mbox{.}(2009)]%
        {rahman2009optimized}
\bibfield{author}{\bibinfo{person}{Md~Mainur Rahman}, \bibinfo{person}{Guenther Ruhe}, {and} \bibinfo{person}{Thomas Zimmermann}.} \bibinfo{year}{2009}\natexlab{}.
\newblock \showarticletitle{Optimized assignment of developers for fixing bugs an initial evaluation for eclipse projects}. In \bibinfo{booktitle}{\emph{2009 3rd International Symposium on Empirical Software Engineering and Measurement}}. IEEE, \bibinfo{pages}{439--442}.
\newblock


\bibitem[Remil et~al\mbox{.}(2024)]%
        {remil2024aiops}
\bibfield{author}{\bibinfo{person}{Youcef Remil}, \bibinfo{person}{Anes Bendimerad}, \bibinfo{person}{Romain Mathonat}, {and} \bibinfo{person}{Mehdi Kaytoue}.} \bibinfo{year}{2024}\natexlab{}.
\newblock \showarticletitle{Aiops solutions for incident management: Technical guidelines and a comprehensive literature review}.
\newblock \bibinfo{journal}{\emph{arXiv preprint arXiv:2404.01363}} (\bibinfo{year}{2024}).
\newblock


\bibitem[Remil et~al\mbox{.}(2021)]%
        {remil2021interpretable}
\bibfield{author}{\bibinfo{person}{Youcef Remil}, \bibinfo{person}{Anes Bendimerad}, \bibinfo{person}{Marc Plantevit}, \bibinfo{person}{C{\'e}line Robardet}, {and} \bibinfo{person}{Mehdi Kaytoue}.} \bibinfo{year}{2021}\natexlab{}.
\newblock \showarticletitle{Interpretable summaries of black box incident triaging with subgroup discovery}. In \bibinfo{booktitle}{\emph{2021 IEEE 8TH international Conference on Data Science and Advanced Analytics (DSAA)}}. IEEE, \bibinfo{pages}{1--10}.
\newblock


\bibitem[Robertson-Steel(2006)]%
        {robertson2006evolution}
\bibfield{author}{\bibinfo{person}{Iain Robertson-Steel}.} \bibinfo{year}{2006}\natexlab{}.
\newblock \showarticletitle{Evolution of triage systems}.
\newblock \bibinfo{journal}{\emph{Emergency medicine journal}} \bibinfo{volume}{23}, \bibinfo{number}{2} (\bibinfo{year}{2006}), \bibinfo{pages}{154--155}.
\newblock


\bibitem[Sabor et~al\mbox{.}(2017)]%
        {sabor2017durfex}
\bibfield{author}{\bibinfo{person}{Korosh~Koochekian Sabor}, \bibinfo{person}{Abdelwahab Hamou-Lhadj}, {and} \bibinfo{person}{Alf Larsson}.} \bibinfo{year}{2017}\natexlab{}.
\newblock \showarticletitle{Durfex: a feature extraction technique for efficient detection of duplicate bug reports}. In \bibinfo{booktitle}{\emph{2017 IEEE international conference on software quality, reliability and security (QRS)}}. IEEE, \bibinfo{pages}{240--250}.
\newblock


\bibitem[Sadlek et~al\mbox{.}(2025)]%
        {sadlek2025severity}
\bibfield{author}{\bibinfo{person}{Luk{\'a}{\v{s}} Sadlek}, \bibinfo{person}{Muhammad~Mudassar Yamin}, \bibinfo{person}{Pavel {\v{C}}eleda}, {and} \bibinfo{person}{Basel Katt}.} \bibinfo{year}{2025}\natexlab{}.
\newblock \showarticletitle{Severity-based triage of cybersecurity incidents using kill chain attack graphs}.
\newblock \bibinfo{journal}{\emph{Journal of Information Security and Applications}}  \bibinfo{volume}{89} (\bibinfo{year}{2025}), \bibinfo{pages}{103956}.
\newblock


\bibitem[Sajedi~Badashian et~al\mbox{.}(2016)]%
        {sajedi2016crowdsourced}
\bibfield{author}{\bibinfo{person}{Ali Sajedi~Badashian}, \bibinfo{person}{Abram Hindle}, {and} \bibinfo{person}{Eleni Stroulia}.} \bibinfo{year}{2016}\natexlab{}.
\newblock \showarticletitle{Crowdsourced Bug Triaging: Leveraging Q{\&}A Platforms for Bug Assignment}. In \bibinfo{booktitle}{\emph{Fundamental Approaches to Software Engineering}}, \bibfield{editor}{\bibinfo{person}{Perdita Stevens} {and} \bibinfo{person}{Andrzej W{\k{a}}sowski}} (Eds.). \bibinfo{publisher}{Springer Berlin Heidelberg}, \bibinfo{address}{Berlin, Heidelberg}, \bibinfo{pages}{231--248}.
\newblock
\showISBNx{978-3-662-49665-7}


\bibitem[Sajedi-Badashian and Stroulia(2020a)]%
        {sajedi2020guidelines}
\bibfield{author}{\bibinfo{person}{Ali Sajedi-Badashian} {and} \bibinfo{person}{Eleni Stroulia}.} \bibinfo{year}{2020}\natexlab{a}.
\newblock \showarticletitle{Guidelines for evaluating bug-assignment research}.
\newblock \bibinfo{journal}{\emph{Journal of Software: Evolution and Process}} \bibinfo{volume}{32}, \bibinfo{number}{9} (\bibinfo{year}{2020}), \bibinfo{pages}{e2250}.
\newblock


\bibitem[Sajedi-Badashian and Stroulia(2020b)]%
        {sajedi2020vocabulary}
\bibfield{author}{\bibinfo{person}{Ali Sajedi-Badashian} {and} \bibinfo{person}{Eleni Stroulia}.} \bibinfo{year}{2020}\natexlab{b}.
\newblock \showarticletitle{Vocabulary and time based bug-assignment: A recommender system for open-source projects}.
\newblock \bibinfo{journal}{\emph{Software: Practice and Experience}} \bibinfo{volume}{50}, \bibinfo{number}{8} (\bibinfo{year}{2020}), \bibinfo{pages}{1539--1564}.
\newblock


\bibitem[Samir et~al\mbox{.}(2023)]%
        {samir2023improving}
\bibfield{author}{\bibinfo{person}{Mina Samir}, \bibinfo{person}{Nada Sherief}, {and} \bibinfo{person}{Walid Abdelmoez}.} \bibinfo{year}{2023}\natexlab{}.
\newblock \showarticletitle{Improving bug assignment and developer allocation in software engineering through interpretable machine learning models}.
\newblock \bibinfo{journal}{\emph{Computers}} \bibinfo{volume}{12}, \bibinfo{number}{7} (\bibinfo{year}{2023}), \bibinfo{pages}{128}.
\newblock


\bibitem[Sarkar et~al\mbox{.}(2019)]%
        {sarkar2019improving}
\bibfield{author}{\bibinfo{person}{Aindrila Sarkar}, \bibinfo{person}{Peter~C Rigby}, {and} \bibinfo{person}{B{\'e}la Bartalos}.} \bibinfo{year}{2019}\natexlab{}.
\newblock \showarticletitle{Improving bug triaging with high confidence predictions at ericsson}. In \bibinfo{booktitle}{\emph{2019 IEEE International Conference on Software Maintenance and Evolution (ICSME)}}. IEEE, \bibinfo{pages}{81--91}.
\newblock


\bibitem[Sepahvand et~al\mbox{.}(2023)]%
        {sepahvand2023using}
\bibfield{author}{\bibinfo{person}{Reza Sepahvand}, \bibinfo{person}{Reza Akbari}, \bibinfo{person}{Behnaz Jamasb}, \bibinfo{person}{Sattar Hashemi}, {and} \bibinfo{person}{Omid Boushehrian}.} \bibinfo{year}{2023}\natexlab{}.
\newblock \showarticletitle{Using word embedding and convolution neural network for bug triaging by considering design flaws}.
\newblock \bibinfo{journal}{\emph{Science of Computer Programming}}  \bibinfo{volume}{228} (\bibinfo{year}{2023}), \bibinfo{pages}{102945}.
\newblock


\bibitem[Servant and Jones(2012)]%
        {servant2012whosefault}
\bibfield{author}{\bibinfo{person}{Francisco Servant} {and} \bibinfo{person}{James~A. Jones}.} \bibinfo{year}{2012}\natexlab{}.
\newblock \showarticletitle{WhoseFault: Automatic developer-to-fault assignment through fault localization}. In \bibinfo{booktitle}{\emph{2012 34th International Conference on Software Engineering (ICSE)}}. \bibinfo{pages}{36--46}.
\newblock
\href{https://doi.org/10.1109/ICSE.2012.6227208}{doi:\nolinkurl{10.1109/ICSE.2012.6227208}}


\bibitem[Shao et~al\mbox{.}(2008)]%
        {shao2008efficient}
\bibfield{author}{\bibinfo{person}{Qihong Shao}, \bibinfo{person}{Yi Chen}, \bibinfo{person}{Shu Tao}, \bibinfo{person}{Xifeng Yan}, {and} \bibinfo{person}{Nikos Anerousis}.} \bibinfo{year}{2008}\natexlab{}.
\newblock \showarticletitle{Efficient ticket routing by resolution sequence mining}. In \bibinfo{booktitle}{\emph{Proceedings of the 14th ACM SIGKDD International Conference on Knowledge Discovery and Data Mining}} (Las Vegas, Nevada, USA) \emph{(\bibinfo{series}{KDD '08})}. \bibinfo{publisher}{Association for Computing Machinery}, \bibinfo{address}{New York, NY, USA}, \bibinfo{pages}{605–613}.
\newblock
\showISBNx{9781605581934}
\href{https://doi.org/10.1145/1401890.1401964}{doi:\nolinkurl{10.1145/1401890.1401964}}


\bibitem[Shokripour et~al\mbox{.}(2013)]%
        {shokripour2013so}
\bibfield{author}{\bibinfo{person}{Ramin Shokripour}, \bibinfo{person}{John Anvik}, \bibinfo{person}{Zarinah~M Kasirun}, {and} \bibinfo{person}{Sima Zamani}.} \bibinfo{year}{2013}\natexlab{}.
\newblock \showarticletitle{Why so complicated? simple term filtering and weighting for location-based bug report assignment recommendation}. In \bibinfo{booktitle}{\emph{2013 10th working conference on mining software repositories (MSR)}}. IEEE, \bibinfo{pages}{2--11}.
\newblock


\bibitem[Shokripour et~al\mbox{.}(2015)]%
        {shokripour2015time}
\bibfield{author}{\bibinfo{person}{Ramin Shokripour}, \bibinfo{person}{John Anvik}, \bibinfo{person}{Zarinah~M Kasirun}, {and} \bibinfo{person}{Sima Zamani}.} \bibinfo{year}{2015}\natexlab{}.
\newblock \showarticletitle{A time-based approach to automatic bug report assignment}.
\newblock \bibinfo{journal}{\emph{Journal of Systems and Software}}  \bibinfo{volume}{102} (\bibinfo{year}{2015}), \bibinfo{pages}{109--122}.
\newblock


\bibitem[Singh and Singh(2023)]%
        {singh2023empirical}
\bibfield{author}{\bibinfo{person}{Neetu Singh} {and} \bibinfo{person}{Sandeep~Kumar Singh}.} \bibinfo{year}{2023}\natexlab{}.
\newblock \showarticletitle{An Empirical Assessment of the Performance of Multi-Armed Bandits and Contextual Multi-Armed Bandits in Handling Cold-Start Bugs}. In \bibinfo{booktitle}{\emph{Proceedings of the 2023 Fifteenth International Conference on Contemporary Computing}}. \bibinfo{pages}{750--758}.
\newblock


\bibitem[Singh and Singh(2025)]%
        {singh2025navigating}
\bibfield{author}{\bibinfo{person}{Neetu Singh} {and} \bibinfo{person}{Sandeep~Kumar Singh}.} \bibinfo{year}{2025}\natexlab{}.
\newblock \showarticletitle{Navigating bug cold start with contextual multi-armed bandits: an enhanced approach to developer assignment in software bug repositories}.
\newblock \bibinfo{journal}{\emph{Automated Software Engineering}} \bibinfo{volume}{32}, \bibinfo{number}{2} (\bibinfo{year}{2025}), \bibinfo{pages}{39}.
\newblock


\bibitem[Su et~al\mbox{.}(2023)]%
        {su2023still}
\bibfield{author}{\bibinfo{person}{Yanqi Su}, \bibinfo{person}{Zheming Han}, \bibinfo{person}{Zhipeng Gao}, \bibinfo{person}{Zhenchang Xing}, \bibinfo{person}{Qinghua Lu}, {and} \bibinfo{person}{Xiwei Xu}.} \bibinfo{year}{2023}\natexlab{}.
\newblock \showarticletitle{Still confusing for bug-component triaging? Deep feature learning and ensemble setting to rescue}. In \bibinfo{booktitle}{\emph{2023 IEEE/ACM 31st International Conference on Program Comprehension (ICPC)}}. IEEE, \bibinfo{pages}{316--327}.
\newblock


\bibitem[Su et~al\mbox{.}(2021)]%
        {su2021reducing}
\bibfield{author}{\bibinfo{person}{Yanqi Su}, \bibinfo{person}{Zhenchang Xing}, \bibinfo{person}{Xin Peng}, \bibinfo{person}{Xin Xia}, \bibinfo{person}{Chong Wang}, \bibinfo{person}{Xiwei Xu}, {and} \bibinfo{person}{Liming Zhu}.} \bibinfo{year}{2021}\natexlab{}.
\newblock \showarticletitle{Reducing bug triaging confusion by learning from mistakes with a bug tossing knowledge graph}. In \bibinfo{booktitle}{\emph{2021 36th IEEE/ACM International Conference on Automated Software Engineering (ASE)}}. IEEE, \bibinfo{pages}{191--202}.
\newblock


\bibitem[Sun et~al\mbox{.}(2017)]%
        {sun2017enhancing}
\bibfield{author}{\bibinfo{person}{Xiaobing Sun}, \bibinfo{person}{Hui Yang}, \bibinfo{person}{Xin Xia}, {and} \bibinfo{person}{Bin Li}.} \bibinfo{year}{2017}\natexlab{}.
\newblock \showarticletitle{Enhancing developer recommendation with supplementary information via mining historical commits}.
\newblock \bibinfo{journal}{\emph{Journal of Systems and Software}}  \bibinfo{volume}{134} (\bibinfo{year}{2017}), \bibinfo{pages}{355--368}.
\newblock


\bibitem[Sun et~al\mbox{.}(2025)]%
        {sun2025tixfusion}
\bibfield{author}{\bibinfo{person}{Yongqian Sun}, \bibinfo{person}{Bowen Hao}, \bibinfo{person}{Xiaotian Wang}, \bibinfo{person}{Chenyu Zhao}, \bibinfo{person}{Yongxin Zhao}, \bibinfo{person}{Binpeng Shi}, \bibinfo{person}{Shenglin Zhang}, \bibinfo{person}{Qiao Ge}, \bibinfo{person}{Wenhu Li}, \bibinfo{person}{Hua Wei}, {and} \bibinfo{person}{Dan Pei}.} \bibinfo{year}{2025}\natexlab{}.
\newblock \showarticletitle{LLM-Augmented Ticket Aggregation for Low-cost Mobile OS Defect Resolution}. In \bibinfo{booktitle}{\emph{Proceedings of the 33rd ACM International Conference on the Foundations of Software Engineering}} (Clarion Hotel Trondheim, Trondheim, Norway) \emph{(\bibinfo{series}{FSE Companion '25})}. \bibinfo{publisher}{Association for Computing Machinery}, \bibinfo{address}{New York, NY, USA}, \bibinfo{pages}{215–226}.
\newblock
\showISBNx{9798400712760}
\href{https://doi.org/10.1145/3696630.3728547}{doi:\nolinkurl{10.1145/3696630.3728547}}


\bibitem[Sun et~al\mbox{.}(2024)]%
        {sun2024art}
\bibfield{author}{\bibinfo{person}{Yongqian Sun}, \bibinfo{person}{Binpeng Shi}, \bibinfo{person}{Mingyu Mao}, \bibinfo{person}{Minghua Ma}, \bibinfo{person}{Sibo Xia}, \bibinfo{person}{Shenglin Zhang}, {and} \bibinfo{person}{Dan Pei}.} \bibinfo{year}{2024}\natexlab{}.
\newblock \showarticletitle{Art: A unified unsupervised framework for incident management in microservice systems}. In \bibinfo{booktitle}{\emph{Proceedings of the 39th IEEE/ACM International Conference on Automated Software Engineering}}. \bibinfo{pages}{1183--1194}.
\newblock


\bibitem[Sureka et~al\mbox{.}(2015)]%
        {sureka2015decision}
\bibfield{author}{\bibinfo{person}{Ashish Sureka}, \bibinfo{person}{Himanshu Singh}, \bibinfo{person}{Manjunat Bagewadi}, \bibinfo{person}{Abhishek Mitra}, {and} \bibinfo{person}{Rohit Karanth}.} \bibinfo{year}{2015}\natexlab{}.
\newblock \showarticletitle{A Decision Support Platform for Guiding a Bug Triage for Resolver Recommendation Using Textual and Non-Textual Features.}. In \bibinfo{booktitle}{\emph{QuASoQ/WAWSE/CMCE@ APSEC}}. \bibinfo{pages}{23--30}.
\newblock


\bibitem[Tamrawi et~al\mbox{.}(2011)]%
        {tamrawi2011bugzie}
\bibfield{author}{\bibinfo{person}{Ahmed Tamrawi}, \bibinfo{person}{Tung~Thanh Nguyen}, \bibinfo{person}{Jafar~M. Al-Kofahi}, {and} \bibinfo{person}{Tien~N. Nguyen}.} \bibinfo{year}{2011}\natexlab{}.
\newblock \showarticletitle{Fuzzy set and cache-based approach for bug triaging}. In \bibinfo{booktitle}{\emph{Proceedings of the 19th ACM SIGSOFT Symposium and the 13th European Conference on Foundations of Software Engineering}} (Szeged, Hungary) \emph{(\bibinfo{series}{ESEC/FSE '11})}. \bibinfo{publisher}{Association for Computing Machinery}, \bibinfo{address}{New York, NY, USA}, \bibinfo{pages}{365–375}.
\newblock
\showISBNx{9781450304436}
\href{https://doi.org/10.1145/2025113.2025163}{doi:\nolinkurl{10.1145/2025113.2025163}}


\bibitem[Tian et~al\mbox{.}(2016)]%
        {tian2016learning}
\bibfield{author}{\bibinfo{person}{Yuan Tian}, \bibinfo{person}{Dinusha Wijedasa}, \bibinfo{person}{David Lo}, {and} \bibinfo{person}{Claire Le~Goues}.} \bibinfo{year}{2016}\natexlab{}.
\newblock \showarticletitle{Learning to rank for bug report assignee recommendation}. In \bibinfo{booktitle}{\emph{2016 IEEE 24th International Conference on Program Comprehension (ICPC)}}. IEEE, \bibinfo{pages}{1--10}.
\newblock


\bibitem[Uddin(2024)]%
        {uddin2024novel}
\bibfield{author}{\bibinfo{person}{KM Uddin}.} \bibinfo{year}{2024}\natexlab{}.
\newblock \emph{\bibinfo{title}{A NOVEL BUG TRIAGING STRATEGY USING DEVELOPER RECOMMENDATION AND LOAD BALANCING MODEL}}.
\newblock \bibinfo{thesistype}{Ph.\,D. Dissertation}. \bibinfo{school}{{\copyright} University of Dhaka}.
\newblock


\bibitem[Wang and Li(2021)]%
        {wang2021effective}
\bibfield{author}{\bibinfo{person}{Hongbing Wang} {and} \bibinfo{person}{Qi Li}.} \bibinfo{year}{2021}\natexlab{}.
\newblock \showarticletitle{Effective bug triage based on a hybrid neural network}. In \bibinfo{booktitle}{\emph{2021 28th Asia-Pacific Software Engineering Conference (APSEC)}}. IEEE, \bibinfo{pages}{82--91}.
\newblock


\bibitem[Wang et~al\mbox{.}(2025)]%
        {wang2025fixer}
\bibfield{author}{\bibinfo{person}{Rongcun Wang}, \bibinfo{person}{Xingyu Ji}, \bibinfo{person}{Yuan Tian}, \bibinfo{person}{Senlei Xu}, \bibinfo{person}{Xiaobing Sun}, {and} \bibinfo{person}{Shujuan Jiang}.} \bibinfo{year}{2025}\natexlab{}.
\newblock \showarticletitle{Fixer-level supervised contrastive learning for bug assignment}.
\newblock \bibinfo{journal}{\emph{Empirical Software Engineering}} \bibinfo{volume}{30}, \bibinfo{number}{3} (\bibinfo{year}{2025}), \bibinfo{pages}{76}.
\newblock


\bibitem[Wang et~al\mbox{.}(2024a)]%
        {wang2024empirical}
\bibfield{author}{\bibinfo{person}{Rongcun Wang}, \bibinfo{person}{Xingyu Ji}, \bibinfo{person}{Senlei Xu}, \bibinfo{person}{Yuan Tian}, \bibinfo{person}{Shujuan Jiang}, {and} \bibinfo{person}{Rubing Huang}.} \bibinfo{year}{2024}\natexlab{a}.
\newblock \showarticletitle{An empirical assessment of different word embedding and deep learning models for bug assignment}.
\newblock \bibinfo{journal}{\emph{Journal of Systems and Software}}  \bibinfo{volume}{210} (\bibinfo{year}{2024}), \bibinfo{pages}{111961}.
\newblock


\bibitem[Wang et~al\mbox{.}(2017)]%
        {wang2017app}
\bibfield{author}{\bibinfo{person}{Shance Wang}, \bibinfo{person}{Zhongjie Wang}, \bibinfo{person}{Xiaofei Xu}, {and} \bibinfo{person}{Quan~Z Sheng}.} \bibinfo{year}{2017}\natexlab{}.
\newblock \showarticletitle{App update patterns: How developers act on user reviews in mobile app stores}. In \bibinfo{booktitle}{\emph{International Conference on Service-Oriented Computing}}. Springer, \bibinfo{pages}{125--141}.
\newblock


\bibitem[Wang et~al\mbox{.}(2014)]%
        {wang2014fixercache}
\bibfield{author}{\bibinfo{person}{Song Wang}, \bibinfo{person}{Wen Zhang}, {and} \bibinfo{person}{Qing Wang}.} \bibinfo{year}{2014}\natexlab{}.
\newblock \showarticletitle{FixerCache: unsupervised caching active developers for diverse bug triage}. In \bibinfo{booktitle}{\emph{Proceedings of the 8th ACM/IEEE International Symposium on Empirical Software Engineering and Measurement}} (Torino, Italy) \emph{(\bibinfo{series}{ESEM '14})}. \bibinfo{publisher}{Association for Computing Machinery}, \bibinfo{address}{New York, NY, USA}, Article \bibinfo{articleno}{25}, \bibinfo{numpages}{10}~pages.
\newblock
\showISBNx{9781450327749}
\href{https://doi.org/10.1145/2652524.2652536}{doi:\nolinkurl{10.1145/2652524.2652536}}


\bibitem[Wang et~al\mbox{.}(2021)]%
        {wang2021fast}
\bibfield{author}{\bibinfo{person}{Yaohui Wang}, \bibinfo{person}{Guozheng Li}, \bibinfo{person}{Zijian Wang}, \bibinfo{person}{Yu Kang}, \bibinfo{person}{Yangfan Zhou}, \bibinfo{person}{Hongyu Zhang}, \bibinfo{person}{Feng Gao}, \bibinfo{person}{Jeffrey Sun}, \bibinfo{person}{Li Yang}, \bibinfo{person}{Pochian Lee}, {et~al\mbox{.}}} \bibinfo{year}{2021}\natexlab{}.
\newblock \showarticletitle{Fast outage analysis of large-scale production clouds with service correlation mining}. In \bibinfo{booktitle}{\emph{2021 IEEE/ACM 43rd International Conference on Software Engineering (ICSE)}}. IEEE, \bibinfo{pages}{885--896}.
\newblock


\bibitem[Wang et~al\mbox{.}(2024b)]%
        {wang2024comet}
\bibfield{author}{\bibinfo{person}{Zexin Wang}, \bibinfo{person}{Jianhui Li}, \bibinfo{person}{Minghua Ma}, \bibinfo{person}{Ze Li}, \bibinfo{person}{Yu Kang}, \bibinfo{person}{Chaoyun Zhang}, \bibinfo{person}{Chetan Bansal}, \bibinfo{person}{Murali Chintalapati}, \bibinfo{person}{Saravan Rajmohan}, \bibinfo{person}{Qingwei Lin}, \bibinfo{person}{Dongmei Zhang}, \bibinfo{person}{Changhua Pei}, {and} \bibinfo{person}{Gaogang Xie}.} \bibinfo{year}{2024}\natexlab{b}.
\newblock \showarticletitle{Large Language Models Can Provide Accurate and Interpretable Incident Triage}. In \bibinfo{booktitle}{\emph{2024 IEEE 35th International Symposium on Software Reliability Engineering (ISSRE)}}. \bibinfo{pages}{523--534}.
\newblock
\showISSN{2332-6549}
\href{https://doi.org/10.1109/ISSRE62328.2024.00056}{doi:\nolinkurl{10.1109/ISSRE62328.2024.00056}}


\bibitem[Wei et~al\mbox{.}(2018)]%
        {wei2018enhancing}
\bibfield{author}{\bibinfo{person}{Miaomiao Wei}, \bibinfo{person}{Shikai Guo}, \bibinfo{person}{Rong Chen}, {and} \bibinfo{person}{Jian Gao}.} \bibinfo{year}{2018}\natexlab{}.
\newblock \showarticletitle{Enhancing bug report assignment with an optimized reduction of training set}. In \bibinfo{booktitle}{\emph{International Conference on Knowledge Science, Engineering and Management}}. Springer, \bibinfo{pages}{36--47}.
\newblock


\bibitem[Wei et~al\mbox{.}(2025)]%
        {wei2025improving}
\bibfield{author}{\bibinfo{person}{Wei Wei}, \bibinfo{person}{Haojie Li}, \bibinfo{person}{Xinshuang Ren}, \bibinfo{person}{Feng Jiang}, \bibinfo{person}{Xu Yu}, \bibinfo{person}{Xingyu Gao}, {and} \bibinfo{person}{Junwei Du}.} \bibinfo{year}{2025}\natexlab{}.
\newblock \showarticletitle{Improving bug triage with the bug personalized tossing relationship}.
\newblock \bibinfo{journal}{\emph{Information and Software Technology}}  \bibinfo{volume}{179} (\bibinfo{year}{2025}), \bibinfo{pages}{107642}.
\newblock


\bibitem[Wohlin(2014)]%
        {wohlin2014guidelines}
\bibfield{author}{\bibinfo{person}{Claes Wohlin}.} \bibinfo{year}{2014}\natexlab{}.
\newblock \showarticletitle{Guidelines for snowballing in systematic literature studies and a replication in software engineering}. In \bibinfo{booktitle}{\emph{Proceedings of the 18th international conference on evaluation and assessment in software engineering}}. \bibinfo{pages}{1--10}.
\newblock


\bibitem[Wu et~al\mbox{.}(2022)]%
        {wu2022spatial}
\bibfield{author}{\bibinfo{person}{Hongrun Wu}, \bibinfo{person}{Yutao Ma}, \bibinfo{person}{Zhenglong Xiang}, \bibinfo{person}{Chen Yang}, {and} \bibinfo{person}{Keqing He}.} \bibinfo{year}{2022}\natexlab{}.
\newblock \showarticletitle{A spatial-temporal graph neural network framework for automated software bug triaging}.
\newblock \bibinfo{journal}{\emph{Knowledge-Based Systems}}  \bibinfo{volume}{241} (\bibinfo{year}{2022}), \bibinfo{pages}{108308}.
\newblock


\bibitem[Wu et~al\mbox{.}(2011)]%
        {wu2011drex}
\bibfield{author}{\bibinfo{person}{Wenjin Wu}, \bibinfo{person}{Wen Zhang}, \bibinfo{person}{Ye Yang}, {and} \bibinfo{person}{Qing Wang}.} \bibinfo{year}{2011}\natexlab{}.
\newblock \showarticletitle{Drex: Developer recommendation with k-nearest-neighbor search and expertise ranking}. In \bibinfo{booktitle}{\emph{2011 18th Asia-Pacific Software Engineering Conference}}. IEEE, \bibinfo{pages}{389--396}.
\newblock


\bibitem[Xi et~al\mbox{.}(2019)]%
        {xi2019bug}
\bibfield{author}{\bibinfo{person}{Sheng-Qu Xi}, \bibinfo{person}{Yuan Yao}, \bibinfo{person}{Xu-Sheng Xiao}, \bibinfo{person}{Feng Xu}, {and} \bibinfo{person}{Jian Lv}.} \bibinfo{year}{2019}\natexlab{}.
\newblock \showarticletitle{Bug triaging based on tossing sequence modeling}.
\newblock \bibinfo{journal}{\emph{Journal of Computer Science and Technology}} \bibinfo{volume}{34}, \bibinfo{number}{5} (\bibinfo{year}{2019}), \bibinfo{pages}{942--956}.
\newblock
\href{https://doi.org/10.1007/s11390-019-1953-5}{doi:\nolinkurl{10.1007/s11390-019-1953-5}}


\bibitem[Xia et~al\mbox{.}(2017)]%
        {xia2017topicminer}
\bibfield{author}{\bibinfo{person}{Xin Xia}, \bibinfo{person}{David Lo}, \bibinfo{person}{Ying Ding}, \bibinfo{person}{Jafar~M. Al-Kofahi}, \bibinfo{person}{Tien~N. Nguyen}, {and} \bibinfo{person}{Xinyu Wang}.} \bibinfo{year}{2017}\natexlab{}.
\newblock \showarticletitle{Improving Automated Bug Triaging with Specialized Topic Model}.
\newblock \bibinfo{journal}{\emph{IEEE Transactions on Software Engineering}} \bibinfo{volume}{43}, \bibinfo{number}{3} (\bibinfo{year}{2017}), \bibinfo{pages}{272--297}.
\newblock
\href{https://doi.org/10.1109/TSE.2016.2576454}{doi:\nolinkurl{10.1109/TSE.2016.2576454}}


\bibitem[Xia et~al\mbox{.}(2013)]%
        {xia2013accurate}
\bibfield{author}{\bibinfo{person}{Xin Xia}, \bibinfo{person}{David Lo}, \bibinfo{person}{Xinyu Wang}, {and} \bibinfo{person}{Bo Zhou}.} \bibinfo{year}{2013}\natexlab{}.
\newblock \showarticletitle{Accurate developer recommendation for bug resolution}. In \bibinfo{booktitle}{\emph{2013 20th Working Conference on Reverse Engineering (WCRE)}}. \bibinfo{pages}{72--81}.
\newblock
\href{https://doi.org/10.1109/WCRE.2013.6671282}{doi:\nolinkurl{10.1109/WCRE.2013.6671282}}


\bibitem[Xie et~al\mbox{.}(2013)]%
        {xie2013impact}
\bibfield{author}{\bibinfo{person}{Jialiang Xie}, \bibinfo{person}{Minghui Zhou}, {and} \bibinfo{person}{Audris Mockus}.} \bibinfo{year}{2013}\natexlab{}.
\newblock \showarticletitle{Impact of triage: a study of mozilla and gnome}. In \bibinfo{booktitle}{\emph{2013 ACM/IEEE International Symposium on Empirical Software Engineering and Measurement}}. IEEE, \bibinfo{pages}{247--250}.
\newblock


\bibitem[Xie et~al\mbox{.}(2012)]%
        {xie2012dretom}
\bibfield{author}{\bibinfo{person}{Xihao Xie}, \bibinfo{person}{Wen Zhang}, \bibinfo{person}{Ye Yang}, {and} \bibinfo{person}{Qing Wang}.} \bibinfo{year}{2012}\natexlab{}.
\newblock \showarticletitle{Dretom: Developer recommendation based on topic models for bug resolution}. In \bibinfo{booktitle}{\emph{Proceedings of the 8th international conference on predictive models in software engineering}}. \bibinfo{pages}{19--28}.
\newblock


\bibitem[Xu et~al\mbox{.}(2023)]%
        {xu2023method}
\bibfield{author}{\bibinfo{person}{Yang Xu}, \bibinfo{person}{Chao Liu}, \bibinfo{person}{Yong Li}, \bibinfo{person}{Qiaoluan Xie}, {and} \bibinfo{person}{Hyun-Deok Choi}.} \bibinfo{year}{2023}\natexlab{}.
\newblock \showarticletitle{A Method of Component Prediction for Crash Bug Reports Using Component-Based Features and Machine Learning}. In \bibinfo{booktitle}{\emph{2023 IEEE International Conference on Software Analysis, Evolution and Reengineering (SANER)}}. IEEE, \bibinfo{pages}{773--777}.
\newblock


\bibitem[Xuan et~al\mbox{.}(2014)]%
        {xuan2014towards}
\bibfield{author}{\bibinfo{person}{Jifeng Xuan}, \bibinfo{person}{He Jiang}, \bibinfo{person}{Yan Hu}, \bibinfo{person}{Zhilei Ren}, \bibinfo{person}{Weiqin Zou}, \bibinfo{person}{Zhongxuan Luo}, {and} \bibinfo{person}{Xindong Wu}.} \bibinfo{year}{2014}\natexlab{}.
\newblock \showarticletitle{Towards effective bug triage with software data reduction techniques}.
\newblock \bibinfo{journal}{\emph{IEEE transactions on knowledge and data engineering}} \bibinfo{volume}{27}, \bibinfo{number}{1} (\bibinfo{year}{2014}), \bibinfo{pages}{264--280}.
\newblock


\bibitem[Xuan et~al\mbox{.}(2010)]%
        {xuan2010automatic}
\bibfield{author}{\bibinfo{person}{Jifeng Xuan}, \bibinfo{person}{He Jiang}, \bibinfo{person}{Zhilei Ren}, \bibinfo{person}{Jun Yan}, {and} \bibinfo{person}{Zhongxuan Luo}.} \bibinfo{year}{2010}\natexlab{}.
\newblock \showarticletitle{Automatic Bug Triage using Semi-Supervised Text Classification.}. In \bibinfo{booktitle}{\emph{SEKE}}. \bibinfo{pages}{209--214}.
\newblock


\bibitem[Xuan et~al\mbox{.}(2012)]%
        {xuan2012developer}
\bibfield{author}{\bibinfo{person}{Jifeng Xuan}, \bibinfo{person}{He Jiang}, \bibinfo{person}{Zhilei Ren}, {and} \bibinfo{person}{Weiqin Zou}.} \bibinfo{year}{2012}\natexlab{}.
\newblock \showarticletitle{Developer prioritization in bug repositories}. In \bibinfo{booktitle}{\emph{2012 34th International Conference on Software Engineering (ICSE)}}. \bibinfo{pages}{25--35}.
\newblock
\href{https://doi.org/10.1109/ICSE.2012.6227209}{doi:\nolinkurl{10.1109/ICSE.2012.6227209}}


\bibitem[Yadav et~al\mbox{.}(2024)]%
        {yadav2024developer}
\bibfield{author}{\bibinfo{person}{Asmita Yadav}, \bibinfo{person}{Mohammed Baljon}, \bibinfo{person}{Shailendra Mishra}, \bibinfo{person}{Sandeep~Kumar Singh}, \bibinfo{person}{Sharad Saxena}, {and} \bibinfo{person}{Sunil~Kumar Sharma}.} \bibinfo{year}{2024}\natexlab{}.
\newblock \showarticletitle{Developer load balancing bug triage: Developed load balance}.
\newblock \bibinfo{journal}{\emph{Expert Systems}} \bibinfo{volume}{41}, \bibinfo{number}{6} (\bibinfo{year}{2024}), \bibinfo{pages}{e13006}.
\newblock


\bibitem[Yadav et~al\mbox{.}(2019)]%
        {yadav2019ranking}
\bibfield{author}{\bibinfo{person}{Asmita Yadav}, \bibinfo{person}{Sandeep~Kumar Singh}, {and} \bibinfo{person}{Jasjit~S Suri}.} \bibinfo{year}{2019}\natexlab{}.
\newblock \showarticletitle{Ranking of software developers based on expertise score for bug triaging}.
\newblock \bibinfo{journal}{\emph{Information and Software Technology}}  \bibinfo{volume}{112} (\bibinfo{year}{2019}), \bibinfo{pages}{1--17}.
\newblock


\bibitem[Yang et~al\mbox{.}(2014)]%
        {yang2014towards}
\bibfield{author}{\bibinfo{person}{Geunseok Yang}, \bibinfo{person}{Tao Zhang}, {and} \bibinfo{person}{Byungjeong Lee}.} \bibinfo{year}{2014}\natexlab{}.
\newblock \showarticletitle{Towards Semi-automatic Bug Triage and Severity Prediction Based on Topic Model and Multi-feature of Bug Reports}. In \bibinfo{booktitle}{\emph{2014 IEEE 38th Annual Computer Software and Applications Conference}}. \bibinfo{pages}{97--106}.
\newblock
\href{https://doi.org/10.1109/COMPSAC.2014.16}{doi:\nolinkurl{10.1109/COMPSAC.2014.16}}


\bibitem[Yu et~al\mbox{.}(2024)]%
        {yu2024survey}
\bibfield{author}{\bibinfo{person}{Qingyang Yu}, \bibinfo{person}{Nengwen Zhao}, \bibinfo{person}{Mingjie Li}, \bibinfo{person}{Zeyan Li}, \bibinfo{person}{Honglin Wang}, \bibinfo{person}{Wenchi Zhang}, \bibinfo{person}{Kaixin Sui}, {and} \bibinfo{person}{Dan Pei}.} \bibinfo{year}{2024}\natexlab{}.
\newblock \showarticletitle{A survey on intelligent management of alerts and incidents in IT services}.
\newblock \bibinfo{journal}{\emph{Journal of Network and Computer Applications}}  \bibinfo{volume}{224} (\bibinfo{year}{2024}), \bibinfo{pages}{103842}.
\newblock


\bibitem[Yu et~al\mbox{.}(2021)]%
        {yu2021bug}
\bibfield{author}{\bibinfo{person}{Xu Yu}, \bibinfo{person}{Fayang Wan}, \bibinfo{person}{Junwei Du}, \bibinfo{person}{Feng Jiang}, \bibinfo{person}{Lantian Guo}, {and} \bibinfo{person}{Junyu Lin}.} \bibinfo{year}{2021}\natexlab{}.
\newblock \showarticletitle{Bug triage model considering cooperative and sequential relationship}. In \bibinfo{booktitle}{\emph{International Conference on Wireless Algorithms, Systems, and Applications}}. Springer, \bibinfo{pages}{160--172}.
\newblock


\bibitem[Yu et~al\mbox{.}(2025)]%
        {yu2021triangle}
\bibfield{author}{\bibinfo{person}{Zhaoyang Yu}, \bibinfo{person}{Minghua Ma}, \bibinfo{person}{Xiaoyu Feng}, \bibinfo{person}{Ruomeng Ding}, \bibinfo{person}{Chaoyun Zhang}, \bibinfo{person}{Ze Li}, \bibinfo{person}{Merali Chintalapati}, \bibinfo{person}{Xuchao Zhang}, \bibinfo{person}{Rujia Wang}, \bibinfo{person}{Chetan Bansal}, \bibinfo{person}{Sarvan Rajmohan}, \bibinfo{person}{Qingwei Lin}, \bibinfo{person}{Shenglin Zhang}, \bibinfo{person}{Changhua Pei}, {and} \bibinfo{person}{Dan Pei}.} \bibinfo{year}{2025}\natexlab{}.
\newblock \showarticletitle{Triangle: Empowering Incident Triage with Multi-LLM-Agents}. In \bibinfo{booktitle}{\emph{Proceedings of the 33rd ACM International Conference on the Foundations of Software Engineering}}. ACM.
\newblock


\bibitem[Zeng et~al\mbox{.}(2017)]%
        {zeng2017knowledge}
\bibfield{author}{\bibinfo{person}{Chunqiu Zeng}, \bibinfo{person}{Wubai Zhou}, \bibinfo{person}{Tao Li}, \bibinfo{person}{Larisa Shwartz}, {and} \bibinfo{person}{Genady~Ya Grabarnik}.} \bibinfo{year}{2017}\natexlab{}.
\newblock \showarticletitle{Knowledge guided hierarchical multi-label classification over ticket data}.
\newblock \bibinfo{journal}{\emph{IEEE Transactions on Network and Service Management}} \bibinfo{volume}{14}, \bibinfo{number}{2} (\bibinfo{year}{2017}), \bibinfo{pages}{246--260}.
\newblock


\bibitem[Zhang et~al\mbox{.}(2025a)]%
        {zhang2025allhands}
\bibfield{author}{\bibinfo{person}{Chaoyun Zhang}, \bibinfo{person}{Zicheng Ma}, \bibinfo{person}{Yuhao Wu}, \bibinfo{person}{Shilin He}, \bibinfo{person}{Si Qin}, \bibinfo{person}{Minghua Ma}, \bibinfo{person}{Xiaoting Qin}, \bibinfo{person}{Yu Kang}, \bibinfo{person}{Yuyi Liang}, \bibinfo{person}{Xiaoyu Gou}, {et~al\mbox{.}}} \bibinfo{year}{2025}\natexlab{a}.
\newblock \showarticletitle{Allhands: Ask me anything on large-scale verbatim feedback via large language models}. In \bibinfo{booktitle}{\emph{2025 IEEE 41st International Conference on Data Engineering (ICDE)}}. IEEE, \bibinfo{pages}{43--57}.
\newblock


\bibitem[Zhang et~al\mbox{.}(2020)]%
        {zhang2020machine}
\bibfield{author}{\bibinfo{person}{Jie~M Zhang}, \bibinfo{person}{Mark Harman}, \bibinfo{person}{Lei Ma}, {and} \bibinfo{person}{Yang Liu}.} \bibinfo{year}{2020}\natexlab{}.
\newblock \showarticletitle{Machine learning testing: Survey, landscapes and horizons}.
\newblock \bibinfo{journal}{\emph{IEEE Transactions on Software Engineering}} \bibinfo{volume}{48}, \bibinfo{number}{1} (\bibinfo{year}{2020}), \bibinfo{pages}{1--36}.
\newblock


\bibitem[Zhang et~al\mbox{.}(2018)]%
        {zhang2018empirical}
\bibfield{author}{\bibinfo{person}{Li Zhang}, \bibinfo{person}{Jia-Hao Tian}, \bibinfo{person}{Jing Jiang}, \bibinfo{person}{Yi-Jun Liu}, \bibinfo{person}{Meng-Yuan Pu}, {and} \bibinfo{person}{Tao Yue}.} \bibinfo{year}{2018}\natexlab{}.
\newblock \showarticletitle{Empirical research in software engineering—a literature survey}.
\newblock \bibinfo{journal}{\emph{Journal of Computer Science and Technology}}  \bibinfo{volume}{33} (\bibinfo{year}{2018}), \bibinfo{pages}{876--899}.
\newblock


\bibitem[Zhang et~al\mbox{.}(2016a)]%
        {zhang2016towards}
\bibfield{author}{\bibinfo{person}{Tao Zhang}, \bibinfo{person}{Jiachi Chen}, \bibinfo{person}{Geunseok Yang}, \bibinfo{person}{Byungjeong Lee}, {and} \bibinfo{person}{Xiapu Luo}.} \bibinfo{year}{2016}\natexlab{a}.
\newblock \showarticletitle{Towards more accurate severity prediction and fixer recommendation of software bugs}.
\newblock \bibinfo{journal}{\emph{Journal of Systems and Software}}  \bibinfo{volume}{117} (\bibinfo{year}{2016}), \bibinfo{pages}{166--184}.
\newblock


\bibitem[Zhang et~al\mbox{.}(2016b)]%
        {zhang2016survey}
\bibfield{author}{\bibinfo{person}{Tao Zhang}, \bibinfo{person}{He Jiang}, \bibinfo{person}{Xiapu Luo}, {and} \bibinfo{person}{Alvin~T.S. Chan}.} \bibinfo{year}{2016}\natexlab{b}.
\newblock \showarticletitle{A Literature Review of Research in Bug Resolution: Tasks, Challenges and Future Directions}.
\newblock \bibinfo{journal}{\emph{Comput. J.}} \bibinfo{volume}{59}, \bibinfo{number}{5} (\bibinfo{date}{05} \bibinfo{year}{2016}), \bibinfo{pages}{741--773}.
\newblock
\showISSN{0010-4620}
\href{https://doi.org/10.1093/comjnl/bxv114}{doi:\nolinkurl{10.1093/comjnl/bxv114}}


\bibitem[Zhang et~al\mbox{.}(2014)]%
        {zhang2014novel}
\bibfield{author}{\bibinfo{person}{Tao Zhang}, \bibinfo{person}{Geunseok Yang}, \bibinfo{person}{Byungjeong Lee}, {and} \bibinfo{person}{Eng~Keong Lua}.} \bibinfo{year}{2014}\natexlab{}.
\newblock \showarticletitle{A novel developer ranking algorithm for automatic bug triage using topic model and developer relations}. In \bibinfo{booktitle}{\emph{2014 21st Asia-Pacific Software Engineering Conference}}, Vol.~\bibinfo{volume}{1}. IEEE, \bibinfo{pages}{223--230}.
\newblock


\bibitem[Zhang(2020)]%
        {zhang2020efficient}
\bibfield{author}{\bibinfo{person}{Wei Zhang}.} \bibinfo{year}{2020}\natexlab{}.
\newblock \showarticletitle{Efficient bug triage for industrial environments}. In \bibinfo{booktitle}{\emph{2020 IEEE International Conference on Software Maintenance and Evolution (ICSME)}}. IEEE, \bibinfo{pages}{727--735}.
\newblock


\bibitem[Zhang et~al\mbox{.}(2016c)]%
        {zhang2016ksap}
\bibfield{author}{\bibinfo{person}{Wen Zhang}, \bibinfo{person}{Song Wang}, {and} \bibinfo{person}{Qing Wang}.} \bibinfo{year}{2016}\natexlab{c}.
\newblock \showarticletitle{KSAP: An approach to bug report assignment using KNN search and heterogeneous proximity}.
\newblock \bibinfo{journal}{\emph{Information and software technology}}  \bibinfo{volume}{70} (\bibinfo{year}{2016}), \bibinfo{pages}{68--84}.
\newblock


\bibitem[Zhang et~al\mbox{.}(2025b)]%
        {zhang2025btal}
\bibfield{author}{\bibinfo{person}{Yanmei Zhang}, \bibinfo{person}{Yuhang Sun}, \bibinfo{person}{Yi Shi}, \bibinfo{person}{Shujuan Jiang}, {and} \bibinfo{person}{Guan Yuan}.} \bibinfo{year}{2025}\natexlab{b}.
\newblock \showarticletitle{BTAL: An imbalance software bug report triage approach based on BERT-TextCNN}.
\newblock \bibinfo{journal}{\emph{Information and Software Technology}}  \bibinfo{volume}{183} (\bibinfo{year}{2025}), \bibinfo{pages}{107731}.
\newblock


\bibitem[Zhao et~al\mbox{.}(2020a)]%
        {zhao2020understanding}
\bibfield{author}{\bibinfo{person}{Nengwen Zhao}, \bibinfo{person}{Junjie Chen}, \bibinfo{person}{Xiao Peng}, \bibinfo{person}{Honglin Wang}, \bibinfo{person}{Xinya Wu}, \bibinfo{person}{Yuanzong Zhang}, \bibinfo{person}{Zikai Chen}, \bibinfo{person}{Xiangzhong Zheng}, \bibinfo{person}{Xiaohui Nie}, \bibinfo{person}{Gang Wang}, \bibinfo{person}{Yong Wu}, \bibinfo{person}{Fang Zhou}, \bibinfo{person}{Wenchi Zhang}, \bibinfo{person}{Kaixin Sui}, {and} \bibinfo{person}{Dan Pei}.} \bibinfo{year}{2020}\natexlab{a}.
\newblock \showarticletitle{Understanding and handling alert storm for online service systems}. In \bibinfo{booktitle}{\emph{Proceedings of the ACM/IEEE 42nd International Conference on Software Engineering: Software Engineering in Practice}} (Seoul, South Korea) \emph{(\bibinfo{series}{ICSE-SEIP '20})}. \bibinfo{publisher}{Association for Computing Machinery}, \bibinfo{address}{New York, NY, USA}, \bibinfo{pages}{162–171}.
\newblock
\showISBNx{9781450371230}
\href{https://doi.org/10.1145/3377813.3381363}{doi:\nolinkurl{10.1145/3377813.3381363}}


\bibitem[Zhao et~al\mbox{.}(2020b)]%
        {zhao2020alertrank}
\bibfield{author}{\bibinfo{person}{Nengwen Zhao}, \bibinfo{person}{Panshi Jin}, \bibinfo{person}{Lixin Wang}, \bibinfo{person}{Xiaoqin Yang}, \bibinfo{person}{Rong Liu}, \bibinfo{person}{Wenchi Zhang}, \bibinfo{person}{Kaixin Sui}, {and} \bibinfo{person}{Dan Pei}.} \bibinfo{year}{2020}\natexlab{b}.
\newblock \showarticletitle{Automatically and Adaptively Identifying Severe Alerts for Online Service Systems}. In \bibinfo{booktitle}{\emph{IEEE INFOCOM 2020 - IEEE Conference on Computer Communications}}. \bibinfo{pages}{2420--2429}.
\newblock
\href{https://doi.org/10.1109/INFOCOM41043.2020.9155219}{doi:\nolinkurl{10.1109/INFOCOM41043.2020.9155219}}


\bibitem[Zhao et~al\mbox{.}(2019)]%
        {zhao2019unified}
\bibfield{author}{\bibinfo{person}{Yuan Zhao}, \bibinfo{person}{Tieke He}, {and} \bibinfo{person}{Zhenyu Chen}.} \bibinfo{year}{2019}\natexlab{}.
\newblock \showarticletitle{A unified framework for bug report assignment}.
\newblock \bibinfo{journal}{\emph{International Journal of Software Engineering and Knowledge Engineering}} \bibinfo{volume}{29}, \bibinfo{number}{04} (\bibinfo{year}{2019}), \bibinfo{pages}{607--628}.
\newblock


\bibitem[Zheng et~al\mbox{.}(2019)]%
        {zheng2019ifeedback}
\bibfield{author}{\bibinfo{person}{Wujie Zheng}, \bibinfo{person}{Haochuan Lu}, \bibinfo{person}{Yangfan Zhou}, \bibinfo{person}{Jianming Liang}, \bibinfo{person}{Haibing Zheng}, {and} \bibinfo{person}{Yuetang Deng}.} \bibinfo{year}{2019}\natexlab{}.
\newblock \showarticletitle{iFeedback: Exploiting user feedback for real-time issue detection in large-scale online service systems}. In \bibinfo{booktitle}{\emph{2019 34th IEEE/ACM International Conference on Automated Software Engineering (ASE)}}. IEEE, \bibinfo{pages}{352--363}.
\newblock


\bibitem[Zou et~al\mbox{.}(2011)]%
        {zou2011towards}
\bibfield{author}{\bibinfo{person}{Weiqin Zou}, \bibinfo{person}{Yan Hu}, \bibinfo{person}{Jifeng Xuan}, {and} \bibinfo{person}{He Jiang}.} \bibinfo{year}{2011}\natexlab{}.
\newblock \showarticletitle{Towards training set reduction for bug triage}. In \bibinfo{booktitle}{\emph{2011 IEEE 35th annual computer software and applications conference}}. IEEE, \bibinfo{pages}{576--581}.
\newblock


\bibitem[Zou et~al\mbox{.}(2020)]%
        {zou2020how}
\bibfield{author}{\bibinfo{person}{Weiqin Zou}, \bibinfo{person}{David Lo}, \bibinfo{person}{Zhenyu Chen}, \bibinfo{person}{Xin Xia}, \bibinfo{person}{Yang Feng}, {and} \bibinfo{person}{Baowen Xu}.} \bibinfo{year}{2020}\natexlab{}.
\newblock \showarticletitle{How Practitioners Perceive Automated Bug Report Management Techniques}.
\newblock \bibinfo{journal}{\emph{IEEE Transactions on Software Engineering}} \bibinfo{volume}{46}, \bibinfo{number}{8} (\bibinfo{year}{2020}), \bibinfo{pages}{836--862}.
\newblock
\href{https://doi.org/10.1109/TSE.2018.2870414}{doi:\nolinkurl{10.1109/TSE.2018.2870414}}


\end{thebibliography}


\end{document}